\documentclass[]{aa} 
\usepackage{graphicx}
\usepackage{txfonts}
\usepackage[pdfstartview=FitH,      
            breaklinks=true,        
            bookmarksopen=true,     
            bookmarksnumbered=true, 
            ]{hyperref}
\usepackage{aas_macros}
\usepackage{wasysym} 
\usepackage{subfigure}
\usepackage{placeins}
\usepackage{pdflscape}
\usepackage{xcolor}
\usepackage[version=4]{mhchem}
\newcommand{\comment}[1]{}
\usepackage{ulem}
\usepackage{multirow}
\usepackage{makecell}
\usepackage{lscape}

\definecolor{dgreen}{rgb}{0.05,0.5,0.1}
\definecolor{burgundy}{rgb}{0.50,0.00,0.13}

\begin{document}

   \title{The Atmospheres of Rocky Exoplanets}

   \subtitle{II. Influence of surface composition on the diversity of cloud condensates}

   \author{O. Herbort\inst{1,2,3,4}
          \and
          P. Woitke\inst{1,2,3}
          \and
          Ch. Helling\inst{1,2,3,5}
          \and
          A. L. Zerkle\inst{2,4}
          }

   \institute{
Space Research Institute, Austrian Academy of Sciences, Schmiedlstrasse 6, A-8042 Graz, Austria
        \and
Centre for Exoplanet Science, University of St Andrews, North Haugh, St Andrews, KY169SS, UK\\ \email{oh35@st-andrews.ac.uk}
         \and
SUPA, School of Physics \& Astronomy, University of St Andrews, North Haugh, St Andrews, KY169SS, UK
		\and
School of Earth \& Environmental Sciences, University of St Andrews, Irvine Building, St Andrews, KY16 9AL, UK
		\and
TU Graz, Fakult\"at f\"ur Mathematik, Physik und Geod\"asie, Petersgasse 16, A-8010 Graz, Austria
}
   \date{Received 25 06, 2021; accepted 26 11, 2021}

  \abstract
 {Clouds are an integral part of planetary atmospheres, with most planets hosting clouds.
 The understanding of not only the formation, but also the composition of clouds is crucial to the understanding of future observations.
 As observations of the planet's surface will remain very difficult, it is essential to link the observable high atmosphere gas and cloud composition to the surface conditions.\\ 
 We present a fast and simple chemical equilibrium model for the troposphere of rocky exoplanets, which is in chemical and phase equilibrium with the crust.
 The hydrostatic equilibrium atmosphere is built from bottom to top.
 In each atmospheric layer chemical equilibrium is solved and all thermally stable condensates are removed, depleting the atmosphere above in the effected elements.
 These removed condensates build an upper limit for cloud formation and can be separated into high and low temperature condensates.\\
The most important cloud condensates for $1000\,\text{K}\gtrsim T_\mathrm{gas}\gtrsim400\,$K are \ce{KCl}[s], \ce{NaCl}[s], \ce{FeS}[s], \ce{FeS2}[s], \ce{FeO}[s], \ce{Fe2O3}[s], and \ce{Fe3O4}[s].
 For $T_\mathrm{gas}\lesssim400\,$K \ce{H2O}[l,s], \ce{C}[s], \ce{NH3}[s], \ce{NH4Cl}[s], and \ce{NH4SH}[s] are thermally stable, while for even lower temperatures of $T_\mathrm{gas}\lesssim150\,$K \ce{CO2}[s], \ce{CH4}[s], \ce{NH3}[s], and \ce{H2S}[s] become stable.
 The inclusion of clouds with trace abundances results in the thermal stability of a total of 72 condensates for atmospheres with the different surface conditions 
 ($300\,\mathrm{K}\leq T_\mathrm{surf}\leq1000\,\mathrm{K}$ and $p_\mathrm{surf}=1\,\mathrm{bar}, 100\,\mathrm{bar}$).
 The different cloud condensates are not independent of each other, but follow sequences of condensation, which are robust against changes in crust composition, surface pressure, and surface temperature.
 Independent of the existence of water as a crust condensate \ce{H2O}[l,s] is a thermally stable cloud condensate for all investigated elemental abundances.
 However, the water cloud base depends on the hydration level of the crust.
 Therefore, the detection of water condensates alone does not necessarily imply stable water on the surface, even if the temperature could allow for water condensation.}

   \maketitle
%
\section{Introduction}\label{sec:Intro}
The surfaces of rocky planets, moons, and exoplanets show a remarkable diversity of physical and chemical conditions.
Surfaces range from bare rocks (e.g. Mercury or LHS~3844~b \citep{Kreidberg2019}), frozen or liquid water (e.g. Europa or Earth), and magma oceans, covering the entire planet during their formation \citep{Nikolaou2019, Elkins-Tanton2012} or covering only the dayside of highly irradiated, tidally locked planets \citep[magma ponds][]{Kite2016}.
Examples would be exoplanets like 55 Cnc e or CoRoT 7b \citep{2009A&A...506..287L, Leger2011}.
Although there is no rocky planet larger than Earth in our solar system, such planets, referred to as Super Earths, have been detected in multiple stellar systems at various distances from the host star \citep[e.g.][]{Dittmann2017, Ribas2018, Bluhm2020, Stock2020}.
Besides these super Earths, planets with mass, radius and stellar irradiation more similar to Earth like Proxima Centauri b \citep{Anglada-Escude2016} or the seven rocky planets in the Trappist 1 system \citep{Gillon2017} have been detected.
Some of these (e.g. Trappist 1 d, e, f, and g) might allow for liquid water on their surface.

The composition of the surface rocks on Earth alone is varying substantially from mafic (magnesium and iron rich, e.g. basalts) to felsic (rich in lighter elements, like Si, e.g. granite), depending on location.
This elemental composition is not only dependent on the planet itself, but also on the stellar system.
Elemental ratios in the near by stars are varying by orders of magnitude with respect to the solar values \citep{Hinkel2014}.
Therefore a substantial variation of the composition of the surface rocks for exoplanets can be expected.
However, through outgassing and deposition, the surface conditions of rocky planets are coupled to the gas composition and cloud formation in the atmosphere, which can be observed \citep[see][]{Grenfell2020}.

Present day observational techniques do not allow observations of the surfaces of rocky exoplanets in more detail than a large-scale temperature map \citep[see e.g.][]{Demory2016a, Hammond2017}.
Furthermore, global aspects of planets like the presence of a cloud cover on GJ~1214~b \citep{Kreidberg2014} or the ruling out of a hydrogen dominated extended atmosphere for the two innermost planets of the Trappist 1 system \citep{DeWit2016} are possible.
Transmission spectroscopy can provide insights into the optically thin part of the atmosphere.
This allows the detection of gas phase abundances as well as some cloud properties from absorption or scattering features \citep{Samra2020}.
This is, however, currently only possible for a very limited number of rocky planets, due to their small size.
Tentative detections of the gas phase molecules HCN \citep[55 Cnc e][]{Tsiaras2016} and \ce{H2O} \citep[LHS~1140~b][]{Edwards2020} have also been suggested.
For gas giants, the situation is different and various atoms and molecules have been detected: 
He \citep{Nortmann2018},
Na \citep{Seidel2019},
K and Li \citep{Chen2018},
CO \citep{Barman2015, Sheppard2017},
\ce{CH4} \citep{Barman2015},
\ce{H2O} \citep{Barman2015, Alonso-Floriano2019},
\ce{HCN}, \ce{NH3}, \ce{C2H2} \citep{Giacobbe2021}
Fe and Ti \citep{Hoeijmakers2018},
TiO \citep{2021Chen}.

The James Webb Space Telescope (JWST) \citep{Gardner2006} will enable the detection of a handful of atmospheric molecules by a sequence of repeated transit observations for rocky exoplanets \citep[see e.g.][]{Komacek2020, Wunderlich2020a}.
Planned space telescopes like ARIEL \citep{Tinetti2018}, LUVOIR \citep{Bolcar2016} and HabEx \citep{Stahl2020} are designed to characterise atmospheres of rocky exoplanets in the habitable zone.
The 30\,m class telescopes, like the E-ELT \citep{Gilmozzi2007}, will also provide the opportunity to observe some atmospheric molecules \citep[see ][]{2019BAAS...51c.162L}.
These instruments may allow the combined detection of atmospheric composition, and strict limits on surface temperature and surface pressure, which could result in the detection of liquid water on exoplanets and potentially a planet that could be habitable to life \textit{as we know it}.

Emission spectra can in principle provide direct clues about the surface composition \citep{Hu2012a, Mansfield2019}, in particular if polarimetry was used.
However, this will be strongly affected by clouds being present \citep{Rossi2017, Rossi2018}.
In case of full cloud coverage, like on Venus, the view on the surface will be obstructed, and hence information can only be deduced from the top of the cloud layer, rather than the planet surface.
Therefore it is not only important to better understand the fundamental physics of cloud formation \citep[see][]{Helling2019}, but also in how far the chemical composition of the clouds can be linked to the surface composition and surface conditions ($p_\text{surf}, T_\text{surf}$).

Atmospheres of hot rocky planets have been modeled as vaporised rock compositions \citep[e.g.][]{Miguel2011, Schaefer2012, Kite2016}. \citet{Thompson2021} present experimental compositions of gas outgassed from heated  chondritic material.
For hot rocky planets, the chemical composition of the clouds can drastically differ from the clouds known in our solar system, with models suggesting a composition of condensed minerals \citep[e.g][]{Mahapatra2017}.
These minerals are also found to contribute to the cloud composition of highly irradiated gas giants \citep[e.g][]{Ackerman2001, Wakeford2015, Helling2020, Woitke2020, Samra2020}.
The interaction between the interior of a rocky planet and its atmosphere is a matter of recent research especially focusing on the atmospheres of young rocky exoplanets in contact with magma oceans and volcanic outgassing \citep[e.g.][]{2014E&PSL.403..307G, Bower2019, Katyal2019, Ortenzi2020, Graham2020, Lichtenberg2021}.

In this work, we aim to better understand the chemical diversity of possible cloud material compositions under the varying element abundances in the atmospheres of rocky exoplanets.
We want to establish a link between the planetary surface and the cloud material composition in the atmosphere above.
Although this relation will be degenerate (see Appendix~\ref{sec:Degen}), it will be a step towards constraining the surface conditions with future observations of atmospheric and cloud composition.
We will focus on the lower atmosphere, the troposphere, where the atmospheric gas interacts chemically with the planet surface, and is then mixed upwards by convection to form the clouds.

First, we introduce our atmospheric model in Sect.~\ref{sec:atmosphere_model}, before we show the results of one example model in Sect.~\ref{ssec:BSEmodel}.
We describe how the resulting cloud materials depend on the assumed crust element abundances in Sect.~\ref{ssec:elementabundances} and what that implies for the atmospheric gas composition in Sect.~\ref{ssec:atmo}.
The link between water clouds and water as a stable condensate on the planet surface is discussed in Sect.~\ref{ssec:surfacewater}.
The impact of surface pressure on the results is shown in Sect.~\ref{ssec:press}.
In Sect.~\ref{sec:planets}, we discuss the implications of this model to planets that will become observable in the near future.
We conclude with a summary of our findings in Sect.~\ref{sec:discussion}.

\section{Method: Atmospheric model}\label{sec:atmosphere_model}
In the work presented here, we develop a model for the lower atmosphere (troposphere) linked to the surface of a rocky planet. 
That is, a planet with an overall density consistent with a rocky body ($\rho \approx\rm 5\,g/cm^3$) and a surface which can be either solid or molten.
The atmosphere and surface are considered to be in chemical and phase equilibrium and the atmosphere is build from bottom to top based on the surface atmosphere interaction.

\subsection{Polytropic atmosphere}\label{ssec:polyatm}
\begin{figure}
\centering
\includegraphics[width = 1.0\linewidth]{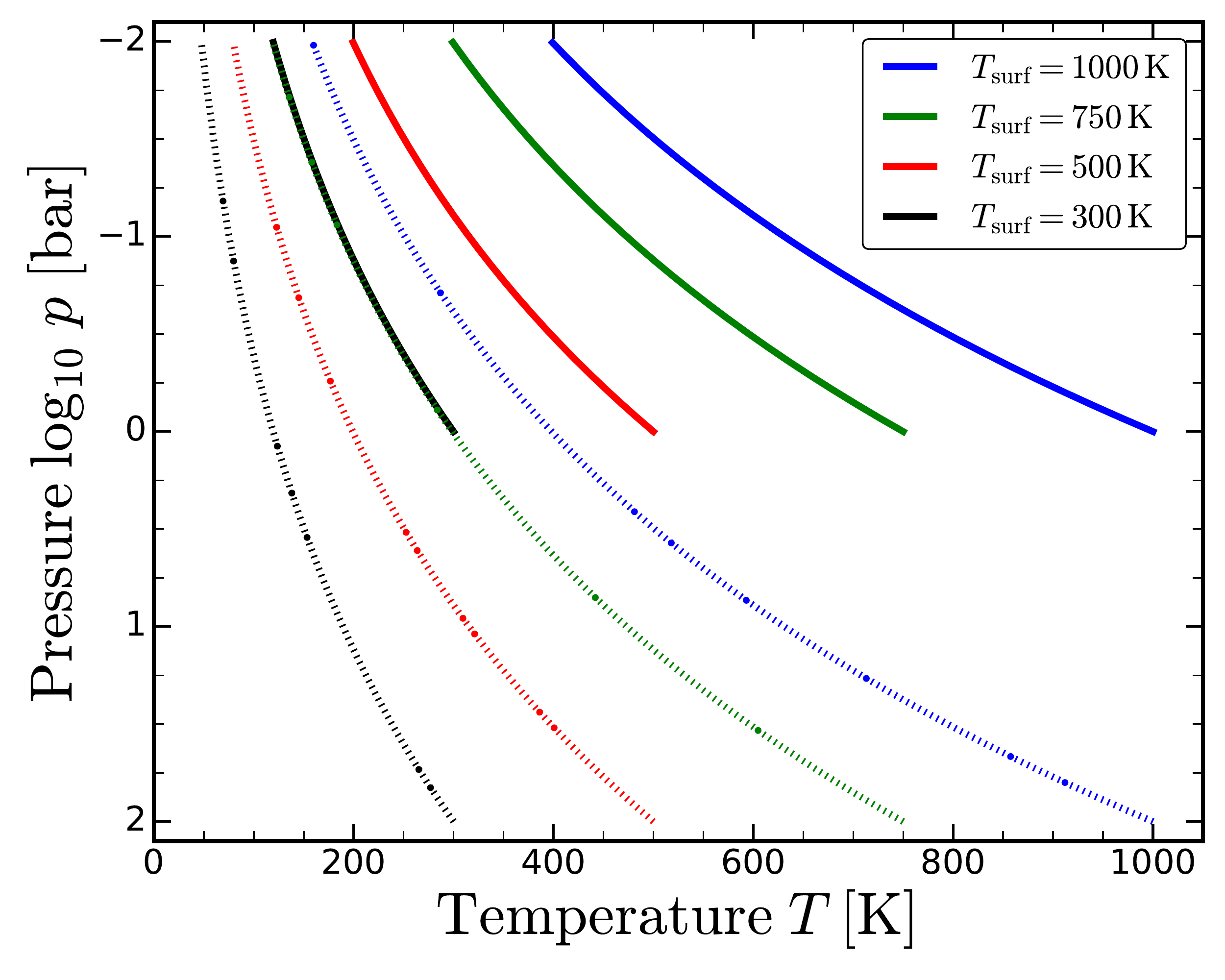}
\caption{Pressure-temperature diagrams for various models investigated in this work. Models for surface pressure of 1\,bar and 100\,bar are shown with solid and dotted lines, respectively. The line colour corresponds to the surface temperature. All models are calculated up to 10\,mbar.}
\label{fig:Tpprofiles}
\end{figure}

The atmosphere is considered to be in hydrostatic equilibrium,
\begin{align}
\frac{\text{d}p}{\text{d}z} = -\rho g
\end{align}
with the atmospheric pressure $p$, atmospheric height $z$, gas mass density $\rho$ and gravitational acceleration $g$.
The latter is deduced from the assumed planetary mass $M_{\rm{P}}$ and radius $R_{\rm{P}}$ by
\begin{align}
g = \frac{GM_{\rm{P}}}{R_{\rm{P}}^2},
\end{align}
where $G$ is the gravitational constant.
The extend of the atmosphere is small compared to the planetary radius and thus we treat $g$ as constant. 
Temperature $T$ and pressure $p$ are assumed to be related by the barotropic equation of state,
\begin{align}
T = \text{const}\,\cdot\,p^\kappa. \label{eq:TpropKappa}
\end{align}
$\kappa$ is related to the change of temperature with increasing height in the troposphere, the lapse rate $\text{d}T/\text{d}z$. 
This assumption is motivated by observations of the lapse rate in atmospheres of solar system bodies (Table \ref{tab:atmospheres}).
Using Eq.~\ref{eq:TpropKappa} and the ideal gas law
\begin{align}
p = \frac{\rho}{\mu} k_\text{B} T,
\end{align}
where $\mu$ is the mean molecular weight and $k_\text{B}$ is the Boltzmann constant, we get
\begin{align}
\frac{\text{d}T}{\text{d}z} &= \frac{\text{d}T}{\text{d}p} \frac{\text{d}p}{\text{d}z} = \text{const}\,\cdot\,(-\rho g)\kappa  p^{\kappa -1} = (-\rho g)\,\kappa\,\frac{T}{p}  = -\kappa g \frac{\mu}{k_\text{B}}.\label{eq:dTdz}
\end{align}
Rearranging for $\kappa$,
\begin{align}
\kappa = \frac{k_\text{B}}{\mu g}\frac{\text{d}T}{\text{d}z}.
\end{align}
The polytropic index $\gamma$ is given by
\begin{align}
\gamma = \frac{1}{1-\kappa}.
\end{align}
For an adiabatic atmosphere, the polytropic index is given by
\begin{align}
\gamma_\text{adiabat} = \frac{c_p}{c_v} \stackrel{\text{ideal gas}}{\approx} \frac{f+2}{f}\ ,
\label{eq:adiabate}
\end{align}
with $f$ being the number of degrees of freedom of the gas species. For a cold gas consisting of linear molecules such as \ce{H2}, \ce{CO2} or \ce{N2}, the number of degree of freedom is $f\!=\!5$ and thus $\gamma_\text{adiabat} \approx 1.4$. 
However, the latter part of Eq.~(\ref{eq:adiabate}) only holds true, if the gas is an ideal gas and thus energy is only stored in its translational and rotational degrees of freedom.
In real atmospheres the species can react with each other, which releases chemical binding energy as function of height, for example due to condensation, in which case the atmospheric structure will be flatter than an adiabate.
Table~\ref{tab:atmospheres} provides an overview of multiple atmospheres of planets and moons in our solar system.
Although the rocky bodies in our solar system show a large range in $\text{d}T/\text{d}z$, the polytropic indices are only ranging from $\gamma = 1.114$ to $\gamma = 1.282$ for the rocky bodies.
The gas giant Jupiter on the other hand shows a higher polytropic index with $\gamma = 1.441$.
Therefore, we assume $\gamma\!=\!1.25$ in all our models for rocky exoplanets instead of its adiabatic value.
The resulting atmospheric profiles can be seen in Fig.~\ref{fig:Tpprofiles}.
Furthermore, the difference in mean molecular weight for the different atmospheres can be seen in Table~\ref{tab:atmospheres}, underlining it's potential for atmospheric characterisation.

The effect of deviations from adiabatic profiles due to condensation is sometimes referred to as moist convection, opposed to dry convection, when considering the condensation of water clouds.
However, similar physics is at work for different condensates.
Under dry conditions no condensates form, and the lapse rate is large, close to its adiabatic value. 
Phase transitions result in the release of latent heat (enthalpy of vaporisation), which heats up the surrounding gas, resulting in a smaller lapse rate. 
Calculations show that the release of latent heat is negligible for mineral material condensation in a H$_2$-rich gas of solar composition \citep{Woitke2003}, because the number of condensible molecules per gas particle is $<\!10^{-4}$, as is true in Jupiter's atmosphere.
However, if substantial amounts of the atmosphere can condense, the difference may become significant.
In the Earth atmosphere, the condensible fraction is about $10^{-2}$, which results in lapse rates of (10-30)\,K/km and  (3.5-6.5)\,K/km, for dry and moist convection, respectively \citep[see e.g.][]{Pierce1998, Muralikrishna2017}.

\begin{table}
\caption{Comparison of the lower atmospheres of multiple bodies in the solar system.}
\label{tab:atmospheres}
\centering
\vspace*{-2mm}
\begin{tabular}{cccccc}
\hline
  & Earth & Venus& Mars& Titan & Jupiter\\ \hline
$T_{\rm{surf}}$ [K]&288&740$^\ast$&210$^\cap$&94$^\dagger$&\\
$p_{\rm{surf}}$ [bar]&1&93$^\ast$&0.006$^\cup$&1.5$^\dagger$&\\
$g$ [m/s$^2$]&9.81&8.87$^\ast$&3.72&1.35&24.8\\
$\mu$ [amu]&29&43.5&43.5&27.4&2.3\\
$\frac{\text{d}T}{\text{d}z}$ [K/km] & 6.$5^\diamond$ & $7.8^+$ & $2^\triangle$ & $1.0^\natural$ & 2.1\\
Valid to [bar] &0.2&$0.23^+$&$3\cdot10^{-4}\nabla$&$0.3^\forall$&$0.3^\S$\\
$\kappa$&0.197&0.166&0.102&0.220&0.306\\
$\gamma$&1.245&1.199&1.114&1.282&1.441\\
\end{tabular}
\tablebib{($\diamond$)~\citet{Muralikrishna2017}; ($\ast$)~\citet{Basilevsky2003}; ($\dagger$)~\citet{2005Natur.438..785F}; ($\cup$)~\citet{Haberle2015}; ($\cap$)~\citet{Smith2008}; (+) \citet{Limaye2017}; ($\nabla$) \citet{Holstein-Rathlou2016}; ($\forall$) \citet{Schinder2012}; ($\S$) \citet{Seiff1998}  }
\end{table}

\begin{figure}
\centering
\includegraphics[width=.40\textwidth]{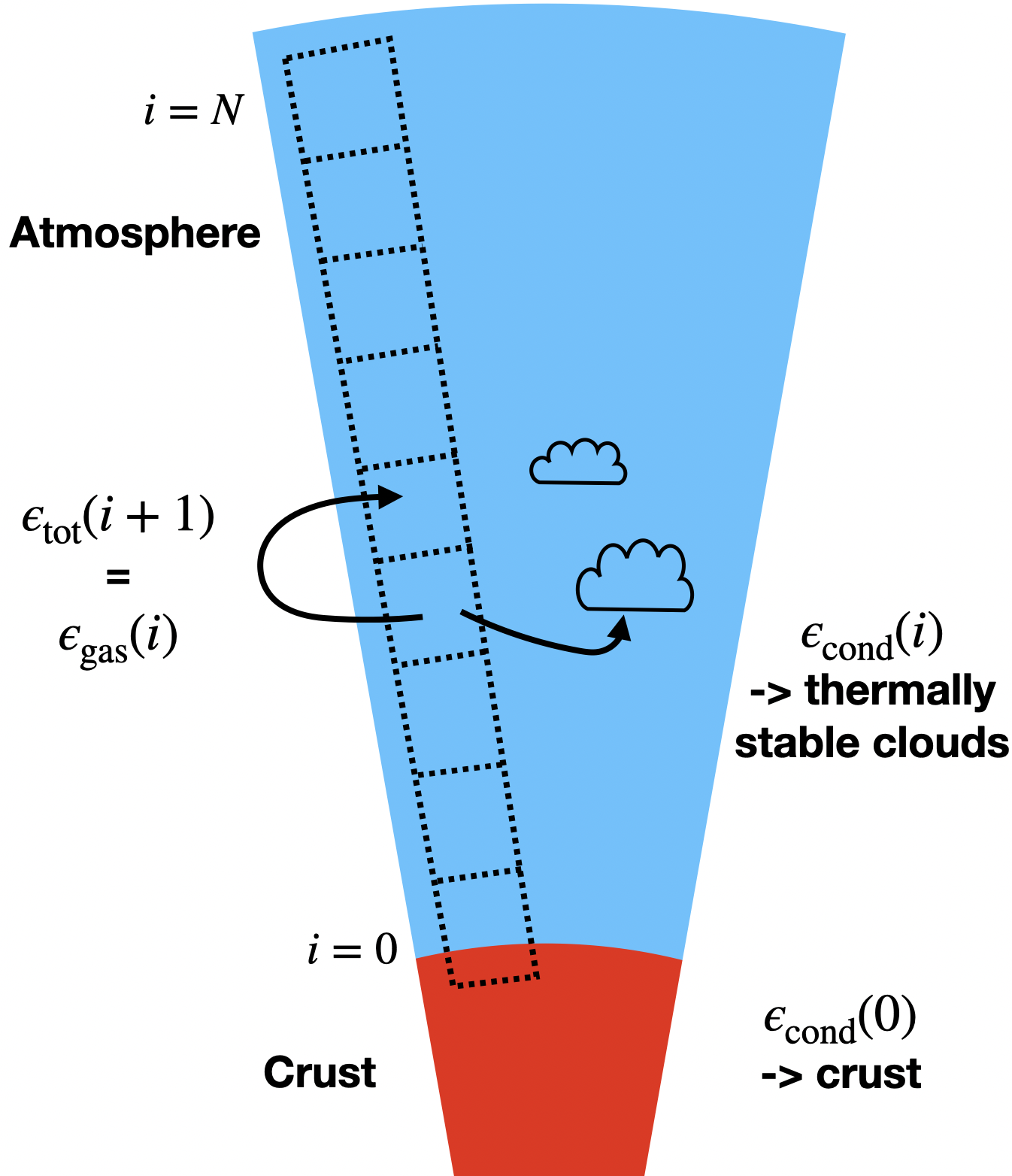}
\caption{Sketch of the bottom to top {\sc GGchem} model for an atmosphere with cloud condensation. Chemical and phase equilibrium is solved in each atmospheric layer, and the condensates are interpreted as clouds. The condensed element abundances $\epsilon_{\rm cond}$ are removed, and the remaining gas phase element abundances in layer $i$ are used for the total element abundances $\epsilon_\mathrm{tot}$ in the next atmospheric layer $i+1$.}
\label{fig:AutoStruc}
\end{figure}

\begin{table*}[!ht]
\caption{Comparison of different element abundances given in \% mass fractions used in this work.}
\label{tab:abundances}
{\centering
\vspace*{-2mm}
\begin{tabular}{c|cccccccccc}
\hline
	&	BSE	    &	BSE12	&	BSE15	&	CC	    &	MORB	&	CI	    &		PWD 	&					Earth	&	Archean	&	solar	\\
	  &1&$1\ast$&$1\ast$&2&3&4&$5\dagger$&$1\nabla$&$1\ast$&6\\ \hline
H	&	0.006	&	1.198	&	1.457	&	0.045	&	0.023	&	1.992	&		        &					0.3341	&	2.309	&	98.4	\\
C	&	0.006	&	0.0054	&	0.005	&	0.199	&	0.019	&	3.520	&		1.2		&					1.957	&	0.0052	&	0.316	\\
N	&	8.8E-05	&	7.9E-05	&	7.7E-05	&	0.006	&	5.5E-05	&	0.298	&	            &					0.02533	&	7.6E-05	&	0.092	\\
O	&	44.42	&	49.280	&	50.320	&	47.20	&	44.5	&	46.420	&		41.0	&					50.100	&	49.880	&	0.765	\\
F	&	0.002	&	0.0018	&	0.0017	&	0.053	&	0.017	&	0.0059	&		        &					0.04816	&	0.0017	&	0.000067\\
Na	&	0.29	&	0.260	&	0.253	&	2.36	&	2.012	&	0.505	&		        &					2.145	&	0.251	&	0.0039	\\
Mg	&	22.01	&	19.69	&	19.180	&	2.20	&	4.735	&	9.790	&		18.0	&					1.999	&	19.010	&	0.094	\\
Al	&	2.12	&	1.897	&	1.847	&	7.96	&	8.199	&	0.860	&		0.27	&					7.234	&	1.831	&	0.0074	\\
Si	&	21.61	&	19.33	&	18.830	&	28.80	&	23.62	&	10.820	&		18.0	&					26.17	&	18.670	&	0.089	\\
P	&	0.008	&	0.0072	&	0.0070	&	0.076	&	0.057	&	0.098	&		0.22	&					0.06906	&	0.0069	&	0.00078	\\
S	&	0.027	&	0.024	&	0.024	&	0.070	&	0.110	&	5.411	&		3.3		&					0.06361	&	0.023	&	0.041	\\
Cl	&	0.004	&	0.0036	&	0.0035	&	0.047	&	0.014	&	0.071	&		        &					0.04271	&	0.0035	&	0.0011	\\
K	&	0.02	&	0.018	&	0.017	&	2.14	&	0.152	&	0.055	&		        &					1.945	&	0.017	&	0.00041	\\
Ca	&	2.46	&	2.201	&	2.143	&	3.85	&	8.239	&	0.933	&		6.9		&					3.499	&	2.125	&	0.0086	\\
Ti	&	0.12	&	0.107	&	0.105	&	0.401	&	0.851	&	0.046	&	            &					0.3644	&	0.104	&	0.00042	\\
Cr	&	0.29	&	0.260	&	0.253	&	0.013	&	0.033	&	0.268	&		        &					0.01181	&	0.251	&	0.00222	\\
Mn	&	0.11	&	0.098	&	0.096	&	0.072	&	0.132	&	0.195	&		        &					0.06543	&	0.095	&	0.0014	\\
Fe	&	6.27	&	5.610	&	5.463	&	4.32	&	7.278	&	18.710	&		10		&					3.926	&	5.416	&	0.172	\\ \hline
sum	&	99.77	&	99.99	&	100.00	&	99.81	&	99.99	&	100.00	&		99.73	&					100.00	&	100.00	&	100.00	\\ \hline
\end{tabular}}
\\*[1mm]$\ast$: Element abundances based on BSE abundances, but altered by the addition of \ce{H2O}, see Sect.~\ref{ssec:input}\\
$\dagger$: The PWD data is completed by taking the missing element abundances from BSE, MORB, or CI as indicated.\\
$\nabla$: The element abundances are created by a fit to Earth atmospheric concentrations, see Appendix~\ref{Sect:EarthModel}..\\
{(1)~Bulk Silicate Earth: \citet{Schaefer2012}; (2)~Continental Crust: \citet{Schaefer2012}; (3)~Mid Oceanic Ridge Basalt: \citet{Arevalo2010}; (4)~CI chondrite: \citet{Lodders2009}; (5)~Polluted White Dwarf: \citet{Melis2016}; (6)~solar: \citet{Asplund2009}}
\end{table*}

\subsection{Modelling approach}\label{ssec:cloudmodel}
In order to model the atmosphere of a rocky exoplanet, we assume a polytropic atmosphere in hydrostatic equilibrium as described in Eq.(\ref{eq:dTdz}) with the polytropic index $\gamma = 1.25$ (Sect.~\ref{ssec:polyatm}), and calculate the abundances of gas and condensed species in chemical and phase equilibrium at each point in the atmosphere.

In each atmospheric layer, we use the chemical and phase equilibrium solver {\sc GGchem} \citep{2018A&A...614A...1W} to determine the abundances of the thermally stable condensates.
These condensates are then removed from the model, until saturation is achieved.
This way, we determine the depletion of element by condensation with increasing height, following a parcel of gas which is moving upwards in the atmosphere. 
We assume that the condensation and removal by precipitation is fast enough to ensure that all supersaturation ratios obey $S\leq1$ at any point in the atmosphere.

This approach is similar in nature to those in previous works by for example \citet{Lewis1969}, \citet{Lewis1970} and \citet{Weidenschilling1973}.
These works have used a similar approach of atmospheric equilibrium calculation with removal of thermally stable condensate to investigate cloud condensates and gas compositions.
They especially focused on gas giants as well as ice giants with including the major volatile elements (esp. H, He, C, N and O).
In the model presented here, we include further elements (see Sec.~\ref{ssec:input}) into our model to also investigate the stability of metal bearing condensates.
The inclusion of the most common rock forming elements also allows the investigation of the planetary crust in contact with the atmosphere.
This allows us to use this approach for rocky exoplanets.

A further difference to the previous studies is the assumed atmospheric structure.
The older models \citep[e.g.][]{Lewis1969} include a feedback from the cloud condensation to the atmospheric $p-T$ Structure.
In order to keep our model as quick and simple as possible, we do not include any of these feedbacks and keep the polytropic index constant throughout the atmosphere.

The base of our atmospheric model for rocky planets is at the bottom of the atmosphere, where we calculate the chemical composition of the crust and the gas above it in chemical and phase equilibrium \citep[similar to][]{Herbort2020}.
These results only depend on the total element abundances $\epsilon_\text{tot}$, the surface pressure $p_\text{surf}$, and the surface temperature $T_\text{surf}$.
The gas phase composition of the boundary layer ($i=0$) is a by-product of this computation.
The resulting element composition of the crust is denoted by $\epsilon_\text{cond}(0)$ and the element composition of the gas in contact with the crust as $\epsilon_\text{gas}(0)$, see \citet{Herbort2020}.
We then continue to model the atmosphere layer by layer, from bottom to top.
The gas phase element abundances in one atmospheric layer $i$ are used for the total element abundances of the layer above it $\epsilon_\text{tot}(i+1)\!=\!\epsilon_\text{gas}(i)$, where $i\!=\!\{1,...,N-1\}$ is a layer index and $N$ the total number of atmospheric layers.
We use $N=300$ throughout the models presented in this work.
Figure~\ref{fig:AutoStruc} provides a sketch of our modelling approach.

With this simple approach we provide a fast atmospheric model to identify potential cloud condensates and to discuss at which heights these condensates might occur.
We do not include kinetic cloud formation, nucleation rates, photochemistry, release of latent heat, nor kinetic chemical in this model.
Also geological effects (volcanism, subduction etc.) and biological rates are not included.

An important result of this modelling procedure are the number densities of the condensed species $n_\text{cond}(z)$.
These number densities reflect the amount of condensates that fall out as we lift the gas parcel by $\Delta z$.
According to our approximation of an instant and complete removal of all condensates, $n_\text{cond}$ is expected to be an estimate for the actual cloud densities, where that removal is slower and incomplete.
The calculation of the actual cloud densities would require a proper kinetic cloud model, including nucleation, mixing, growth, and gravitational settling rates, which goes beyond the scope of this paper.
Instead, we use the normalised number density ($n_\text{cond}/n_\text{tot}$) of the condensed species ($n_\text{cond}$) with respect to the total gas density ($n_\text{tot}$) as an indicator for a cloud species to be relevant or not.
We have experimented with different threshold concentrations and use $10^{-10}$ for the investigation of dominating cloud condensates (esp. Sect.~\ref{sssec:mainclouds}) and $10^{-20}$ for trace condensates that allow an investigation of chemical connections of different condensates (esp. Sect.~\ref{sssec:clouddiversity}).
Regions in ($p_\mathrm{gas}, T_\mathrm{gas}$) where these thresholds are overcome indicate the positions of the cloudy regions.
Our results are hence presented in Sect.~\ref{sec:results} for these choices of these threshold value.

One intrinsic result of this model is that for a given cloud condensate, the highest number density of condensates is found when a condensate becomes thermally stable (cloud base).
For the atmosphere above, the number density decreases subsequently, resulting in a characteristic shark fin like cloud distribution.
This can also be seen in works like \citet{Atreya1999}.
\citet{Ackerman2001} introduced a factor $f_\text{rain}$ to their models to reproduce a more realistic cloud formation model by allowing replenishment to the lower atmospheric layers and different supersaturation levels.
We do not include a similar factor to our model, as the scope of this work is to investigate which condensate can be thermally stable in general.
The information on which condensates are thermally stable can subsequently be used for future kinetic cloud formation models.

All our atmospheres are calculated up to $p_\mathrm{gas}=10\,$mbar.
A continuation of the model to lower pressures with an isothermal or inversed temperature profile would produce no further clouds in our model.
The potential condensates would be undersaturated in such regions, because the thermodynamic conditions for condensation would become less and less favourable with constant or increasing temperature.

\subsection{Input parameters} \label{ssec:input}
We use the equilibrium condensation code {\textsc{GGchem}} \citep{2018A&A...614A...1W} and include 18 different elements (H, C, N, O, F, Na, Mg, Al, Si, P, S, Cl, K, Ca, Ti, Cr, Mn, and Fe).
In total 471 gas species and 208 condensed species are included in our models.
The molecular equilibrium constants $k_{p}(T)$ are based on \cite{Stock2008}, \cite{2016A&A...588A..96B}, and the NIST/JANAF database \citep{Chase1982, Chase1986, chase_mono9}.
The $\Delta G_\mathrm{f}^0(T)$ for the condensed phase are taken from {\sc SUPCRTBL} \citep{Zimmer2016} and the {\sc NIST/JANAF} database.
The thermochemical data included in {\sc GGchem} is tested down to temperatures of 100\,K.

In order to investigate a large variety of different planets, various total element abundances $\epsilon_{\rm{tot}}$ are used at the base of the atmosphere. 
These are Bulk Silicate Earth \citep[BSE,][]{Schaefer2012}, Continental Crust \citep[CC,][]{Schaefer2012},  Mid Oceanic Ridge Basalt \citep[MORB,][]{Arevalo2010}, and CI chondrite \citep[CI,][]{Lodders2009}.
We add elemental compositions deduced from the polluted white dwarf SDSS J104341.53+085558.2 \citep{Melis2016}.
The contamination of such white dwarfs is interpreted as the accretion of planets and planetary bodies falling into the white dwarf and is the only current method to constrain the overall element abundances of planets in other stellar systems \citep[see e.g.][]{Farihi2016, Harrison2018, Wilson2019, Bonsor2020}. 
However, not all elements can be measured in the polluted white dwarf, so we complete the dataset with element ratios from BSE, MORB, and CI to create the element abundances PWD BSE, PWD MORB, and PWD CI, respectively.

As in \citet{Herbort2020}, we also include element abundances based on BSE, where the H and O abundances are increased to allow liquid water to be stable at the surface.
Therefore, models with 12\,\% mass fraction and 15\,\% mass fraction added water are included and referred to as BSE12 and BSE15, respectively.
We furthermore include an Archean model, inspired by an early Earth atmosphere, which might have been dominated in reducing species such as \ce{H2} and \ce{CH4} \citep{Catling2020}, before life emerged and ultimately caused the accumulation of atmospheric molecular oxygen during the Great Oxidation Event \citep{Holland2002}.
Experiments by \citep{Miller1953} and \cite{1959Sci...130..245M} used a similar atmospheric composition to study the origin of life.
More recent research suggests that the early Earth had signifiantly smaller amounts of \ce{CH4} in the atmosphere \citep[see e.g.][]{Kasting2014, Catling2020}.
In the exoplanetary context, similar atmospheres could exist, especially for high mass planets, which can sustain their hydrogen envelope and thus host a reducing atmosphere rich in hydrogen \citep{Owen2020}.
In order to investigate an atmosphere similar to the current day Earth, we created a model to represent these conditions. In Appendix~\ref{Sect:EarthModel} we describe this model in detail.
A model with solar abundances from \citet{Asplund2009} is also included.

\section{Results}\label{sec:results}
The resulting properties of our models are presented in this section.
We aim to establish which cloud materials are likely to form over what kind of planetary surface.
In our model, the planetary surface is characterised by the surface pressure ($p_{\rm{surf}}$) the surface temperature ($T_{\rm{surf}}$) and the set of total element abundances ($\epsilon_{\rm{tot}}$).
For the atmospheric part of the model, we have introduced two more parameters, the polytropic index $\gamma$, and the surface gravity $g$.
The composition of the near-crust atmospheric gas (i.e. the lowest point in our model) is determined by chemical and phase equilibrium with the planet surface.

First, we study the results for one specific set of element abundances (Sect.~\ref{ssec:BSEmodel}).
Subsequently, we compare the results for different element abundances and investigate the cloud material (Sect.~\ref{ssec:elementabundances}) and gas composition (Sect.~\ref{ssec:atmo}).
The link between water clouds and water on the surface is studied in Sect.~\ref{ssec:surfacewater}.
In Sect.~\ref{ssec:press} we focus on the influence of surface pressure to the cloud layers.

\begin{figure*}[!t]
\centering
\includegraphics[width = .32\linewidth, page=1]{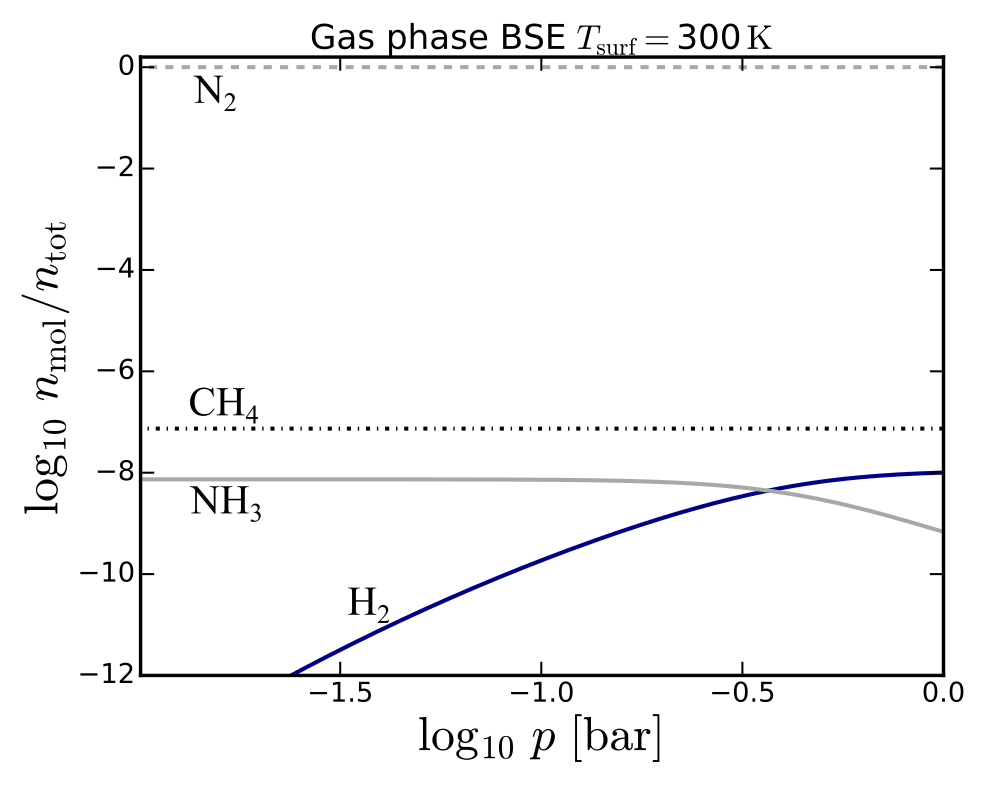}
\includegraphics[width = .32\linewidth, page=1]{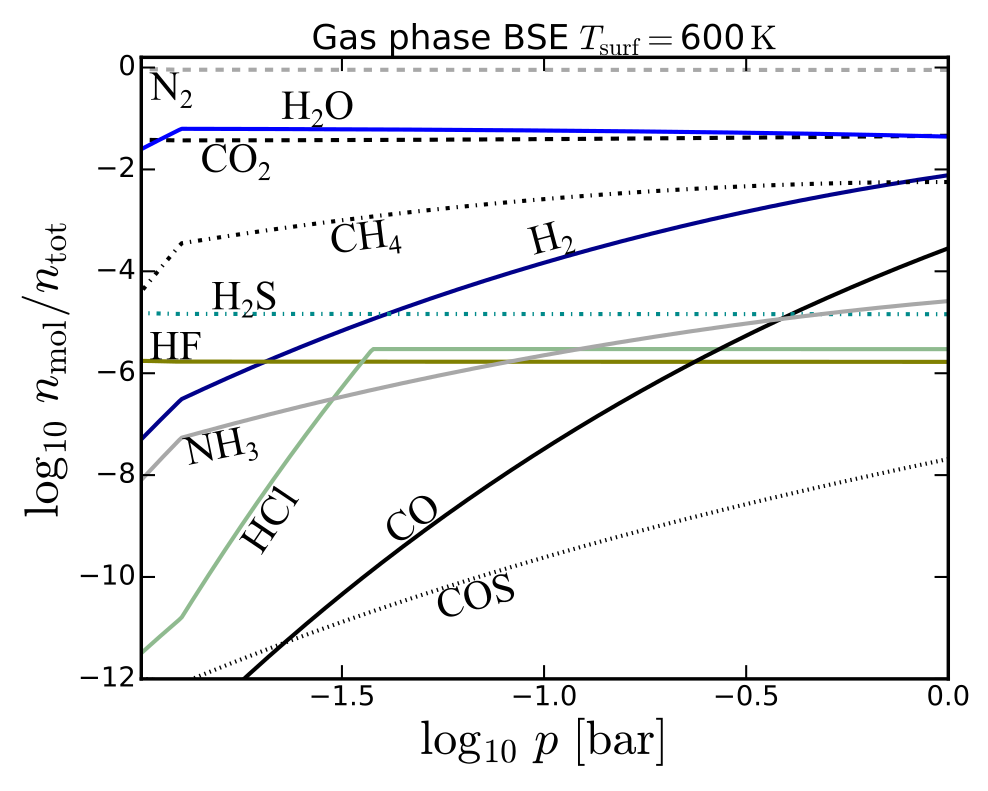}
\includegraphics[width = .32\linewidth, page=1]{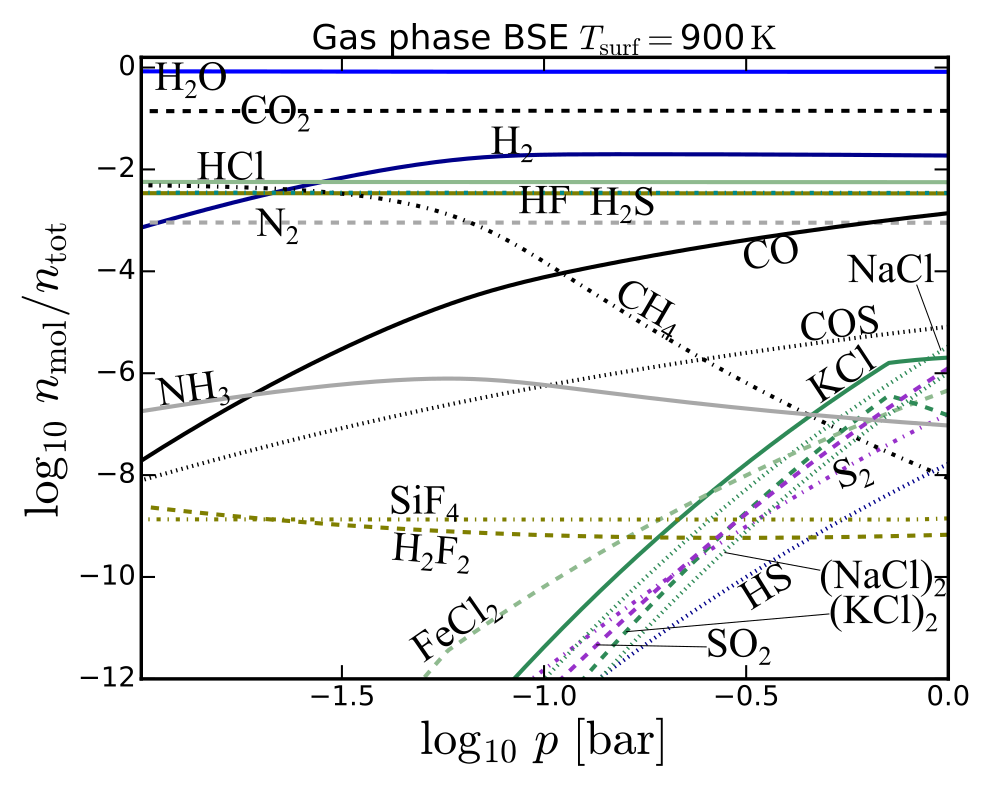}\\
\includegraphics[width = .32\linewidth, page=1]{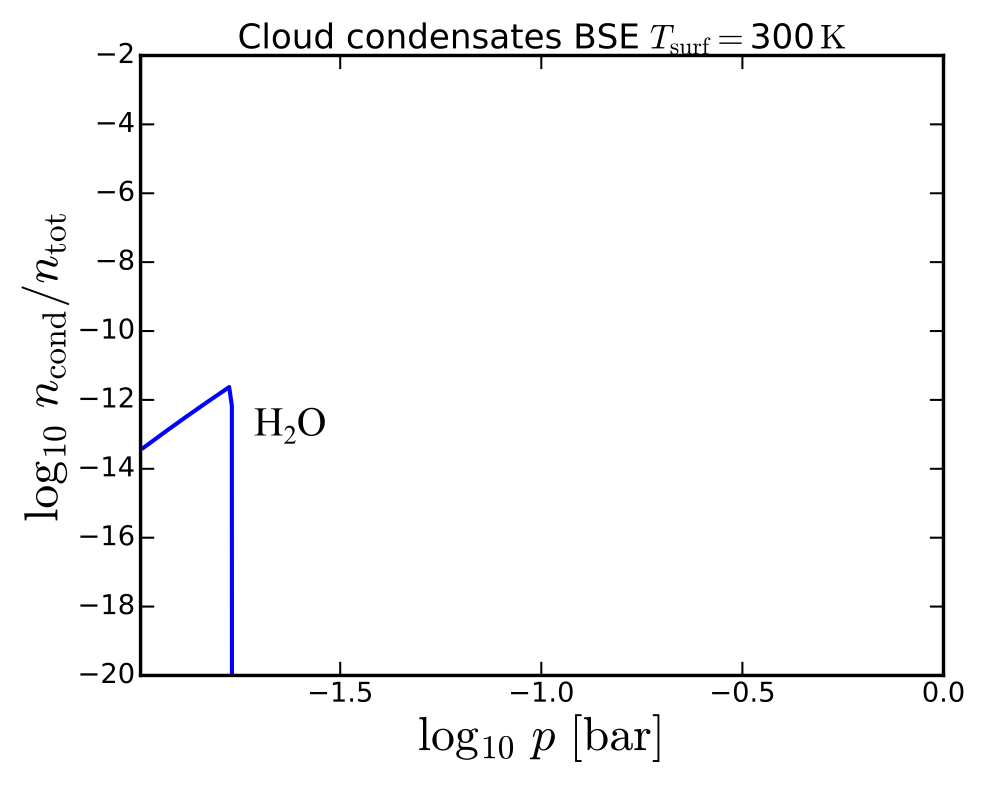}
\includegraphics[width = .32\linewidth, page=1]{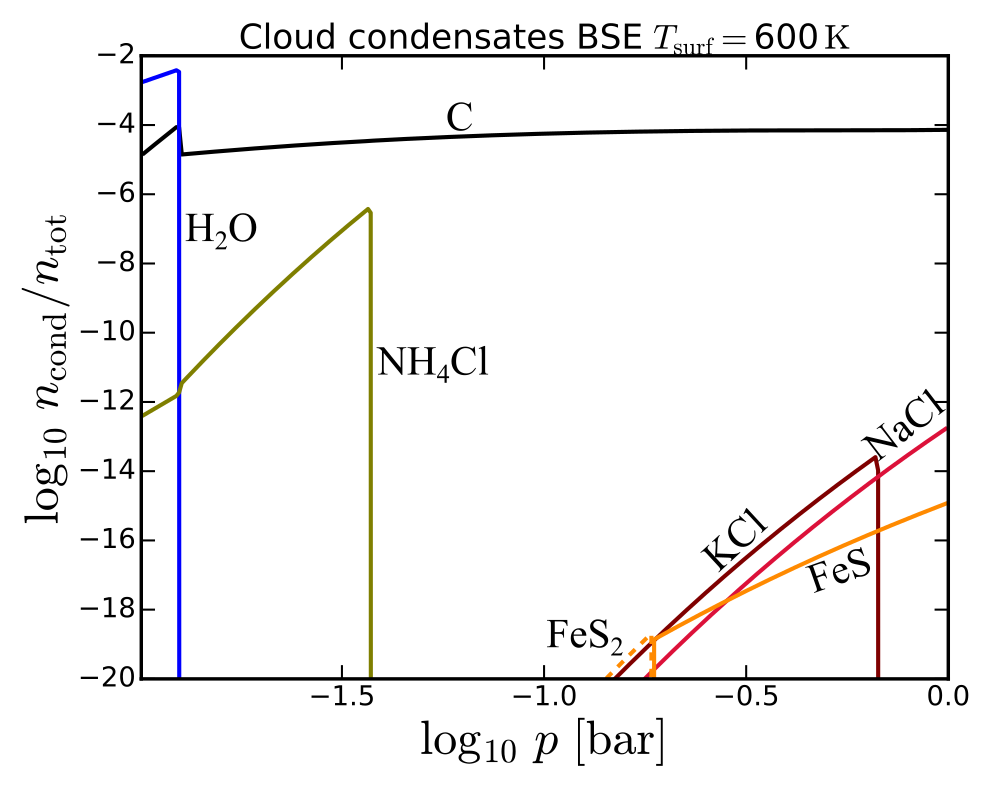}
\includegraphics[width = .32\linewidth, page=1]{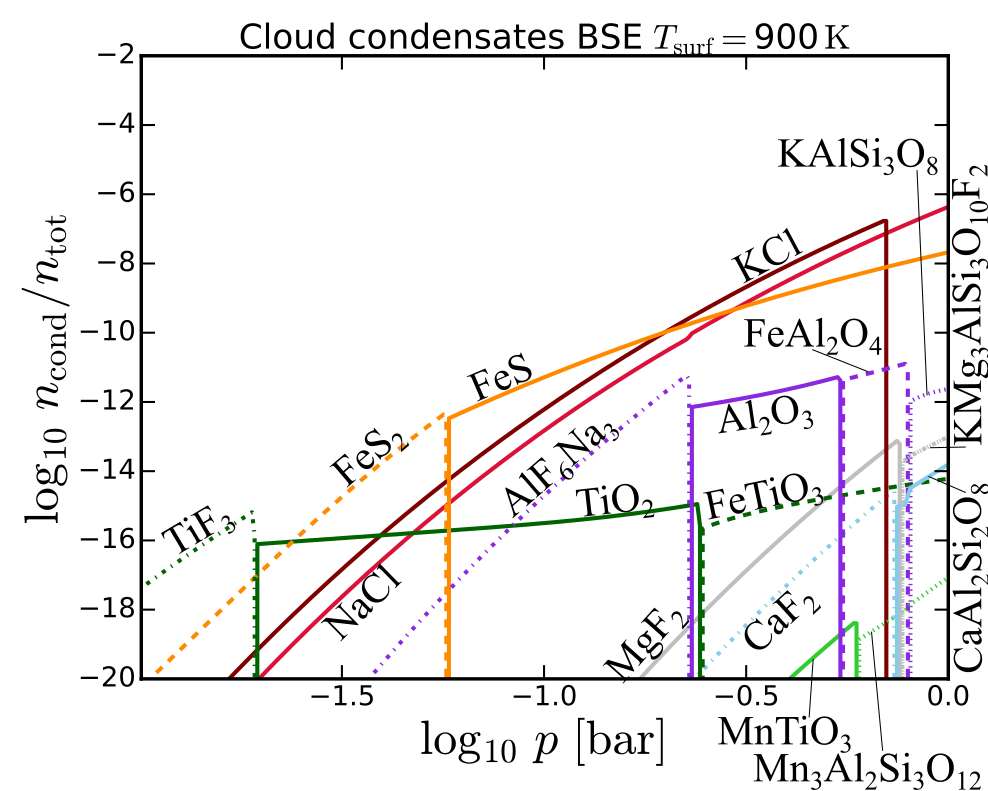}\\
\includegraphics[width = .32\linewidth, page=1]{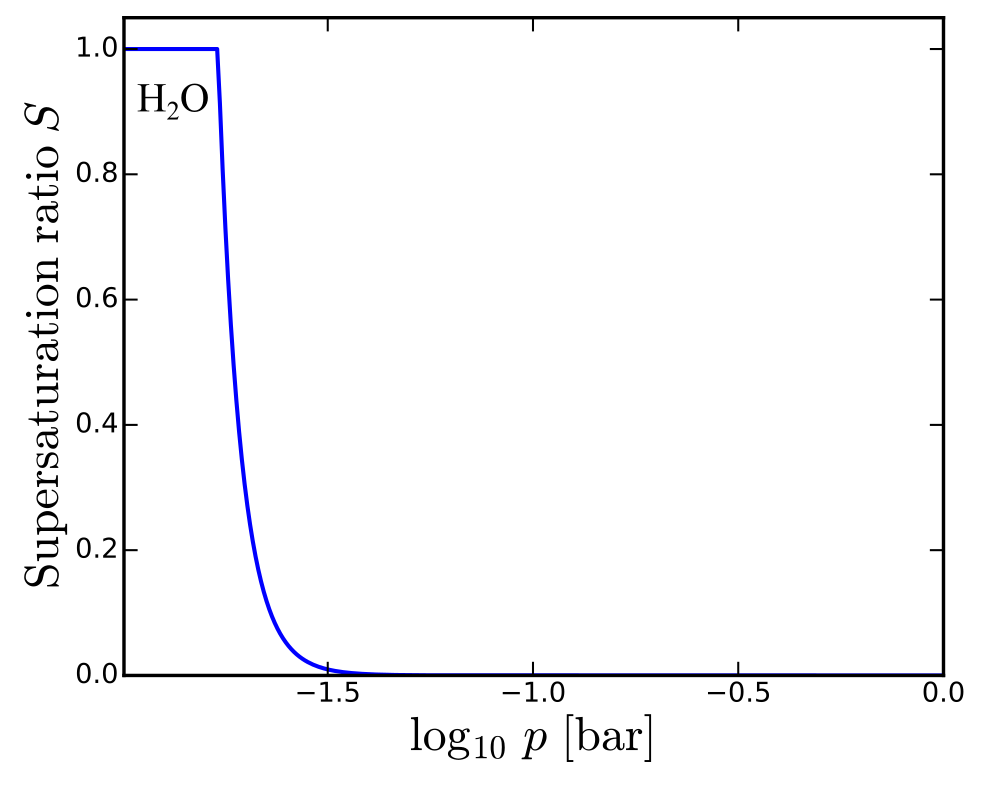}
\includegraphics[width = .32\linewidth, page=1]{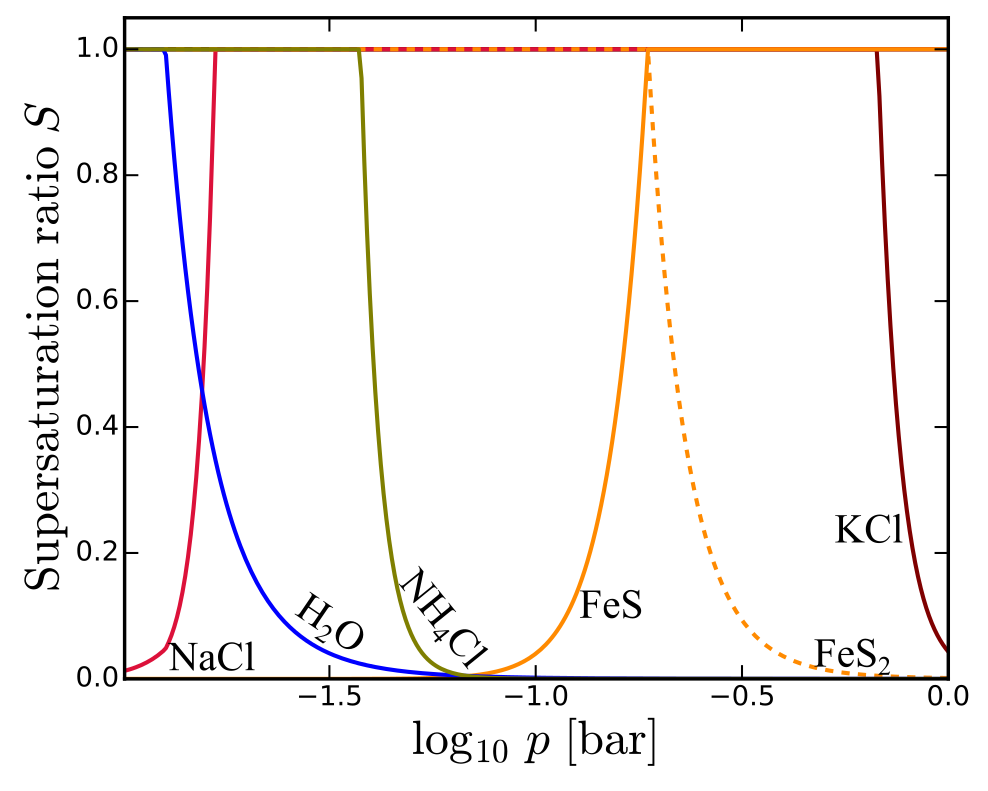}
\includegraphics[width = .32\linewidth, page=1]{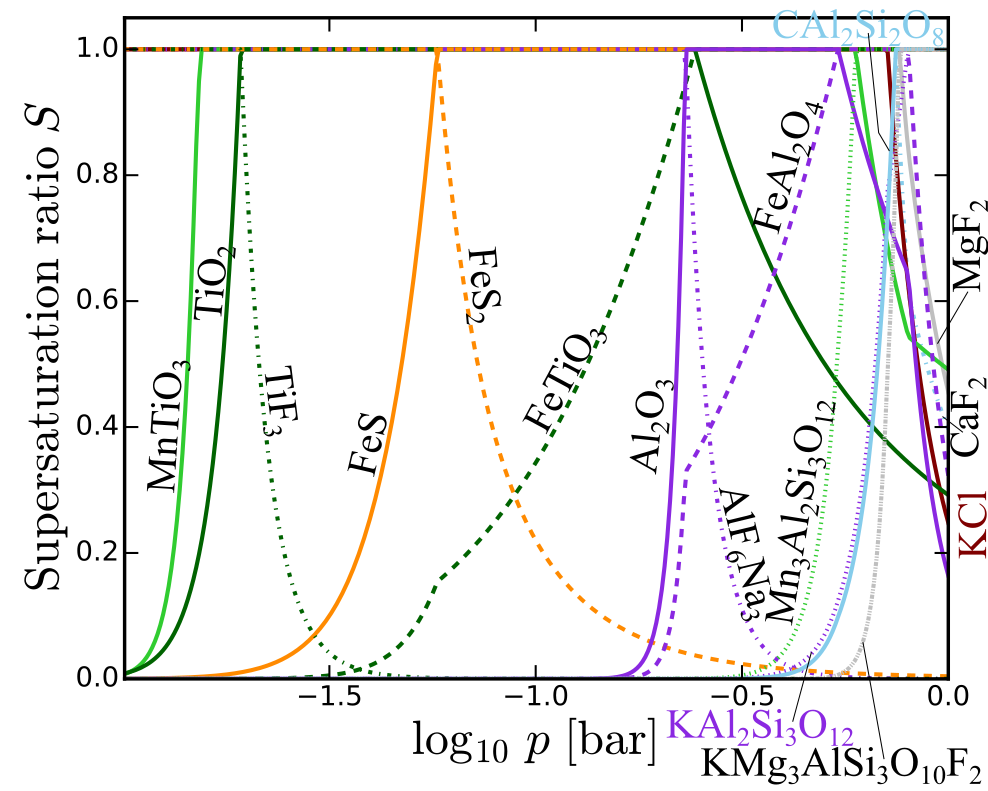}
\caption{Atmospheric models for BSE total element abundances at different surface temperatures. \textbf{Upper panels:} gas composition, \textbf{middle panels:} cloud composition, \textbf{lower panels:} saturation levels. \textbf{Left panels:} $T_\text{surf} = 300\,$K, \textbf{middle panels:} $T_\text{surf} = 600\,$K, \textbf{right panels:} $T_\text{surf} = 900\,$K.
All gas species that reach a concentration of at least 1\,ppb somewhere in the atmosphere are shown.
All cloud species down to a concentration of $10^{-20}$ are shown.
The supersaturation ratios of all condensates in the middle panels are visualised in the respective lower panel.
Condensates belonging to one condensate chain are depicted in one color.}
\label{fig:BSE}
\end{figure*}

\subsection{Bulk Silicate Earth model}\label{ssec:BSEmodel}
The principle behaviour of our model is studied for one example set of element abundances chosen to be Bulk Silicate Earth \citep[BSE,][]{Schaefer2012}, considering different surface temperatures $(T_\text{surf}\!=\!300\,$K, 600\,K, 900\,K).
These three surface temperature represent some of the observed rocky exoplanets, namely Trappist-1~c,d \citep{Delrez2018}, LHS~1478~b \citep{Soto2021}, and K2-228~b \citep{Livingston2018}, respectively.
Tables~\ref{tab:Planets} and \ref{tab:Planets-hot} and Sect.~\ref{sec:planets} provide a link to further known planets and the implications of our models with regard to cloud and surface compositions.
Figure~\ref{fig:BSE} shows the atmospheric composition (upper panels), cloud condensate composition (middle panels) and saturation levels (lower panels) for these three models.

The composition of the gas phase as seen in the upper panels of Fig.~\ref{fig:BSE} is determined by a) the changing local thermodynamic condition ($p_\text{gas}, T_\text{gas}$), resulting in a smooth change in gas phase abundance, and b) the local element abundances.
The local element abundance will be affected by the amount of material condensing ($n_\mathrm{cond}$).
The onset of condensation results in a knee in the gas phase abundances.

For ${T_\text{surf}~=~300\,\mathrm{K}}$  (left side of Fig.~\ref{fig:BSE}, representative of Trappist-1 c,d), the atmosphere is almost a pure \ce{N2} atmosphere with trace amounts of \ce{CH4}, \ce{NH3}, and \ce{H2} with gas phase abundances of $10^{-6}>n_\text{mol}/n_\text{tot}>10^{-9}$.
\ce{H2O}[s] is the only condensate that is more abundant than $n_\text{cond}/n_\text{tot}>10^{-20}$ and is thermally stable in the upper atmosphere.

The $T_\text{surf} = 600\,$K model (middle column in Fig.~\ref{fig:BSE}, representative of LHS~1378~b) is also dominated by \ce{N2}, with \ce{H2O}, \ce{CO2}, \ce{H2}, and \ce{CH4} reaching gas phase abundances of $10^{-1}>n_\text{mol}/n_\text{tot} > 10^{-3}$.
Further 6 molecules reach gas phase abundances with $n_\text{mol}/n_\text{tot}>10^{-9}$.
The main condensates are \ce{H2O}[s], \ce{C}[s], and \ce{NH4Cl}[s].
\ce{H2O}[s] is thermally stable higher in the atmosphere with respect to the $T_\text{surf} = 300\,$K model.
\ce{C}[s] is the only condensate stable throughout the atmosphere with roughly constant normalised number density.
Furthermore \ce{KCl}[s], \ce{NaCl}[s], \ce{FeS}[s], and \ce{FeS2}[s] are present at lower normalised number densities ($n_\text{con}/n_\text{tot}<10^{-12}$). 

The most abundant gas species in the atmosphere of the $T_\text{surf} = 900\,$K model (right side of Fig.~\ref{fig:BSE}, representative of K2-228~b) is \ce{H2O}, followed by \ce{CO2} at $n_\text{mol}/n_\text{tot}\approx10^{-1}$.
Further species with $n_\text{mol}/n_\text{tot}>10^{-3}$ are \ce{H2}, \ce{HCl}, \ce{H2S}, \ce{HF}, and \ce{N2}.
The gas phase abundance of \ce{CO} decreases with height, while \ce{CH4} becomes more abundant.
They are equally abundant at a pressure of $p_{\rm{gas}}\approx0.1$\, bar at abundances of $n_\text{mol}/n_\text{tot}\approx10^{-4}$.
This change from \ce{CO} to \ce{CH4} as carbon bearing molecules is an example for temperature dependent change of gas phase compositions independent of a chance in element abundance.
Further 12 molecules show gas phase abundances of $n_\text{mol}/n_\text{tot}>10^{-9}$ in the $T_\text{surf} = 900\,$K model, underlining the larger diversity in the gas composition for higher $T_\mathrm{gas}$.
The most abundant condensates are \ce{NaCl}[s], \ce{KCl}[s], and \ce{FeS}[s], which are most abundant in the near-crust atmosphere.
Further 14 condensates reach normalised number densities of $n_\text{cond}/n_\text{tot}>10^{-20}$.

As described in \citet{2018A&A...614A...1W} two kinds of condensation occur in the atmosphere.
Type 1: condensation from the gas phase. A new condensate becomes stable and removes elements from the gas phase.
These condensations result in a knee in the gas phase abundances.
Type 2: a transition from one to another condensate without a change in the gas phase, which is characterised by one condensate becoming unstable ($S<1$) while another one becomes stable ($S=1$). 
A type 2 transition links the two effected condensates to a \textit{condensate chain}, a condensation sequence.

The occurring cloud condensate species can be grouped into three different groups.
\begin{itemize}
    \item Group 1: Materials that are stable in the lowest atmospheric layer directly above the surface.\\
    ${T_\text{surf}~=~300\,\mathrm{K}}$: none;\\
    ${T_\text{surf}~=~600\,\mathrm{K}}$: \ce{C}[s], \ce{NaCl}[s], and \ce{FeS}[s];\\
    ${T_\text{surf}~=~900\,\mathrm{K}}$: \ce{NaCl}[s], \ce{FeS}[s], \ce{KAlSi3O8}[s], \ce{FeTiO3}[s], \ce{KMg3AlSi3O10F2}[s], \ce{CaAl2Si2O8}[s], and \ce{Mn3Al2Si3O12}[s].
    \item Group 2: Materials that form by type 1 condensation.\\
    ${T_\text{surf}~=~300\,\mathrm{K}}$: \ce{H2O}[s];\\
    ${T_\text{surf}~=~600\,\mathrm{K}}$:\ce{H2O}[s], \ce{NH4Cl}[s], and \ce{KCl}[s];\\
    ${T_\text{surf}~=~900\,\mathrm{K}}$: \ce{KCl}[s].
    \item Group 3: Materials that form by type 2 condensation.\\
    ${T_\text{surf}~=~300\,\mathrm{K}}$: none;\\
    ${T_\text{surf}~=~600\,\mathrm{K}}$: \ce{FeS2}[s];\\
    ${T_\text{surf}~=~900\,\mathrm{K}}$: \ce{FeS2}[s], \ce{Al2O3}[s], \ce{FeAl2O4} [s], \ce{AlF6Na3}[s], \ce{MgF2}, \ce{CaF2[s]}, \ce{TiO2}[s], \ce{TiF3}[s], and \ce{MnTiO3}[s].
\end{itemize}
Based on these groups, the condensates can be linked to condensate chains, which are listed in the following.\\
gas $\rightarrow$ \ce{H2O}[s]\\
gas $\rightarrow$ \ce{NH4Cl}[s]\\
gas $\rightarrow$ \ce{KCl}[s]\\
crust $\rightarrow$ \ce{C}[s]\\
crust $\rightarrow$ \ce{NaCl}[s]\\
crust $\rightarrow$ \ce{FeS}[s] $\rightarrow$ \ce{FeS2}[s]\\
crust $\rightarrow$ \ce{KAlSi3O8}[s] $\rightarrow$ \ce{Al2O3}[s] $\rightarrow$ \ce{FeAl2O4}[s]$\rightarrow$ \ce{AlF6Na3}[s]\\
crust $\rightarrow$ \ce{KMg3AlSi3O10F2}[s] $\rightarrow$ \ce{MgF2}[s]\\
crust $\rightarrow$ \ce{CaAl2Si2O8}[s] $\rightarrow$ \ce{CaF2}[s]\\
crust $\rightarrow$ \ce{FeTiO3}[s] $\rightarrow$ \ce{TiO2}[s]$\rightarrow$ \ce{TiF3}[s]\\
crust $\rightarrow$ \ce{Mn3Al2Si3O12}[s] $\rightarrow$ \ce{MnTiO3}[s]\\

\subsection{Cloud diversity over diverse rocky surfaces}
\label{ssec:elementabundances}
In this section we study the influence of the surface composition to the thermal stability of cloud species.
The surface composition is determined by the total element abundance ($\epsilon_\mathrm{tot}$, see Table~\ref{tab:abundances} and Sect.\ref{ssec:input}), surface pressure ($p_\mathrm{surf}$) and surface temperature ($T_\mathrm{surf}$).
In this section we keep $p_\text{surf} =1\,$bar for all models.
We cover the range of $T_\text{surf} = 300\,$K to $T_\text{surf} = 1000\,$K.

\begin{figure*}[!t]
\centering
\includegraphics[width = .95\linewidth]{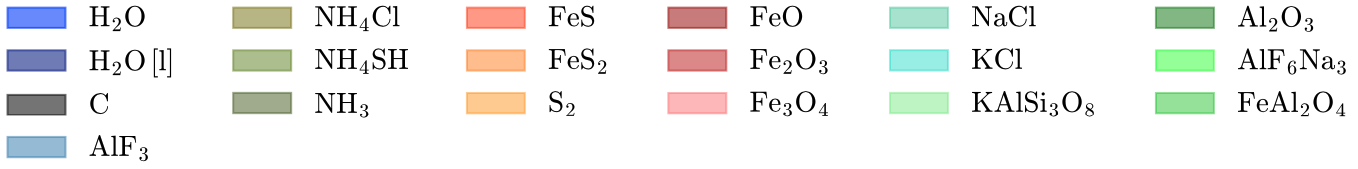}\\
\includegraphics[width = .32\linewidth, page=1]{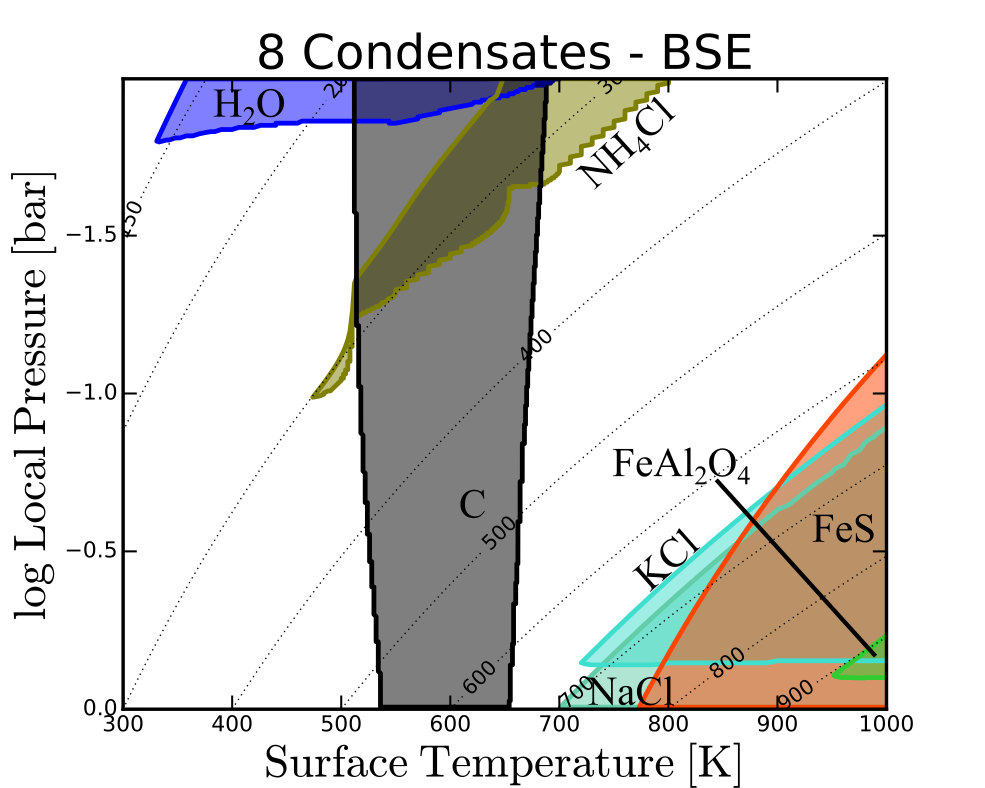}
\includegraphics[width = .32\linewidth, page=1]{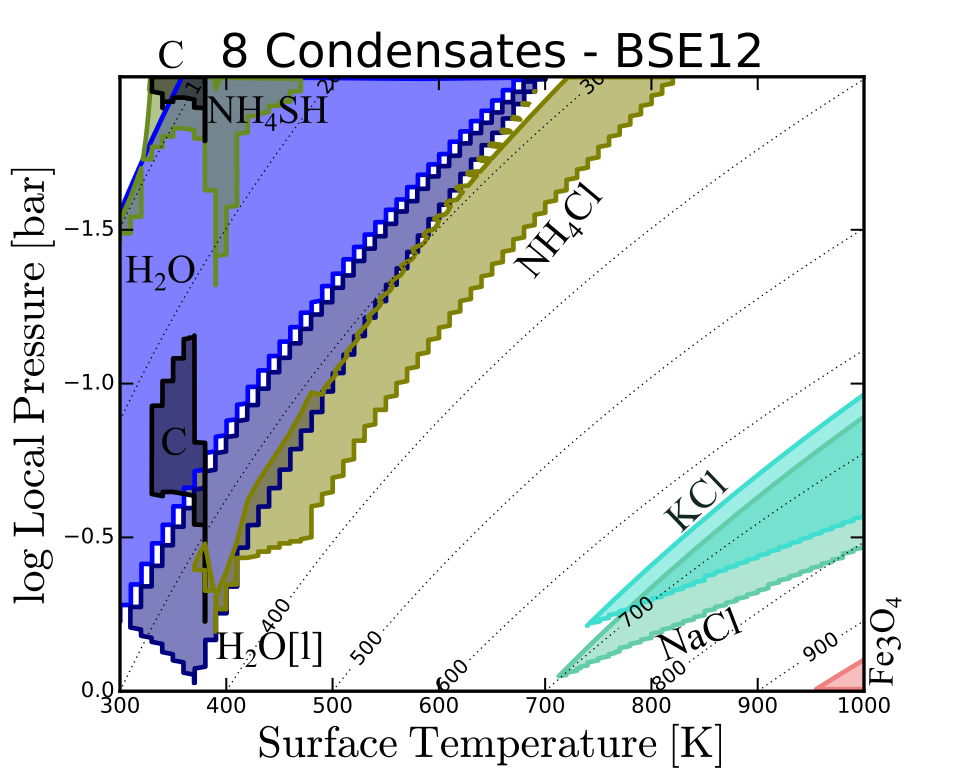}
\includegraphics[width = .32\linewidth, page=1]{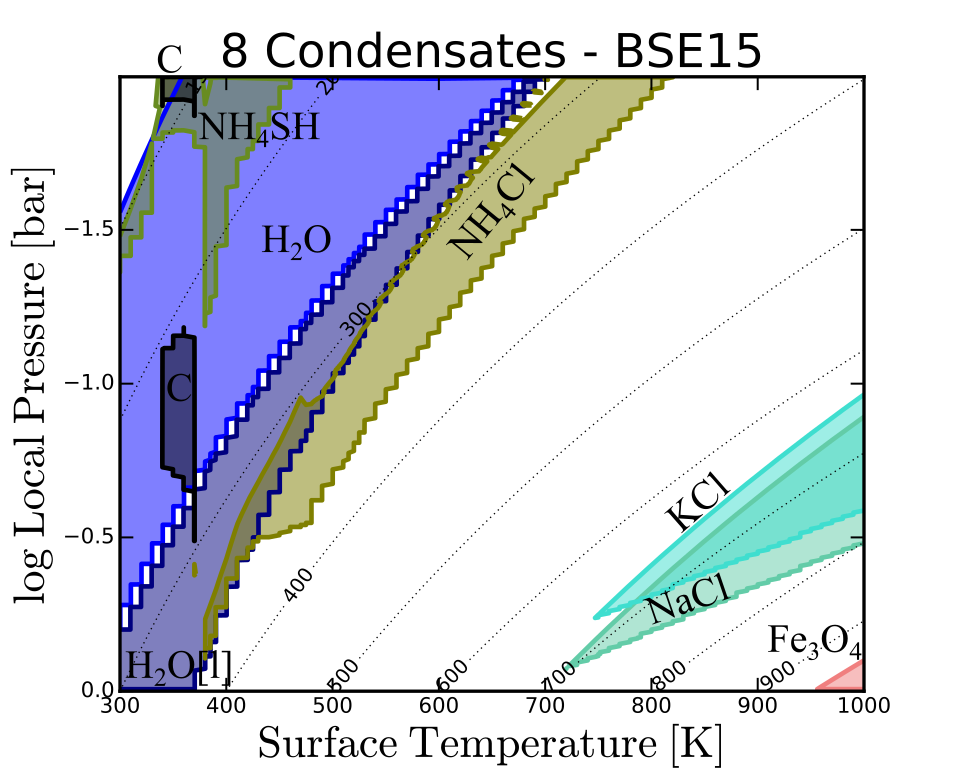}\\
\includegraphics[width = .32\linewidth, page=1]{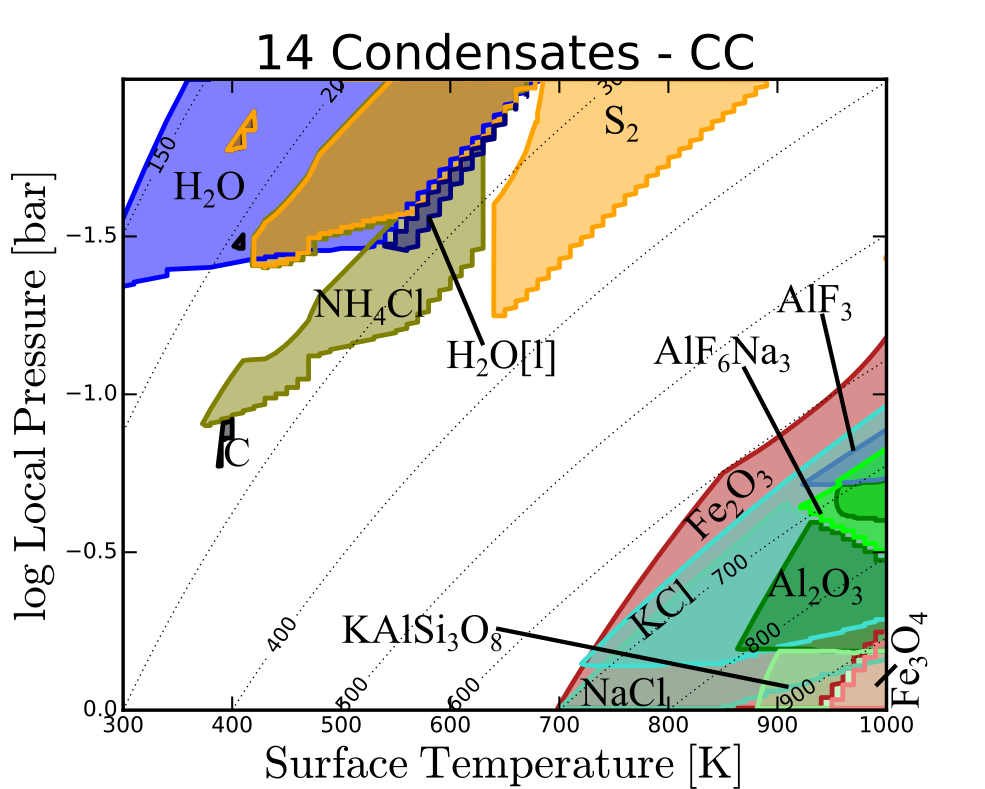}
\includegraphics[width = .32\linewidth, page=1]{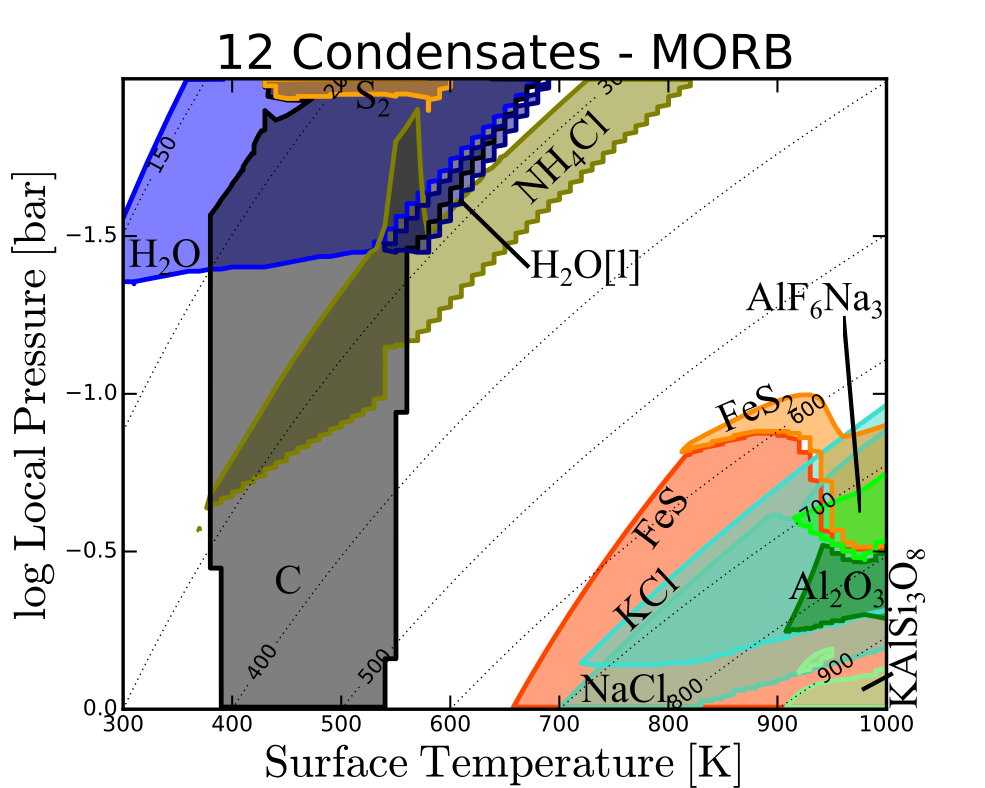}
\includegraphics[width = .32\linewidth, page=1]{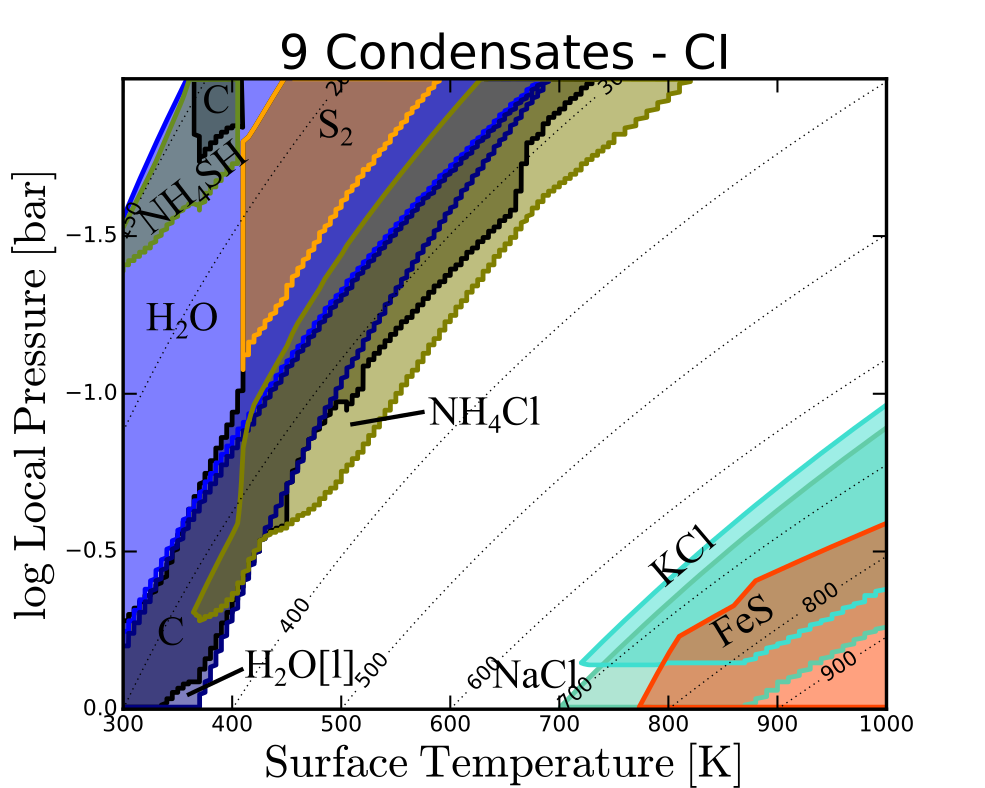}\\
\includegraphics[width = .32\linewidth, page=1]{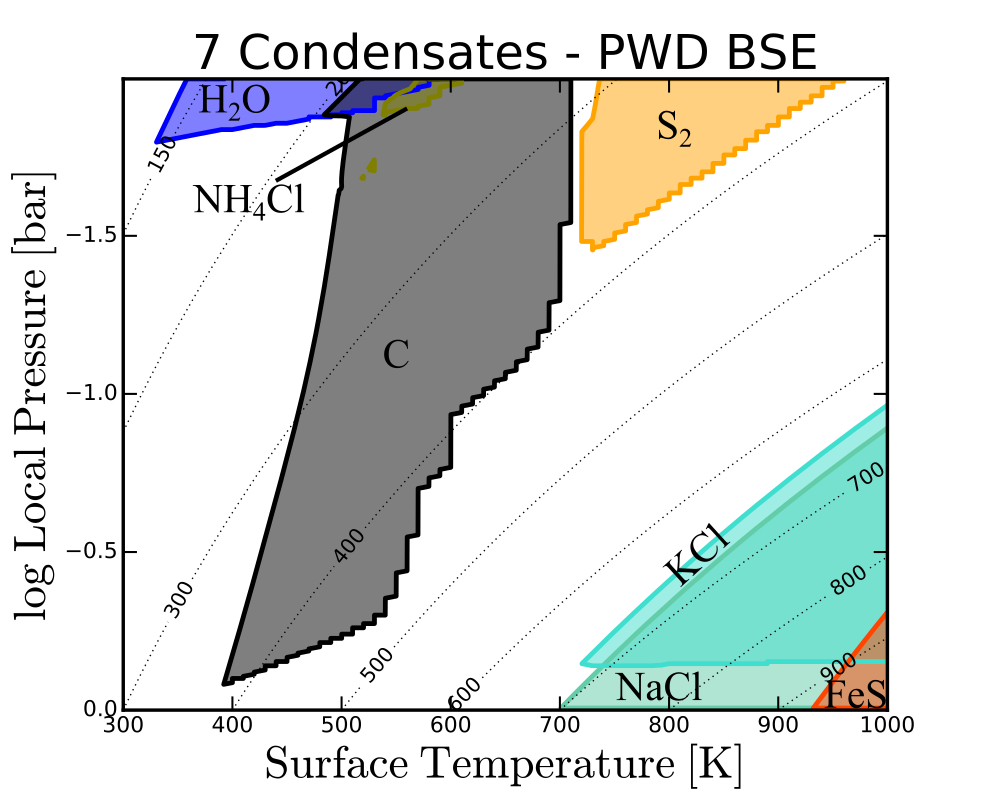}
\includegraphics[width = .32\linewidth, page=1]{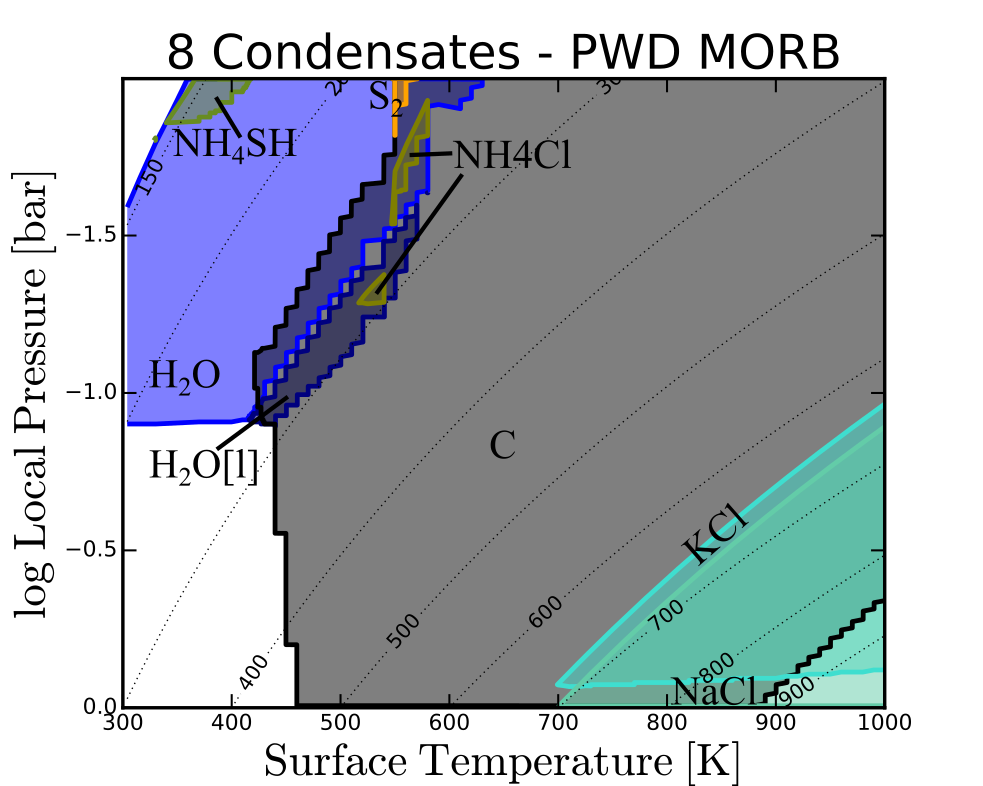}
\includegraphics[width = .32\linewidth, page=1]{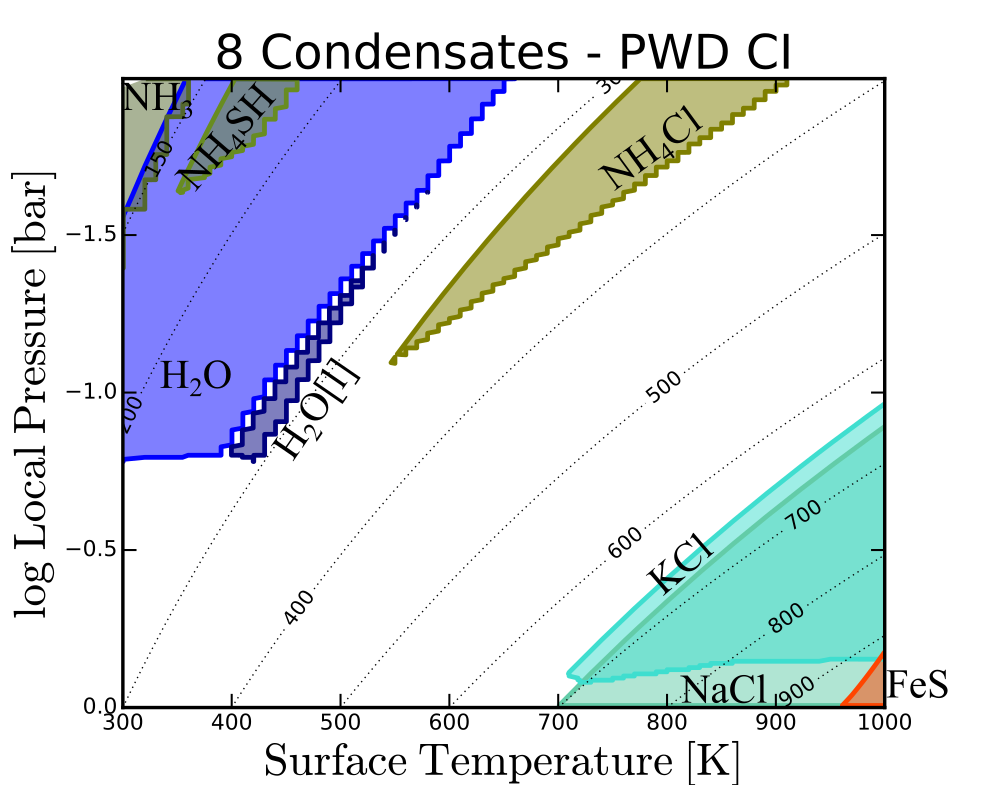}\\
\includegraphics[width = .32\linewidth, page=1]{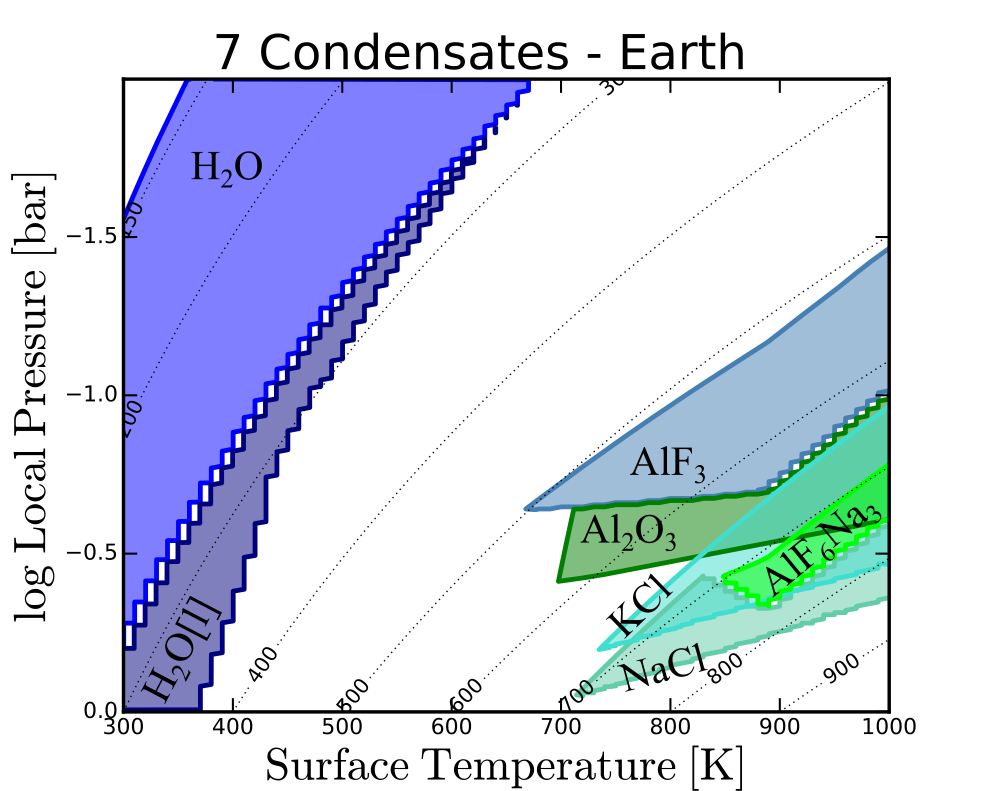}
\includegraphics[width = .32\linewidth, page=1]{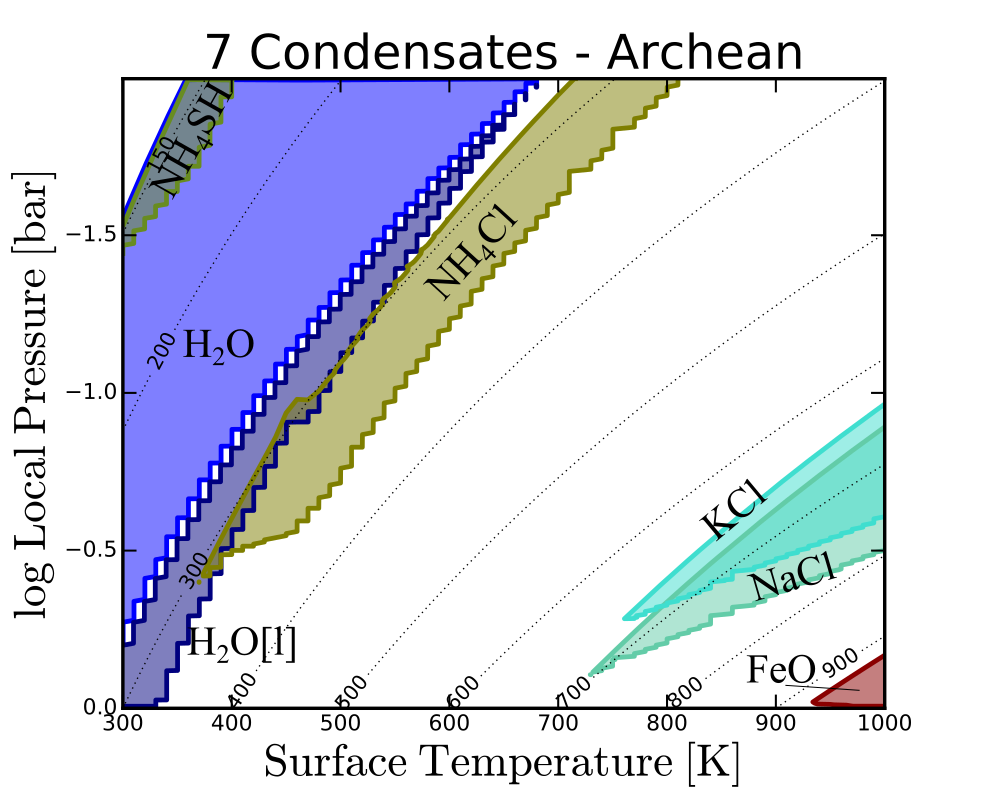}
\includegraphics[width = .32\linewidth, page=1]{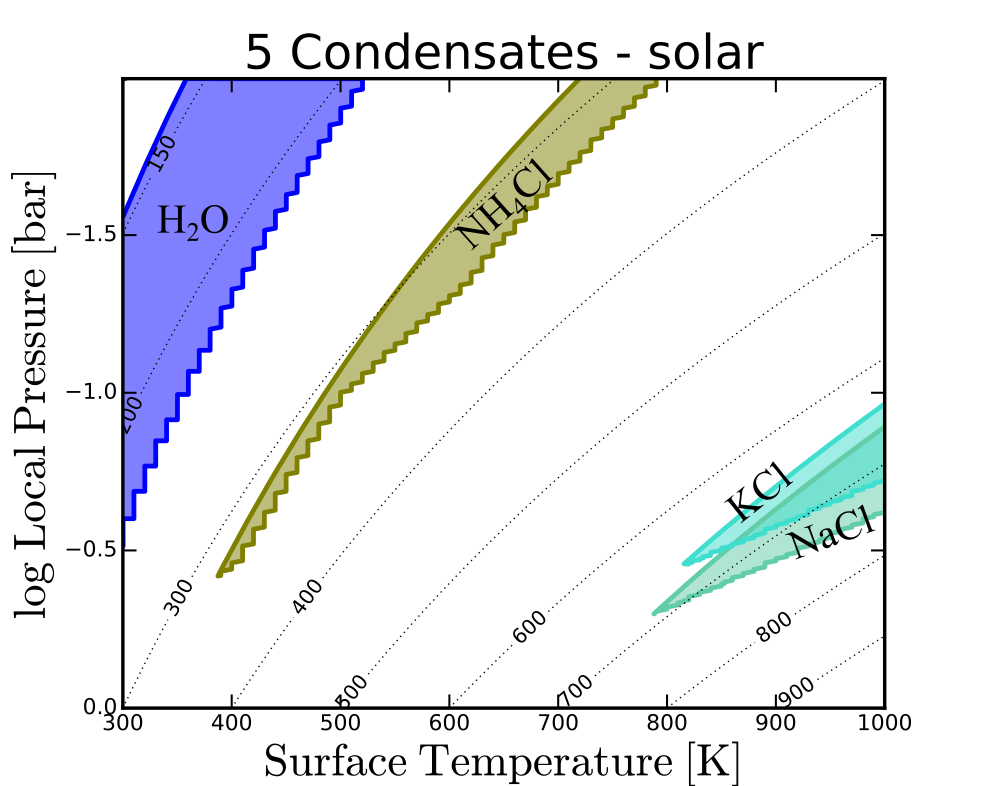}
\caption{All thermally stable condensates with normalised number densities of $n_\text{cond}/n_\text{tot}>10^{-10}$ locally for various elemental abundances are shown.
The liquid phase of condensates is indicated by [l], while all other condensates are solids.
Each column in every subplot is one bottom to top atmospheric model with the corresponding surface temperature on the x axis. 
Models have been calculated by steps of 10\,K in $T_\text{surf}$. 
The dotted lines refer to the the local gas temperature of the atmosphere in Kelvin.
All models are calculated for $p_\text{surf}=1\, \text{bar}$ and $\gamma = 1.25$.
The color scheme for the cloud species is consistent for all element abundances and given in legend above.}
\label{fig:CloudsAll-10}
\end{figure*}

\begin{figure*}[!t]
\centering
\includegraphics[width = .85\linewidth]{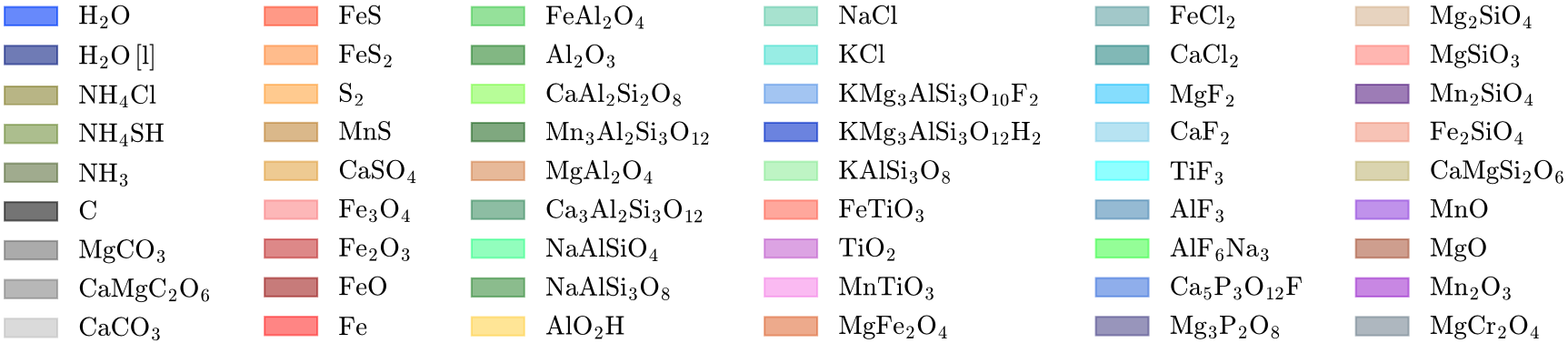}\\
\includegraphics[width = .32\linewidth, page=1]{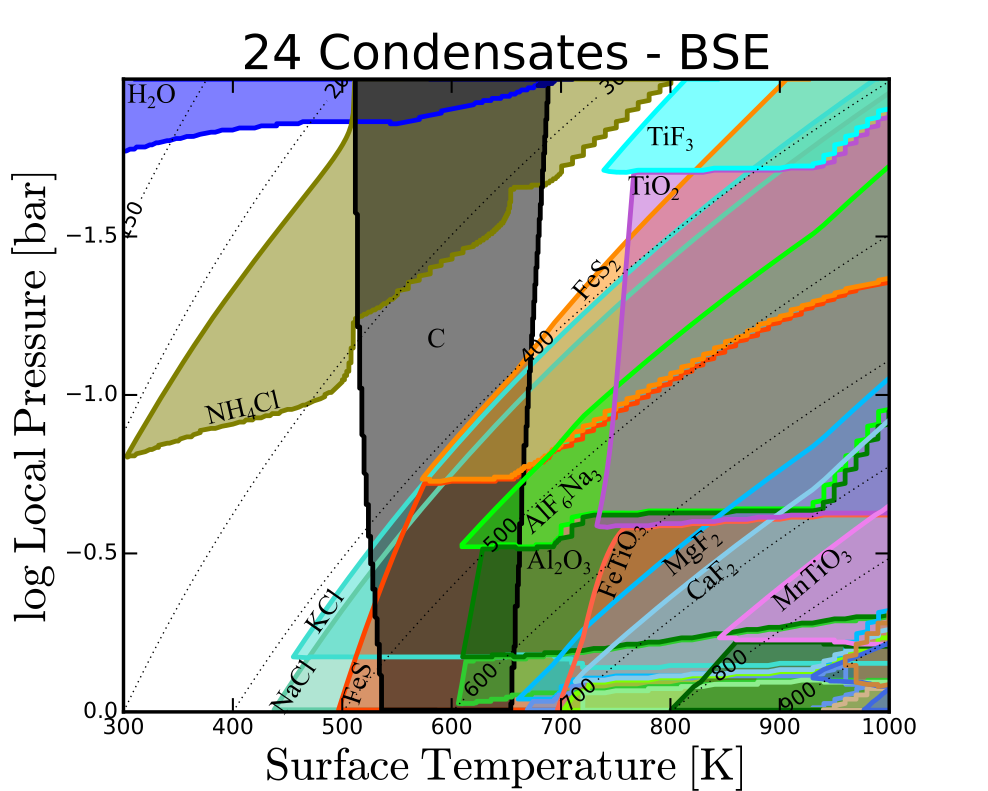}
\includegraphics[width = .32\linewidth, page=1]{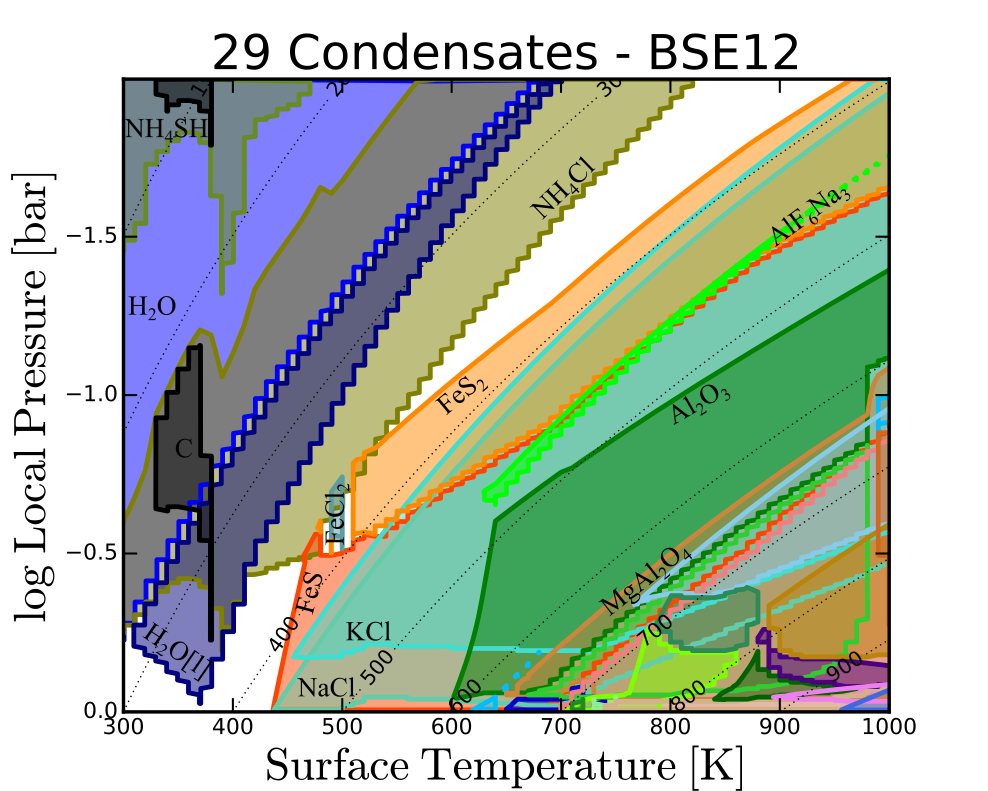}
\includegraphics[width = .32\linewidth, page=1]{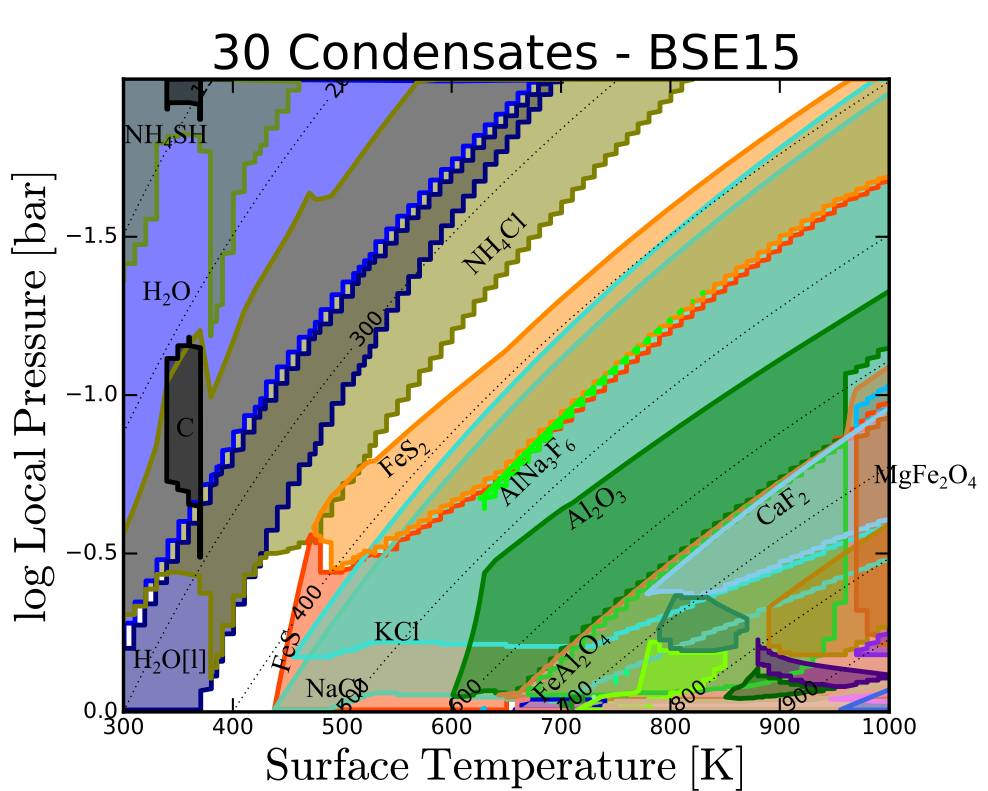}\\
\includegraphics[width = .32\linewidth, page=1]{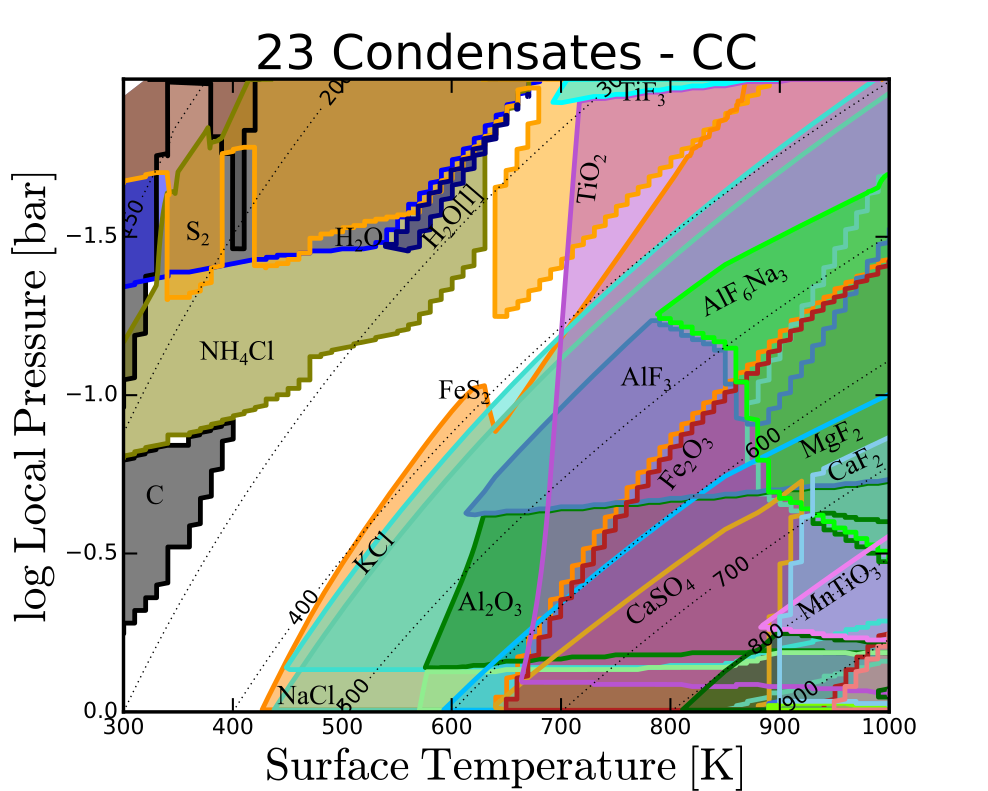}
\includegraphics[width = .32\linewidth, page=1]{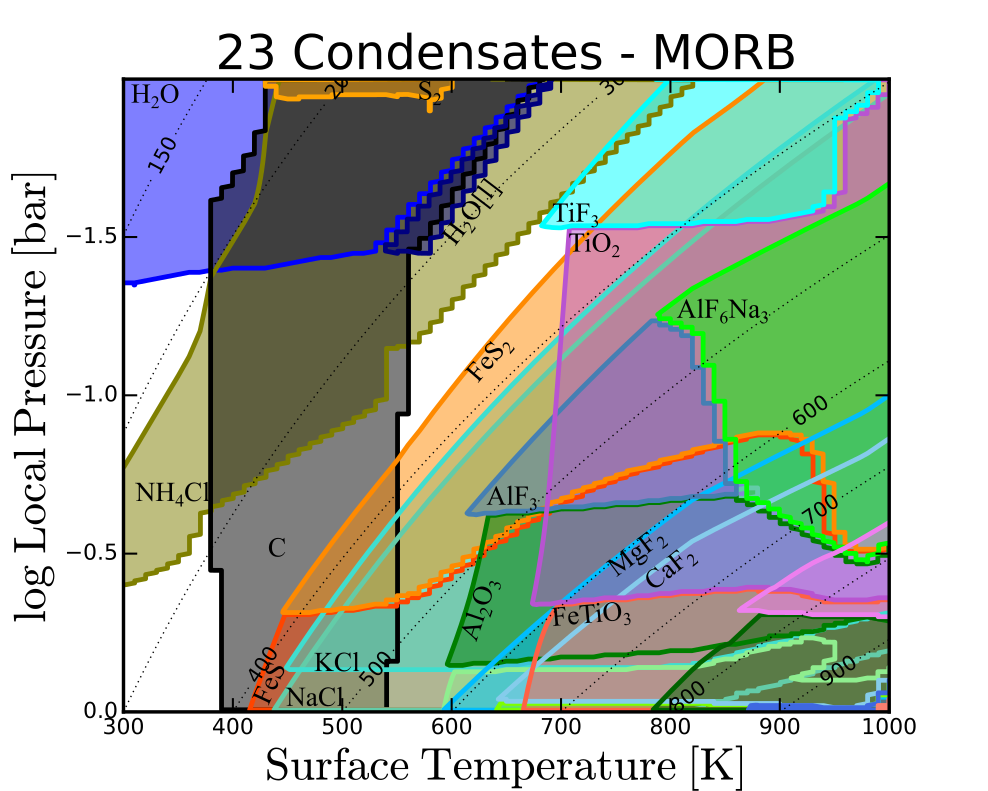}
\includegraphics[width = .32\linewidth, page=1]{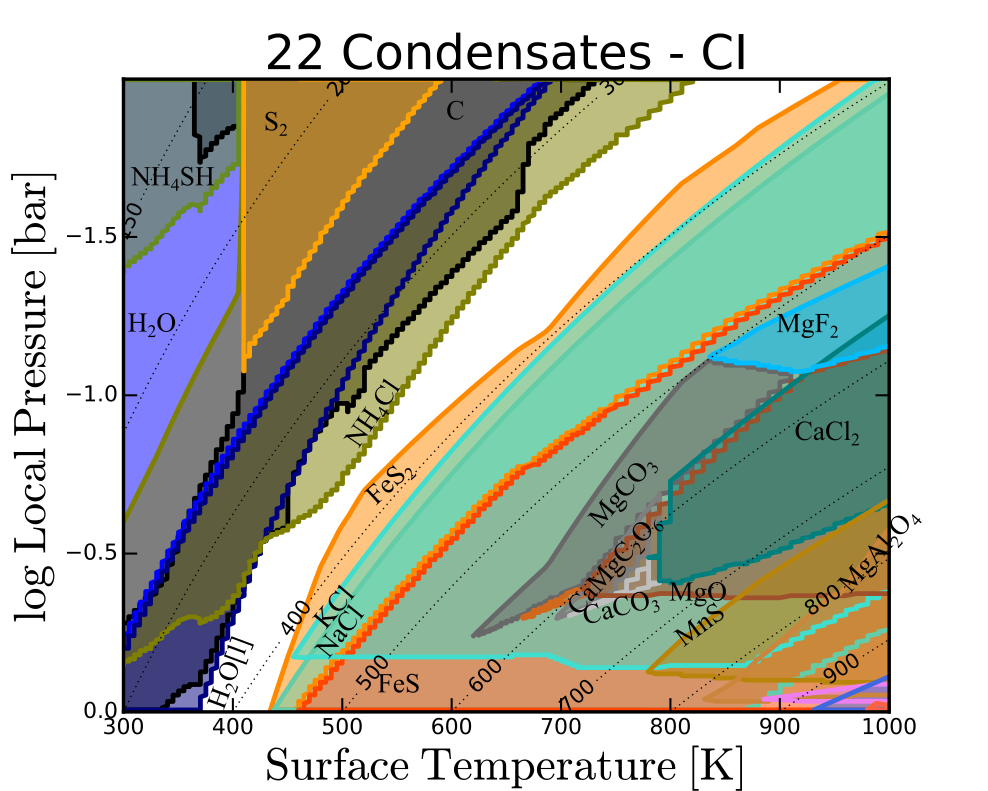}\\
\includegraphics[width = .32\linewidth, page=1]{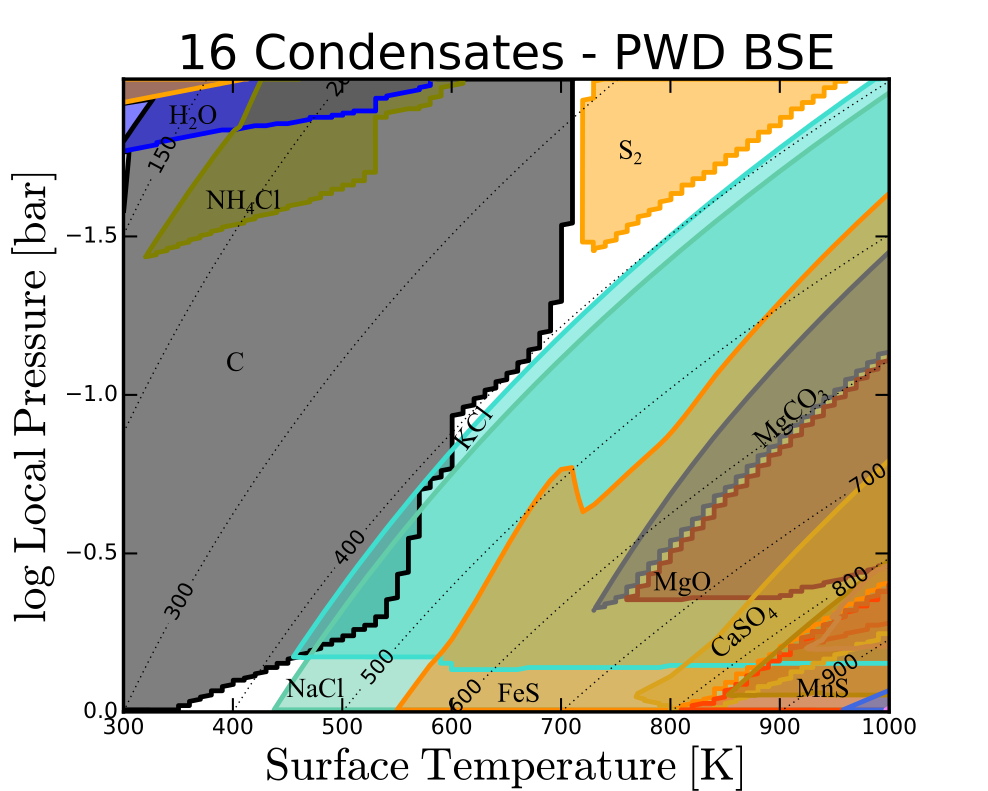}
\includegraphics[width = .32\linewidth, page=1]{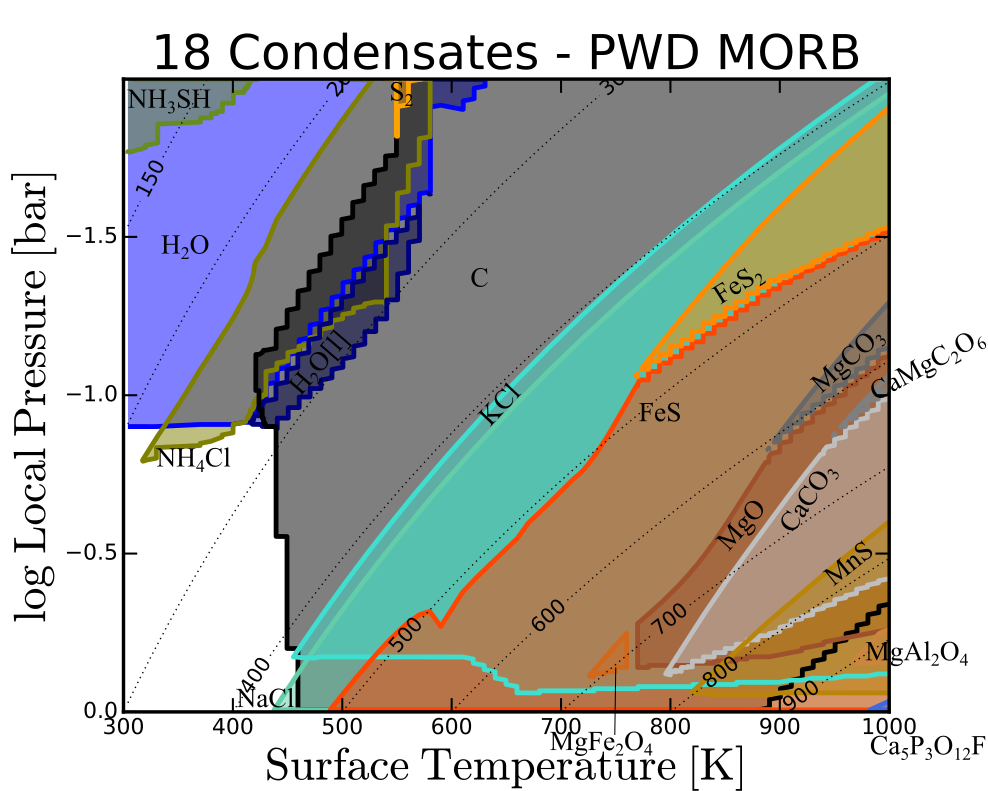}
\includegraphics[width = .32\linewidth, page=1]{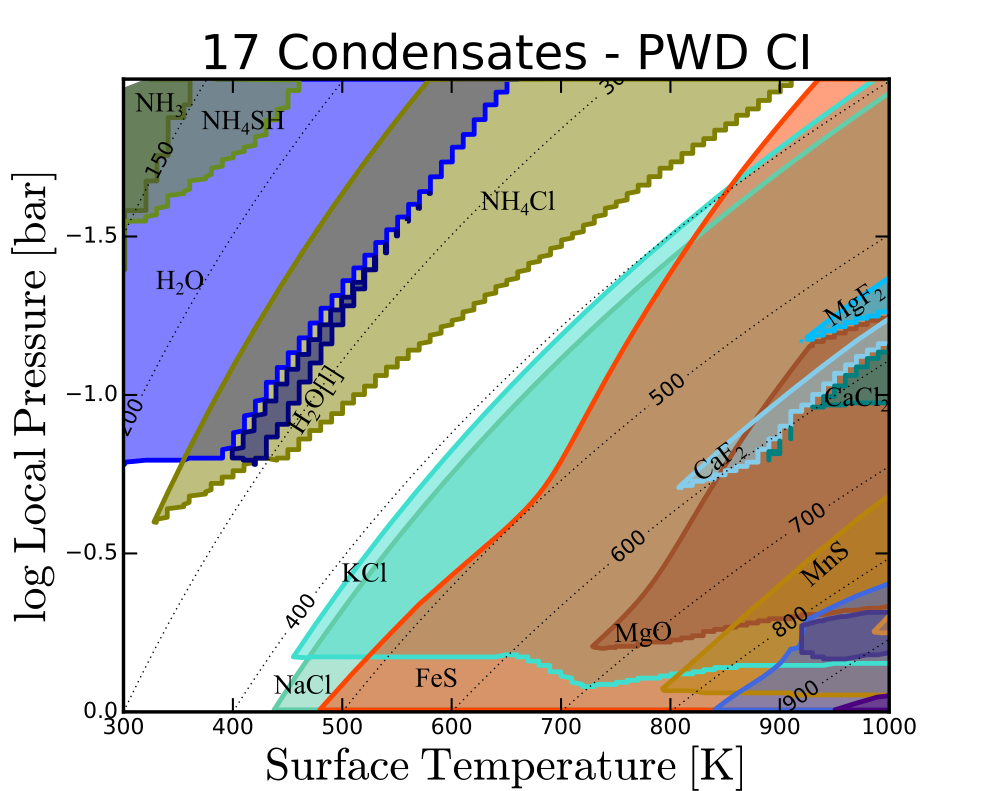}\\
\includegraphics[width = .32\linewidth, page=1]{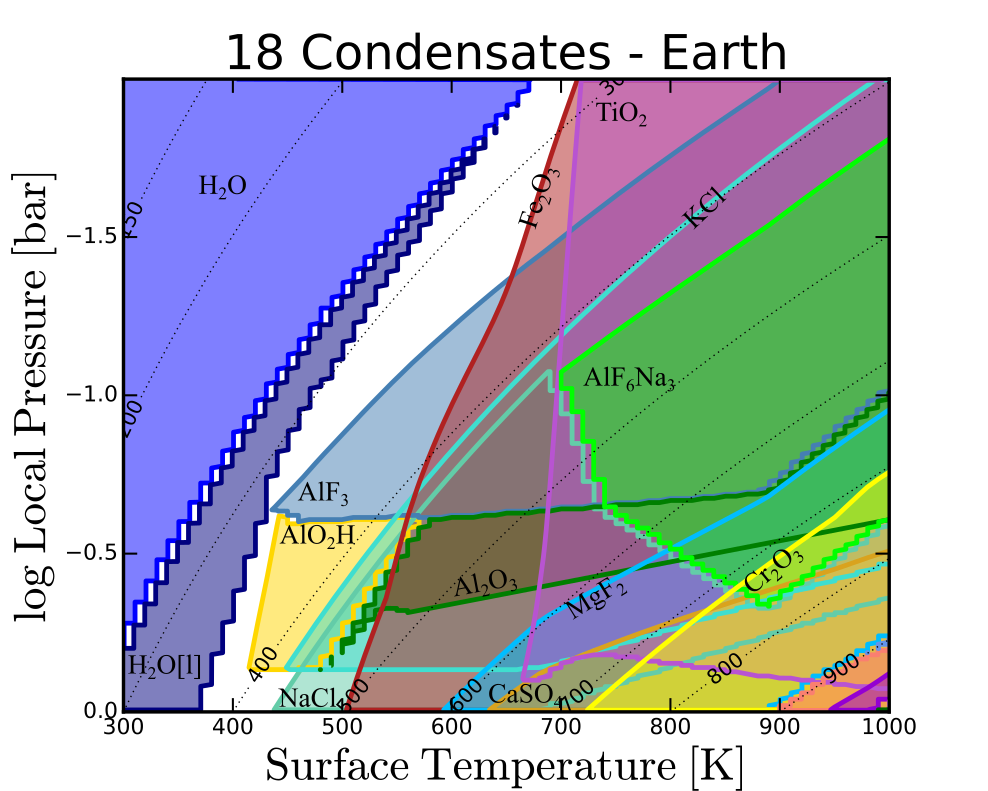}
\includegraphics[width = .32\linewidth, page=1]{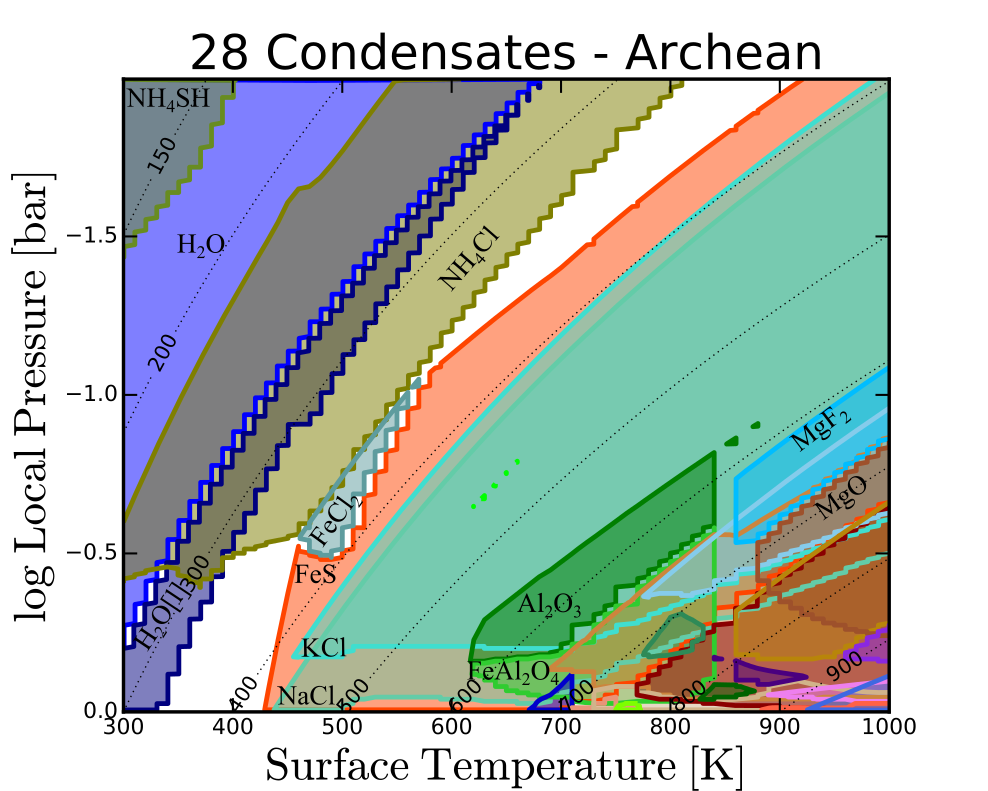}
\includegraphics[width = .32\linewidth, page=1]{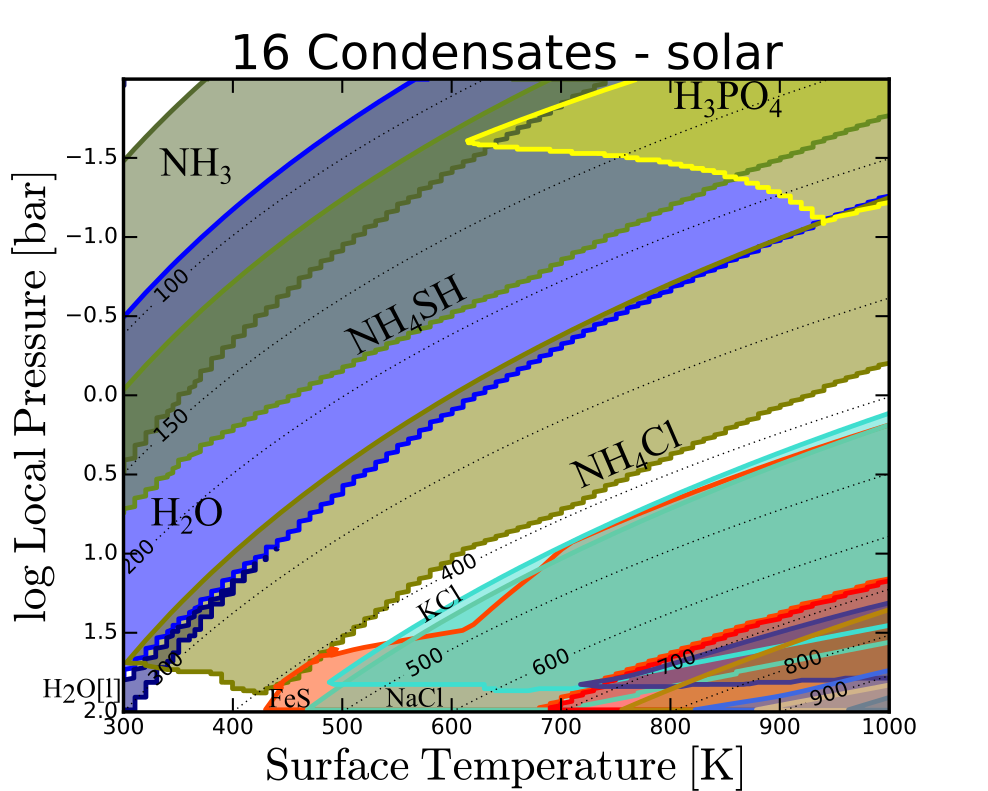}
\caption{As Fig.~\ref{fig:CloudsAll-10}, but with all condensates with normalised number densities higher than $n_\text{cond}/n_\text{tot}>10^{-20}$. The colour scheme is consistent with Fig.~\ref{fig:CloudsAll-10}, but extended to include the further condensates. 
All models are calculated for $p_\text{surf}=100\, \text{bar}$ and $\gamma = 1.25$.}
\label{fig:CloudsAll-20}
\end{figure*}

\subsubsection{Dominating cloud materials for varying crust composition}\label{sssec:mainclouds}
In Fig.~\ref{fig:CloudsAll-10} the most abundant thermally stable cloud condensates are represented as the normalised number density ($n_\text{cond}/n_\text{tot}$) of the number density of composite units ($n_\text{cond}$, see Sec.~\ref{ssec:cloudmodel}) with respect to the total gas density ($n_\text{tot}$) for the different sets of element abundances.
For each thermally stable cloud condensate, the regions, in the $(p_\text{gas},T_\text{surf})$ parameter space where the normalised number density exceeds a given threshold (${n_\text{cond}/n_\text{tot}>10^{-10}}$ for Fig.~\ref{fig:CloudsAll-10}) is considered as that condensate being stable and sufficiently abundant.
This region forms an individual \textit{cloud layer} for each condensate.
The high pressure border of a condensate cloud layer (\textit{cloud base}) is coinciding with the highest abundance of the specific condensate.
For the low pressure boundary of the cloud layer (\textit{cloud top}) two regimes exist:
1) The normalised number density of the cloud condensate drops below the threshold, or 2) the condensate undergoes a type 2 transition such that another condensate becomes stable (\textit{condensate chain}).
Examples for these two regimes are \ce{NH4Cl}[s], \ce{NaCl}[s], and \ce{KCl}[s] for the drop below the threshold and \ce{H2O}[l] to \ce{H2O}[s] or \ce{FeS}[s] to \ce{FeS2}[s] for the type 2 transitions for the atmosphere above a MORB like planetary surface.

Throughout the considered crust compositions, the cloud condensates which reach the overall highest normalised number densities ($n_\text{cond}/n_\text{tot}\approx10^{-2}$) are \ce{H2O}[l,s], \ce{C}[s], and \ce{NH3}[s].
Overall, the individual cloud layers can be separated into two categories, separated by $T_\text{gas}$ into high and low temperature cloud condensates.
For $T_\text{gas}\lesssim400\,$K we find \ce{H2O}[l,s], \ce{C}[s], \ce{S2}[s] and nitrogen containing condensates (\ce{NH3}[s], \ce{NH4Cl}[s], \ce{NH4SH}[s]) thermally stable.
For $T_\text{gas}\gtrsim600\,$K halides (\ce{NaCl}[s], \ce{KCl}[s]), iron sulphides (\ce{FeS}[s], \ce{FeS2}[s]), iron oxides (\ce{FeO}[s], \ce{Fe2O3}[s], \ce{Fe3O4}[s]), and aluminium containing condensates (\ce{Al2O3}[s], \ce{Al6F6Na3}[s], \ce{AlF3}[s], \ce{KAlSi3O8}[s]) are thermally stable and more abundant than $n_\text{cond}/n_\text{tot} > 10^{-10}$. 
The only cloud species to break this separation of high and low temperature clouds at normalised number densities of $n_\text{cond}/n_\text{tot} > 10^{-20}$ is \ce{C}[s] which can be stable at any ($p_\text{gas}, T_\mathrm{gas}$) point investigated in this paper.
Furthermore, \ce{C}[s] shows the unique behaviour of almost vertical cloud base rise with decreasing or increasing surface temperatures (530\,K and 660\,K for BSE, 380\,K and 550\,K for  MORB, respectively), confining the \ce{C}[s] condensation in a specific range of surface temperatures for these models.
However, in the CI model, \ce{C}[s] shows characteristics of the other cold temperature condensates and the cloud base follows the temperature curve.
The cloud base for all other cloud condensates is dependent on the local temperature and therefore diagonal in the $(p_\mathrm{gas}, T_\text{surf})$ space (see Fig.~\ref{fig:CloudsAll-10}), if not constrained by condensate chains.

Towards lower surface temperature the cloud condensates tend to remain stable, but decrease in normalised number density (comparison of Figs.\ref{fig:CloudsAll-10} and \ref{fig:CloudsAll-20}).
Again, a counterexample is \ce{C}[s], which becomes unstable with colder surface temperatures (BSE, MORB, PWD MORB).

For the low temperature cloud materials, \ce{H2O}[l,s] is the dominant cloud constituent, with cloud bases at various pressure levels, depending on the total H and O abundances.
See Sect.~\ref{ssec:surfacewater} for an in depth investigation of the pressure level of the \ce{H2O}[l,s] cloud base.

Independent of the total element abundance, \ce{NaCl}[s] is always thermally stable at slightly higher pressure levels than \ce{KCl}[s].
For all investigated crust compositions, both are stable condensates in the respective atmospheres.
The maximum normalised number density for both halides decreases with decreasing planetary surface temperature.
For $T_\text{surf}\approx700\,$K the maximum normalised number density for these halide cloud species drops below $n_\text{cond}/n_\text{tot}=10^{-10}$ in all of the models.

\begin{figure}[!th]
\centering
\includegraphics[width = .9\linewidth]{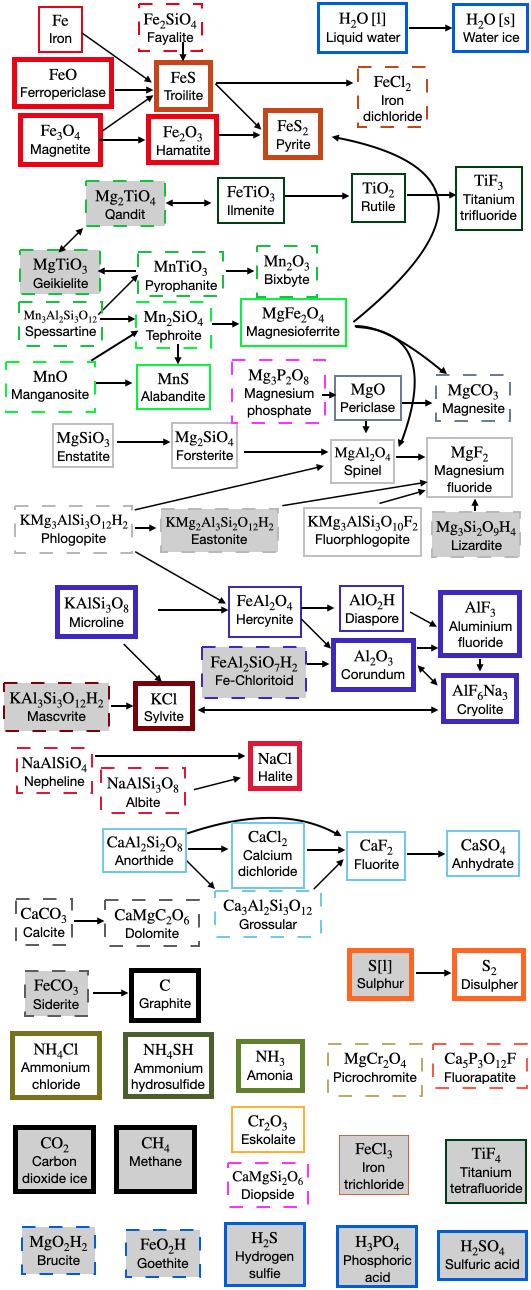}
\caption{Overview of cloud material transitions in all of the models for materials with $n_\text{cond}/n_\text{tot} > 10^{-20}$.
Different box styles correspond to the peak abundance of the condensates as in Table~\ref{tab:cloudspecies} ($n_\text{cond}/n_\text{tot} > 10^{-10}$ (thick), $n_\text{cond}/n_\text{tot} > 10^{-15}$ (thin), and $n_\text{cond}/n_\text{tot} > 10^{-20}$ (dashed)).
The colors represent the different condensate chains.
Along the arrows, the atmospheric pressure is decreasing and the condensates undergo a phase 2 transition \citep{2018A&A...614A...1W}.
Multiple arrows away from a specific condensate indicates that different transitions are found over the various models.
For example \ce{FeS} transitions to \ce{FeS2} in most models, but for BSE12 and Archean it also shows a transition to \ce{FeCl2}.
However, only one of them exists in any one model.
Condensates without an arrow are not part of any condensate chain and do not undergo any phase 2 transition.
Boxes with grey background refer to condensates that have only been found thermally stable in the high pressure models.}
\label{fig:CloudTransitions}
\vspace*{-7mm}
\end{figure}

\begin{table*}[!t]
\caption{A list of all cloud condensates according to their highest normalised number density over all element abundances considered in this work and $300\,\mathrm{K}\geq T_\mathrm{surf}\geq 1000\,\mathrm{K}$. All models with $p_\mathrm{surf}=1\,$bar are included. All low temperature ($T\lesssim 400\,$K) cloud condensate materials are indicated in bold.}
\label{tab:cloudspecies}
\centering
\vspace*{-2mm}
\begin{tabular}{c|ccc}
\hline
    & & &\\[-2.2ex] 
Peak normalised number density	&	$n_\text{cond}/n_\text{tot} > 10^{-10}$	&	$10^{-10} > n_\text{cond}/n_\text{tot} > 10^{-15}$	&	$10^{-15}>n_\text{cond}/n_\text{tot} > 10^{-20}$	\\
\hline
Water    &\textbf{\ce{H2O}, \ce{H2O}[l]} & &\\
Phlogopite & & \ce{KMg3AlSi3O10F2}& \ce{KMg3AlSi3O12H2}\\
Carbon   &\textbf{\ce{C}} & \ce{CaCO3}& \ce{MgCO3}, \ce{CaMgC2O6}\\
Nitrogen  &\textbf{\ce{NH4Cl}, \ce{NH4SH}, \ce{NH3}} & & \\
Sulfur  &\textbf{\ce{S2}}, \ce{FeS}, \ce{FeS2} &\ce{CaSO4}, \ce{MnS} & \\
Phosphorus  & & & \ce{Mg3P2O8}\\
Apatite&&&\ce{Ca5P3O12F}\\
Chlorid  & \ce{NaCl}, \ce{KCl}& \ce{CaCl2}& \ce{FeCl2} \\
Flourine   & \ce{AlF3}, \ce{AlF6Na3}&  \ce{MgF2}, \ce{CaF2}, \ce{TiF3}& \\
Metals &&\ce{Fe}&\\
Hydroxide & &\ce{AlO2H}&\\
Metaloxides  &\makecell{\ce{FeO}, \ce{Fe2O3}, \ce{Fe3O4}\\\ce{FeAl2O4}, \ce{Al2O3} }&
\makecell{\ce{MgO},  \ce{TiO2} , \ce{Cr2O3}\\ \ce{MgFe2O4}, \ce{FeTiO3}, \ce{MgAl2O4} }& 
 \ce{MnO}, \ce{Mn2O3} \ce{MgCr2O4}, \ce{MnTiO3}\\
Silicates   &\ce{KAlSi3O8} & \makecell{\ce{Mg2SiO4}, \ce{MgSiO3}\\ \ce{CaAl2Si2O8}, \ce{Mn3Al2Si3O12}}&
\makecell{\ce{NaAlSiO4}, \ce{NaAlSi3O8}\\ \ce{CaMgSi2O6}, \ce{Ca3Al2Si3O12}\\ \ce{Mn2SiO4}, \ce{Fe2SiO4}} \\
\hline
\end{tabular}
\end{table*}

\subsubsection{Cloud material diversity and transitions}\label{sssec:clouddiversity}
In addition to the most abundant cloud species discussed in Sect.~\ref{sssec:mainclouds} many more cloud condensates are thermally stable for the $(p_\text{gas},T_\text{gas})$ range studied here, but do not reach normalised number densities of $n_\text{cond}/n_\text{tot} > 10^{-10}$.
In order to understand the diversity and chemical connection of potential cloud condensates in exoplanet atmospheres and also potential links of cloud species to the surface conditions ($p_\text{surf},T_\text{surf},$ composition), 
we further investigate thermally stable cloud condensates with lower normalised number densities.
Although these cloud condensates are less likely to be detected, they can help the further understanding of cloud formation, diversity and stability in general.
The thermally stable cloud condensates with $n_\text{cond}/n_\text{tot} > 10^{-20}$ for the different surface compositions can be seen in Fig.~\ref{fig:CloudsAll-20}.
This reveals more condensates that are thermally stable for the high temperature condensate species, while no further low temperature condensate become stable.
However, the gap between the low and high temperature cloud condensates is being filled from the high temperature condensates at relatively low normalised number densities.
The general separation into the two temperatures regimes at a local temperature of $T_\mathrm{gas}\approx400\,$K remains.

In Table~\ref{tab:cloudspecies} all 55 cloud materials that are thermally stable in at least one atmosphere above a rocky surface
with surface temperatures $300\,K\leq T_\text{surf}\leq 1000\,$K and $p_\text{surf}=1\,$bar are listed and sorted with respect to their maximum normalised number density $n_\text{cond}/n_\text{tot}$.
As described in Sect.~\ref{ssec:BSEmodel}, most of these condensates can be arranged in cloud chains by type 2 transitions.
Only 7 condensates (\ce{S2}[s], \ce{C}[s], \ce{NH4Cl}[s], \ce{NH4SH}[s], \ce{NH3}[s], \ce{MgCr2O4}[s], and \ce{Ca5P2O12F}[s]) show no apparent type 2 transition to another condensate and therefore do not belong to any cloud chain.
In Fig.~\ref{fig:CloudTransitions} these condensate chains are visualised. 
Not all of these transitions occur in all models for different element abundances.

\begin{figure*}[!t]
\centering
\includegraphics[width = .32\linewidth, page=1]{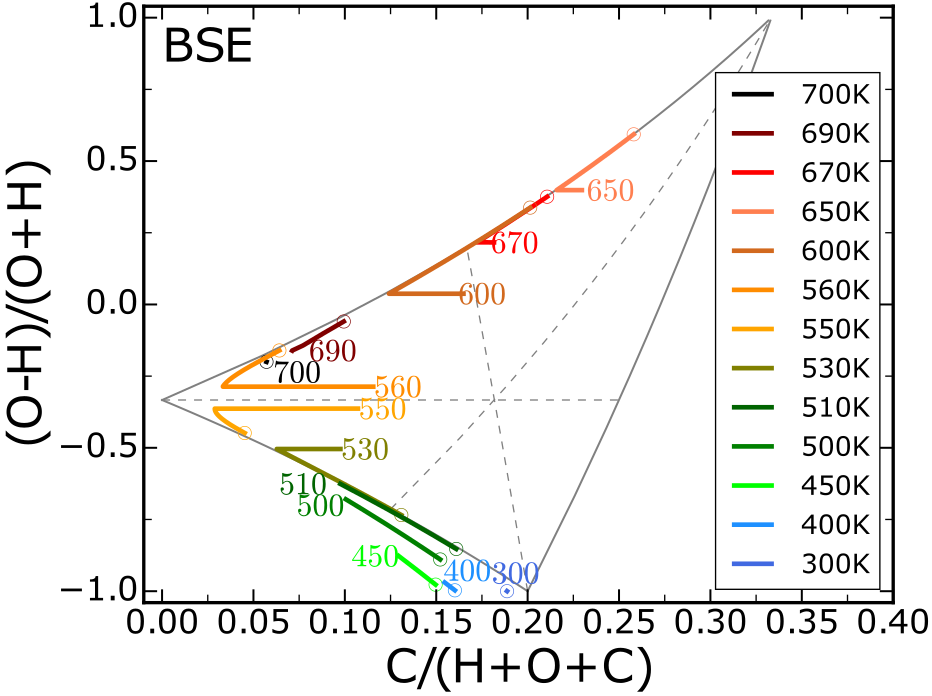}
\includegraphics[width = .32\linewidth, page=1]{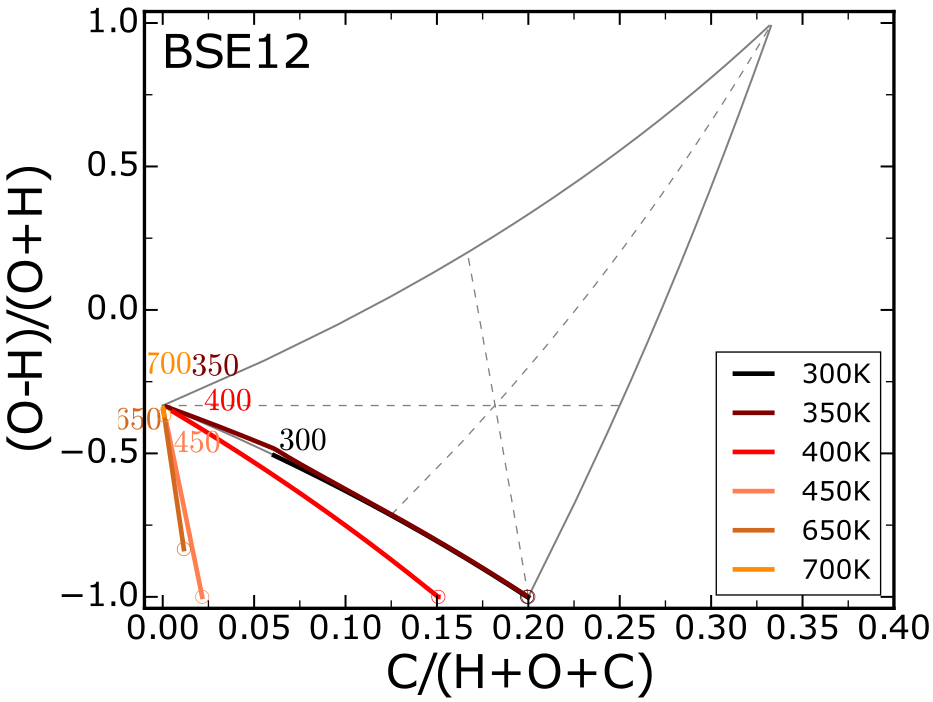}
\includegraphics[width = .32\linewidth, page=1]{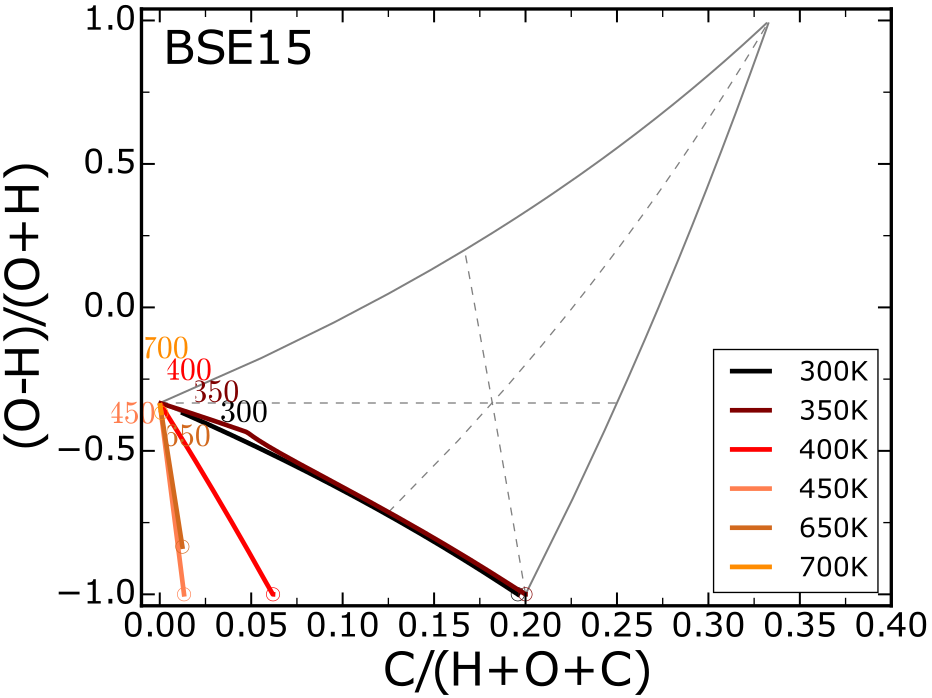}\\
\includegraphics[width = .32\linewidth, page=1]{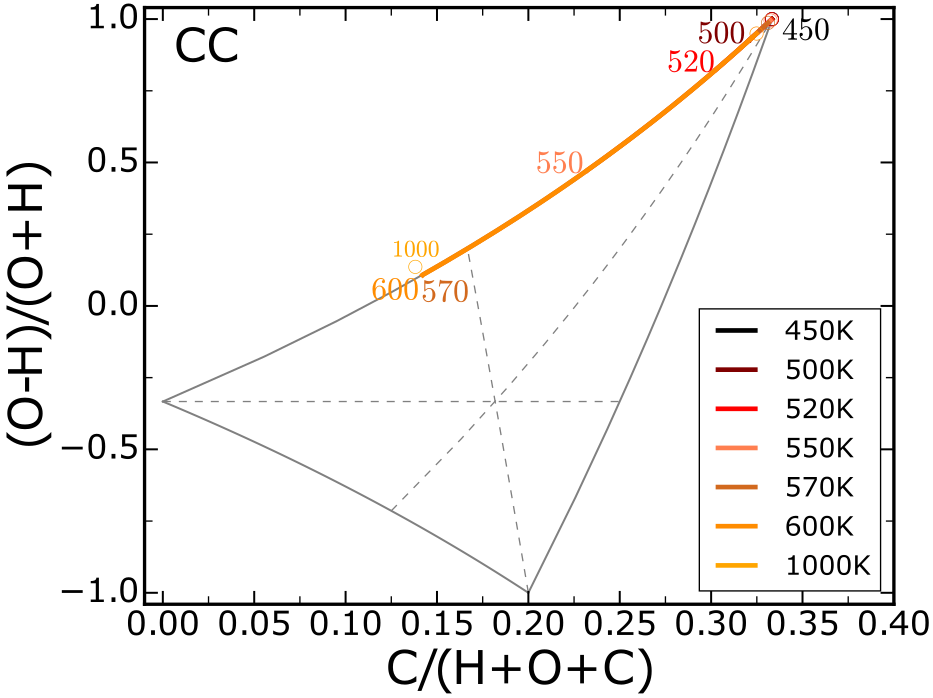}
\includegraphics[width = .32\linewidth, page=1]{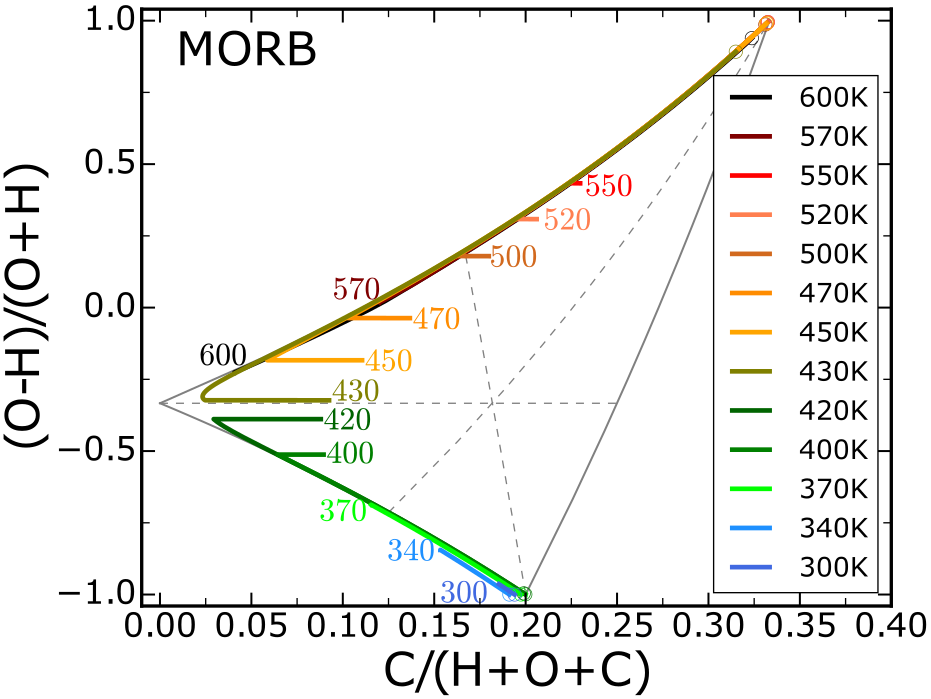}
\includegraphics[width = .32\linewidth, page=1]{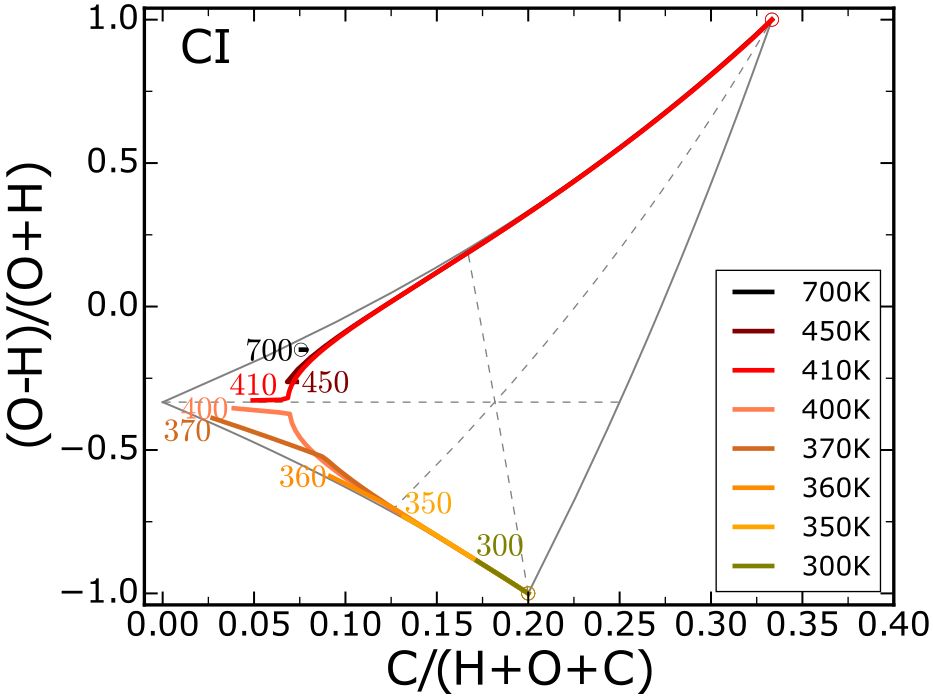}\\
\includegraphics[width = .32\linewidth, page=1]{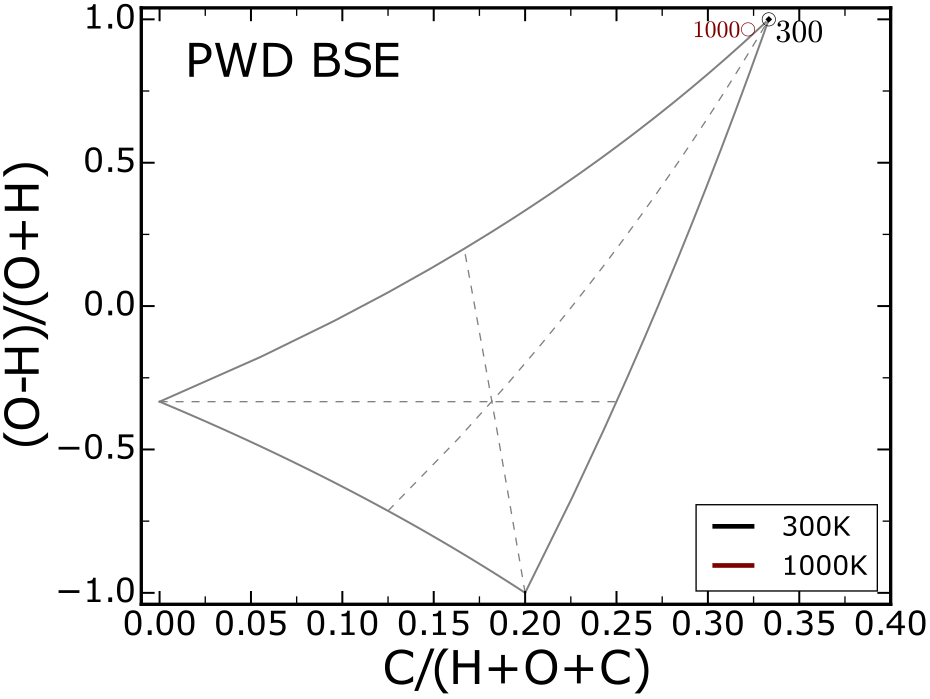}
\includegraphics[width = .32\linewidth, page=1]{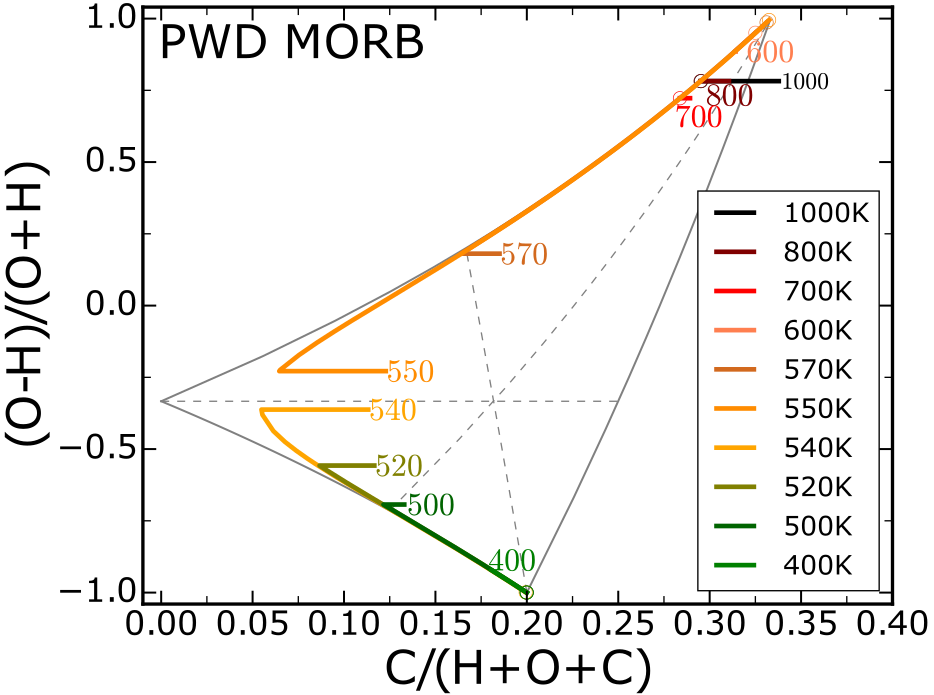}
\includegraphics[width = .32\linewidth, page=1]{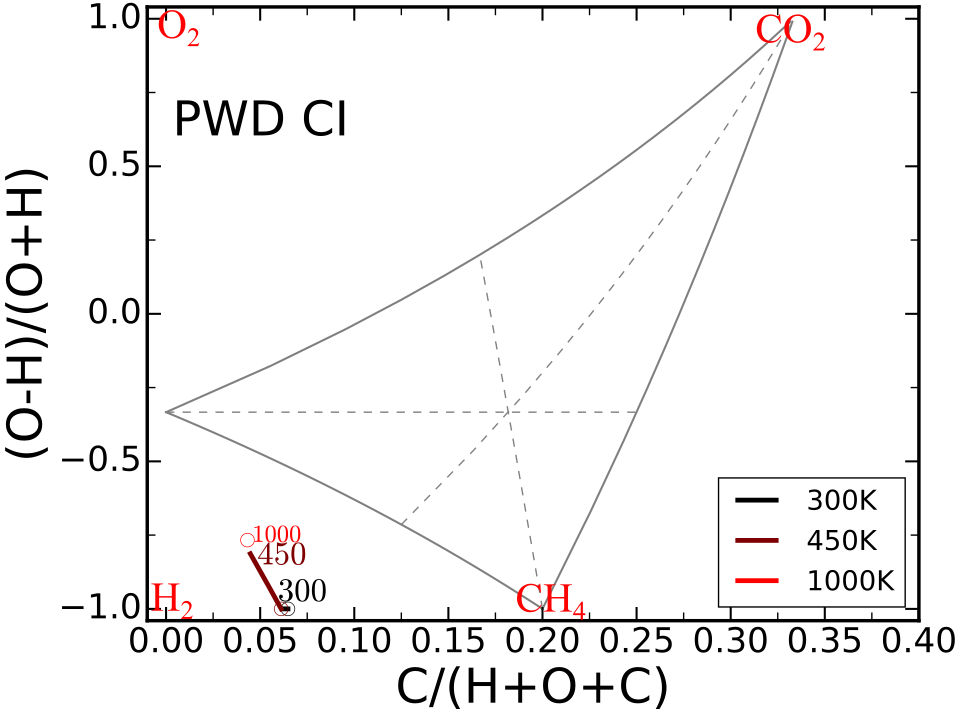}\\
\includegraphics[width = .32\linewidth, page=1]{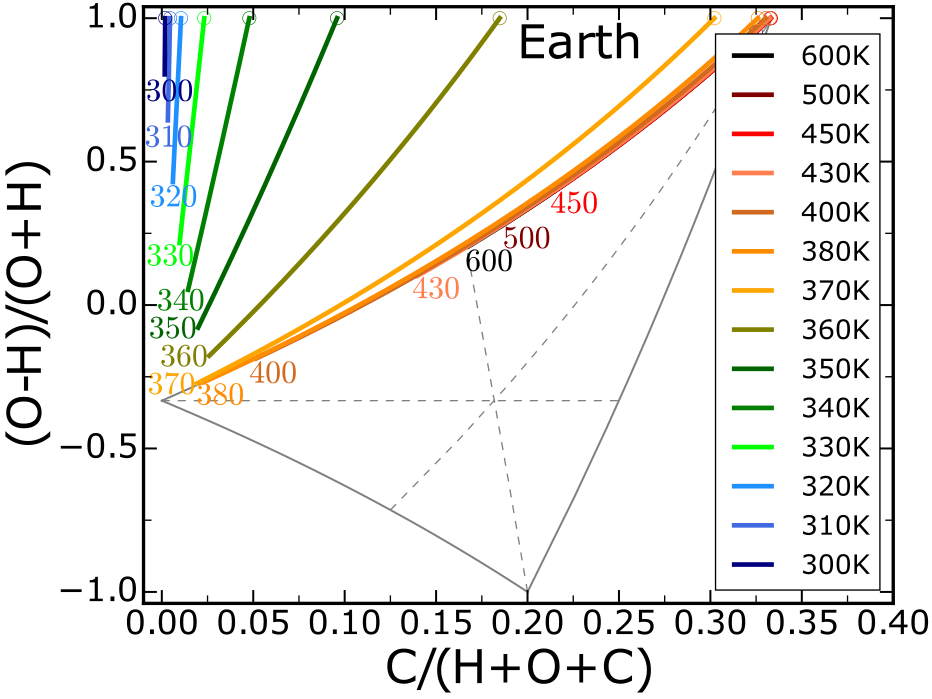}
\includegraphics[width = .32\linewidth, page=1]{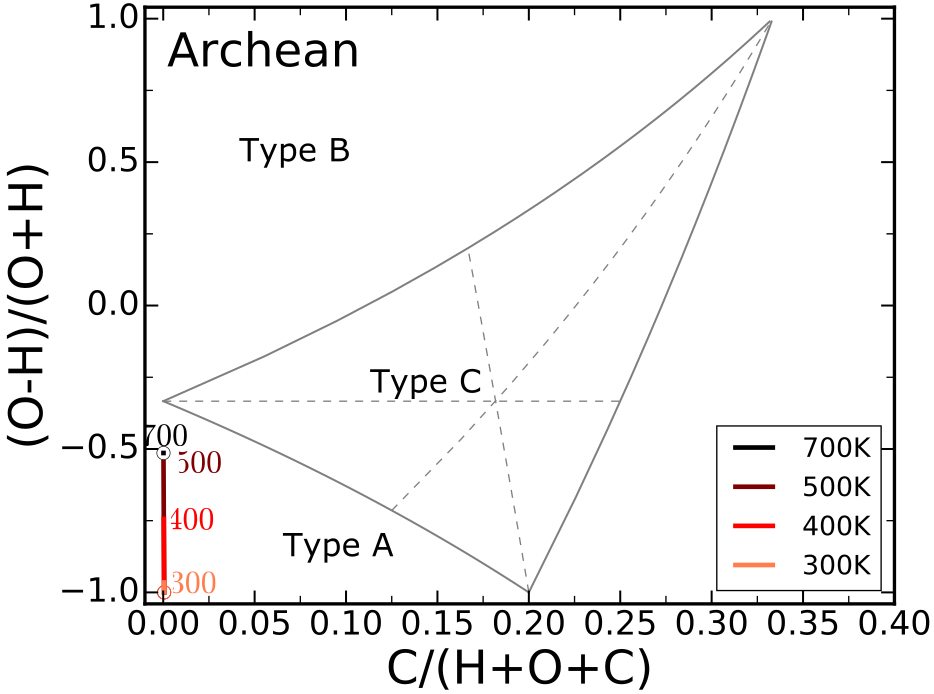}
\includegraphics[width = .32\linewidth, page=1]{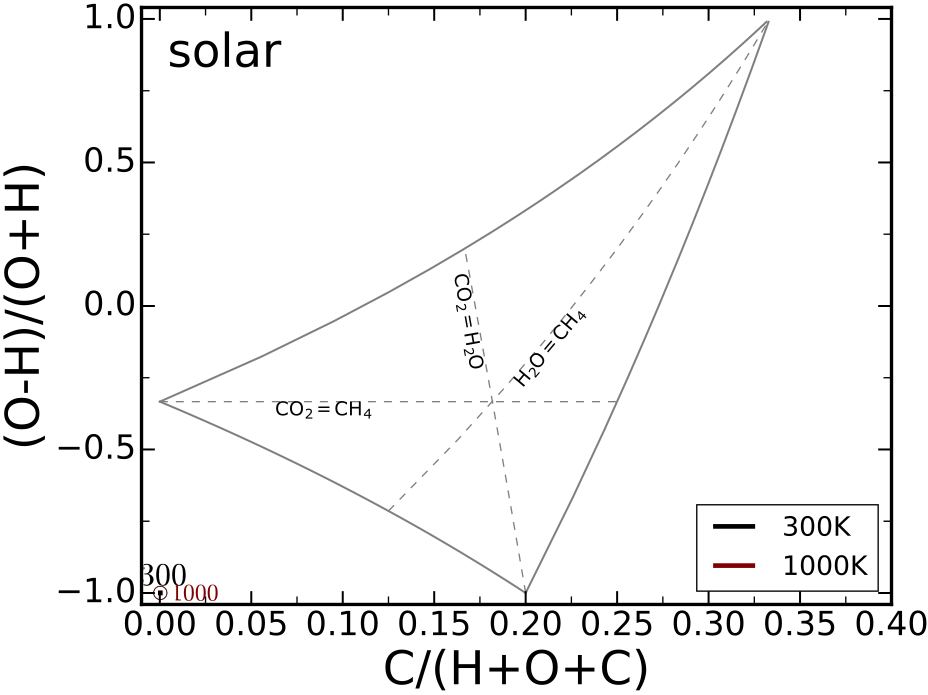}
\caption{Atmospheric compositions with decreasing pressure and various surface temperatures. The base of each atmospheric profile is indicated by the corresponding $T_\text{surf}$ and the top by open circles. The axis of overall redox state measured from O and H abundances as (O-H)/(O+H)  and C content relative to H, O, and C are the same as in \citet{Woitke2020a}. The atmospheric types, atmospheric end members of certain points as well as lines of equal abundances of certain molecules are indicated in the bottom right plots.}
\label{fig:Atmos}
\end{figure*}

\subsection{Atmospheric gas composition}\label{ssec:atmo}
The formation of clouds in an atmosphere removes the effected elements partially from the gas phase and therefore changes the composition of the atmosphere with respect to the atmosphere without condensation.
This results in a depleted atmospheric compositions above the cloud layer.
In this section we investigate the changes in the atmospheric composition with atmospheric height.

\citet{Woitke2020a} introduced an atmospheric classification based on the gas phase abundances of atmospheres only consisting of H, C, N, and O, resulting in three distinctive atmospheric types.
\begin{itemize}
\item Type A atmospheres are hydrogen-rich; they contain \ce{H2O}, \ce{CH4}, \ce{NH3} and either \ce{H2} or \ce{N2}, but no \ce{CO2} and no \ce{O2}.
\item Type B atmospheres are oxygen-rich; they contain \ce{O2}, \ce{H2O}, \ce{CO2} and \ce{N2}, but no \ce{CH4}, no \ce{NH3} and no \ce{H2}.
\item Type C atmospheres show the coexistence of \ce{CH4}, \ce{CO2}, \ce{H2O} and \ce{N2}, with traces of \ce{H2}, but no \ce{O2} and no \ce{CO}.
\end{itemize}

In Fig.~\ref{fig:Atmos}, we show how the gas phase element abundances change with height for rocky planets differing in their surface rock composition and their surface temperatures.
As in \citet{Woitke2020a}, the atmospheric composition is shown in the {(O-H)/(O+H) over C/(H+O+C)} plane, where H, C, and O stand for the respective element abundances.
(O-H)/(O+H) ranges from 1 (only oxygen, no hydrogen, fully oxidising) to -1 (only hydrogen, no oxygen, fully reducing) and therefore is a measure of the overall redox potential of the atmosphere, that is how likely oxidising and reducing reactions are.
C/(H+O+C) is a measure of the carbon fraction in the system, with respect to H, O, and C.
The solid grey lines in Fig.~\ref{fig:Atmos} show the separation of the different atmospheric types (type A below the triangle, type B above and type C inside).
The dashed lines reflect equal abundances of \ce{H2O} and \ce{CO2}, \ce{H2O} and \ce{CH4}, as well as \ce{CH4} and \ce{CO2} (see lower right panel in Fig.~\ref{fig:Atmos}).
Each coloured line in Fig.~\ref{fig:Atmos} corresponds to one of our hydrostatic, polyropic atmospheric models.
The composition at the bottom of the atmosphere is marked by the surface temperature in Fig.~\ref{fig:Atmos}, whereas the top of our atmospheric model at 10\,mbar is marked with an open circle.
This allows to see the changes in the atmospheric composition of the atmosphere.
The cloud condensates that are thermally stable in each of the models can be seen in the respective subplot in Figs.~\ref{fig:CloudsAll-10} and\ref{fig:CloudsAll-20}.

Models with surface temperature $T_\text{surf}\gtrsim 600\,$K show very little change in the parameter space in Fig.~\ref{fig:Atmos} with increasing height, as condensation of for example \ce{KCl}[s] or \ce{NaCl}[s] does not affect the abundances of C, H, and O.
On the other hand cloud condensates like \ce{Al2O3}[s] or \ce{TiO2}[s] do affect the oxygen element abundance.
However, these condensates are limited by the much less abundant elements Al and Ti and therefore the overall oxygen abundance and thus (O-H)/(O+H) does not change significantly.

For surface temperatures $T_\text{surf}\lesssim 600\,$K, the atmospheric elemental abundances drastically change with atmospheric height.
Especially C[s] and \ce{H2O}[l,s] become thermally stable and therefore cause the atmospheres to evolve towards either oxygen or hydrogen rich atmospheres.
For oxygen rich atmospheres, the carbon content can reach up to $1/3$ (pure \ce{CO2}), while hydrogen rich atmospheres can only reach up to a carbon content of $1/5$ (pure \ce{CH4}).
These regions for (O-H)/(O+H) are the only regions in Fig.~\ref{fig:Atmos} that are not effected by condensation of \ce{H2O}[l,s] and C[s] for lower temperatures \citep[see also][]{Woitke2020a}.
In which of these two states an atmosphere evolves, is determined by whether the near-crust atmosphere is over-abundant in O or H with respect to \ce{H2O}, that is whether the atmospheric composition lies above or below (O-H)/(O+H)=-0.33.
This is equivalent to whether \ce{CO2} or \ce{CH4} are more abundant than the other.
This motivates the subtypes of C1 and C2  for type C atmospheres above and below the (O-H)/(O+H)=-0.33 threshold, respectively.
By condensation, the atmosphere cannot change from one to another type C atmosphere.
Thus measurements, that infer a specific atmospheric type in the high atmosphere can constrain the possible atmospheric conditions at the surface.

The removal of condensate results in the atmosphere becoming  either \ce{CO2} dominated (e.g. rocky planets with MORB or CI composition and high $T_\mathrm{surf}$) or \ce{CH4} dominated atmospheres (e.g. rocky planet with CI like composition and low $T_\mathrm{surf}$).
Furthermore, the atmospheres can reach any point with ${\mathrm{(O-H)/(O+H)} = -1.0}$ and varying carbon content, thus ranging from \ce{H2} atmospheres (eg. solar composition, any $T_\mathrm{surf}$) to \ce{CH4} dominated atmospheres (e.g. CI composition, low $T_\mathrm{surf}$).
The intermediate states are also occurring (e.g. PWD CI or BSE12).

Only models for a CC like surface composition with ${T_\text{surf}>600\,\mathrm{K}}$ and Earth abundances with $T_\text{surf}<380\,$K fall into the atmospheric type B.
From calculations by \citet{Woitke2020a} this results in \ce{O2} being present.
However, the models also show the presence of \ce{SO2}, indicating the emergence of subtypes for the type B atmospheres when further elements than H, C, O, and N become of significant importance.
The further investigation of these subtypes will be addressed in a future paper.
However, as Fig.~\ref{fig:Atmos} shows, the impurities by further elements are small enough, so that the atmospheres still fall into the three types shown in \citet{Woitke2020a}.

In general, the atmospheric type does not change with atmospheric height, neither by condensation nor changes in $(p_\text{gas},T_\text{gas})$ in our polytropic atmosphere model.
Contrasting this, the O/H and C/O ratios, which have been used by previous studies to characterise atmospheres of planets \citep[see e.g.][]{Madhusudhan2012, Moses2013, Brewer2017} can change drastically throughout the atmospheres (Figs.~\ref{fig:AtmoOH} and \ref{fig:AtmoOC} in Appendix~\ref{App:Figs}).
This suggests, that the atmospheric types are a good indicator for atmospheres of rocky exoplanets.

In Fig.~\ref{fig:Atmos} it can be seen that the atmospheric composition with increasing height evolves towards (O-H)/(O+H)$\,\to\,\pm\,1$.
Type C atmospheres evolve towards either \ce{CO2} or \ce{CH4} dominated atmospheres. 
Also type A and B atmospheres with a high C content will evolve towards similar compositions. 
Distinguishing these endpoints will need precise measurements of the C, H, and O abundances together with the detection of the specific trace molecules which define the atmospheric type.

Although the atmospheric type does not change {\textit{throughout}} one atmosphere for a given $(p_\mathrm{surf}$, $T_\mathrm{surf})$ and rocky surface composition, changing these parameters can cause the entire atmosphere to change the atmospheric type.
For example the BSE, MORB and CI rocky surface compositions, show that the atmospheric type of the entire atmosphere changes with surface temperature.
This suggests, that planetary atmospheres can change the atmospheric type of the {\textit{entire}} atmosphere over time, for example by  cooling down.

A further indicator for the atmospheric conditions can be some specific cloud types, which are indicative of reducing or oxidising conditions.
For the low temperature cloud condensates our models show that \ce{S2}[s] is only stable under oxidising circumstances, while \ce{NH4SH}[s] is the sulphur condensate under reducing conditions.
\ce{NH4Cl}[s] can exist under both conditions but is constrained by the stability of \ce{NH4SH}[s] in our models.

\begin{figure}[!t]
\centering
\includegraphics[width=0.48\textwidth, page=1]{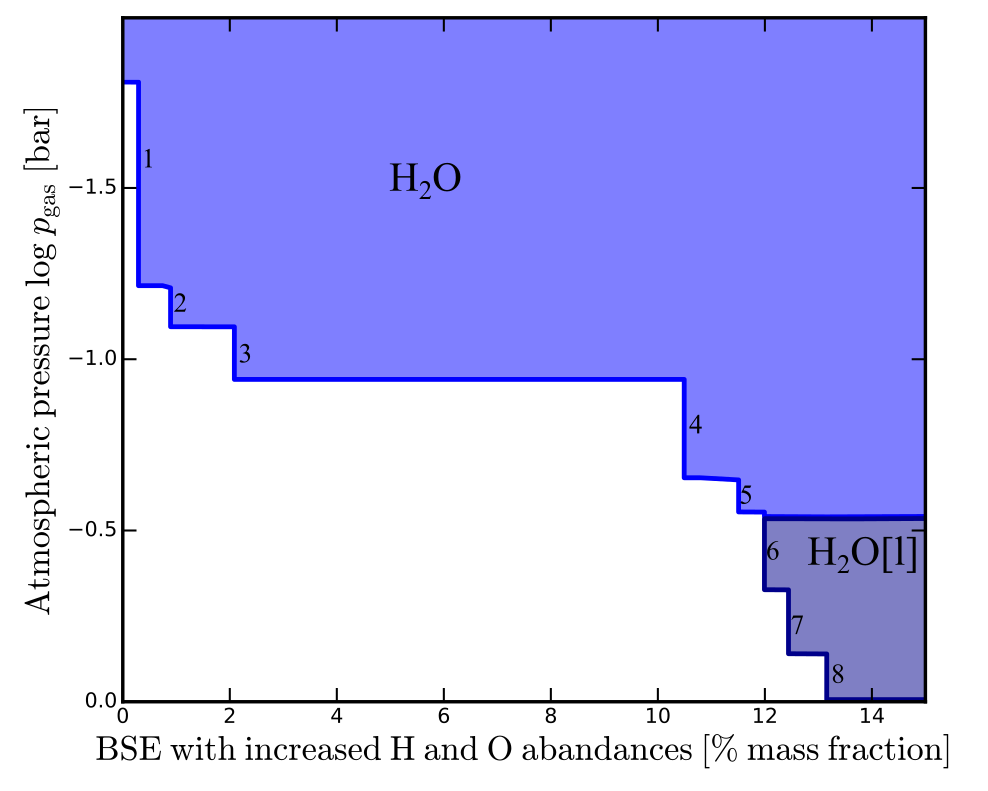}\\
\includegraphics[width=0.48\textwidth, page=1]{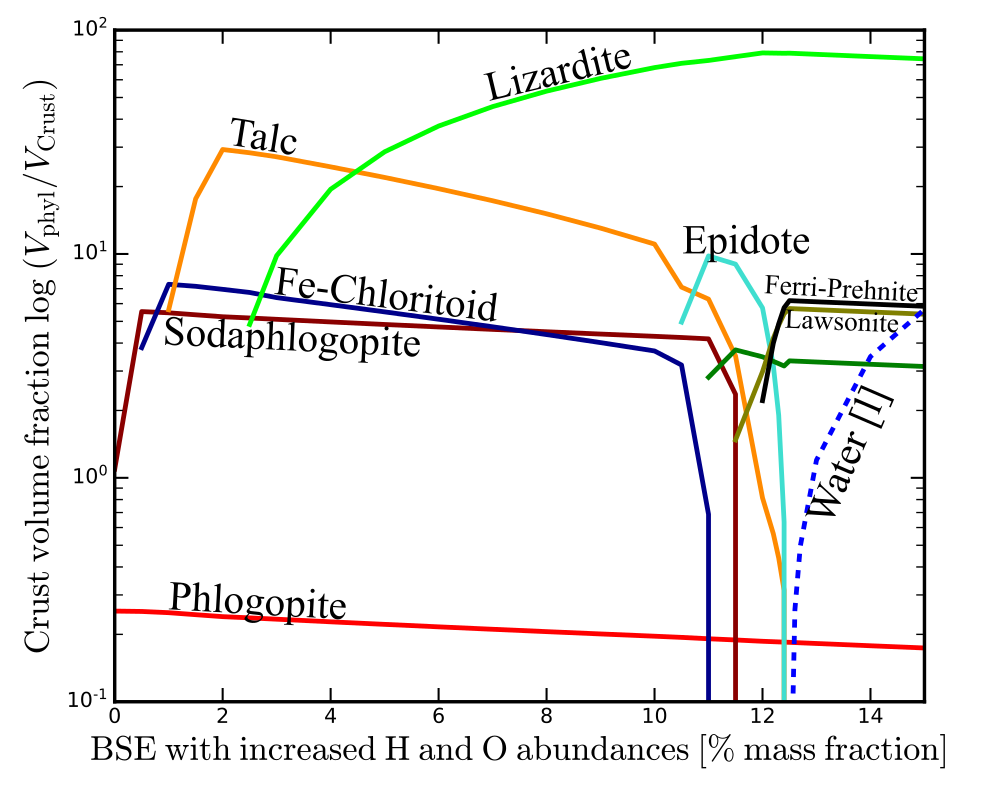}
\caption{Influence of crust hydration on the water cloud base. The model is based on the BSE model, with increasing additional H and O abundances.
The models are calculated for $T_\text{surf} = 350\,$K and $p_\text{surf} = 1\,$bar.
\textbf{Upper panel:} pressure level of the water cloud base.
\textbf{lower panel:} crust volume fraction of different phyllosilicates.
The sum formulas of the different condensate can be seen in Table~\ref{tab:phyllo}.}
\label{fig:Watercloudheight}
\end{figure}

\begin{table}
\caption{Limitation and stability of phyllosilicates in the crust of models with BSE total element abundances and increased H and O abundances. The numbers refer to the cloud base pressure changes in Fig. \ref{fig:Watercloudheight}.}
\label{tab:phyllo}
\centering
\vspace*{-2mm}
 \begin{tabular}{l|ll}
 &limited&stable\\ \hline
1&\ce{NaMg3AlSi3O12H2}[s]&\ce{FeAl2SiO7H2}[s]\\
&Sodaphlogopite&Fe-Chloritoid\\
2&\ce{FeAl2SiO7H2}[s]&\ce{Mg3Si4O12H2}[s]\\
&Fe-Chloritoid&Talc\\
3&\ce{Mg3Si4O12H2}[s]&\ce{Mg3Si2O9H4}[s]\\
&Talc&Lizardite\\
4&&\ce{Ca2FeAl2Si3O13H}[s]\\
&&Epidote\\
5&\ce{Ca2FeAl2Si3O13H}[s]&\ce{Fe3Si2O9H4}[s]\\
&Epidote&Greenalite\\
6&\ce{Fe3Si2O9H4}[s]&\ce{CaAl2Si2O10H4}[s]\\
&Greenalite&Lawsonite\\
7&\ce{Mg3Si2O9H4}[s]&\ce{Mg3Si2O9H4}[s]\\
&Lizardite&Ferri-Prehnite\\
8&\ce{CaAl2Si2O10H4}[s]&\ce{H2O}[l]\\
&Lawsonite&Water\\
&\ce{Ca2FeAlSi3O12H2}[s]&\\
&Ferri-Prehnite&
\end{tabular} 
\end{table}

\subsection{The link between planetary surface hydration and water clouds}\label{ssec:surfacewater}
\citet{Herbort2020} have shown that \ce{H2O}[l,s] will only be thermally stable in the presence of silicate material, if after the formation of phyllosilicates, that is the hydration of the silicates, an excess of H and O prevails.
If this is the case, we call these phyllosilicates {\it saturated}.
Hence, only if all phylosilicates are saturated, water can appear in condensed form which we have shown to be the case for the near-crust atmosphere.
In the following we demonstrate that even if the phyllosilicates that form in the crust of a rocky planet for a given ($p_\mathrm{surf}, T_\mathrm{surf}$) are not saturated in the above sense, \ce{H2O}[l,s] can become thermally stable in the atmosphere stratification that develops above such a crust because of its different thermodynamic conditions where ${p_\mathrm{gas}<p_\mathrm{surf}}$ and ${T_\mathrm{gas}<T_\mathrm{surf}}$.
If \ce{H2O}[l,s] is a stable condensate at the crust, the \ce{H2O}[l,s] cloud base is in the atmosphere crust interaction layer (BSE15, CI, Earth, Archean).
For all other models the cloud base of \ce{H2O}[l,s] is at lower pressures.

Figure~\ref{fig:Watercloudheight} shows the dependence of the local gas pressure level at the cloud base for the water cloud on the hydration of the crust.
We test this dependence by increasing the H and O abundances of the BSE total element abundances in a 2:1 ratio.
The number of added elements is given in percent mass fraction relative to the unaltered BSE abundance.
The local pressure level of the \ce{H2O}[l,s] cloud base does not continuously decrease with increased element abundances, but changes if a given phyllosilicate is saturated.
Coinciding with this, a further phyllosilicate becomes stable. 
With further additional H and O added to the total element abundance, the stable not saturated phyllosilicates incorporate the additional H and O.
In Fig.~\ref{fig:Watercloudheight} these changes are indicated with numbers, and listed in Table \ref{tab:phyllo}.
For change 4, no phyllosilicate reaches is maximum, but Epidote becomes stable.
For change 8, two different condensates reach their limit, while \ce{H2O}[l] (water) becomes a stable condensate and the water cloud is in contact with the surface

This investigation underlines the importance of phyllosilicate on the stability of \ce{H2O}[l,s] at the surface of rocky exoplanets.
We have shown that under the assumption of a hydrostatic equilibrium atmosphere in chemical phase equilibrium with the crust it is possible to have \ce{H2O}[l,s] thermally stable in the atmosphere but not at the crust, although the planetary surface conditions of ($p_\mathrm{surf}, T_\mathrm{surf}$) are consistent with water condensates.

\begin{figure*}
\centering
\includegraphics[width = .95\linewidth]{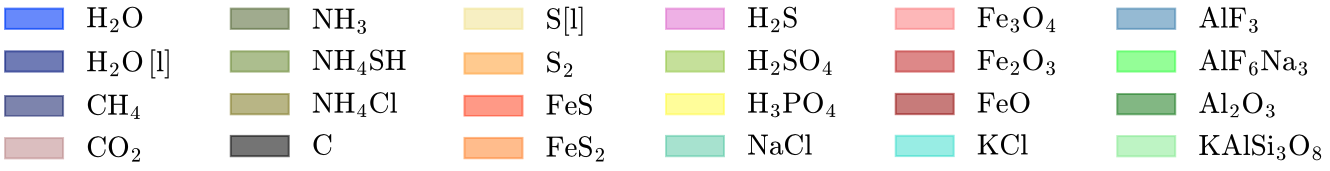}\\
\includegraphics[width = .32\linewidth, page=1]{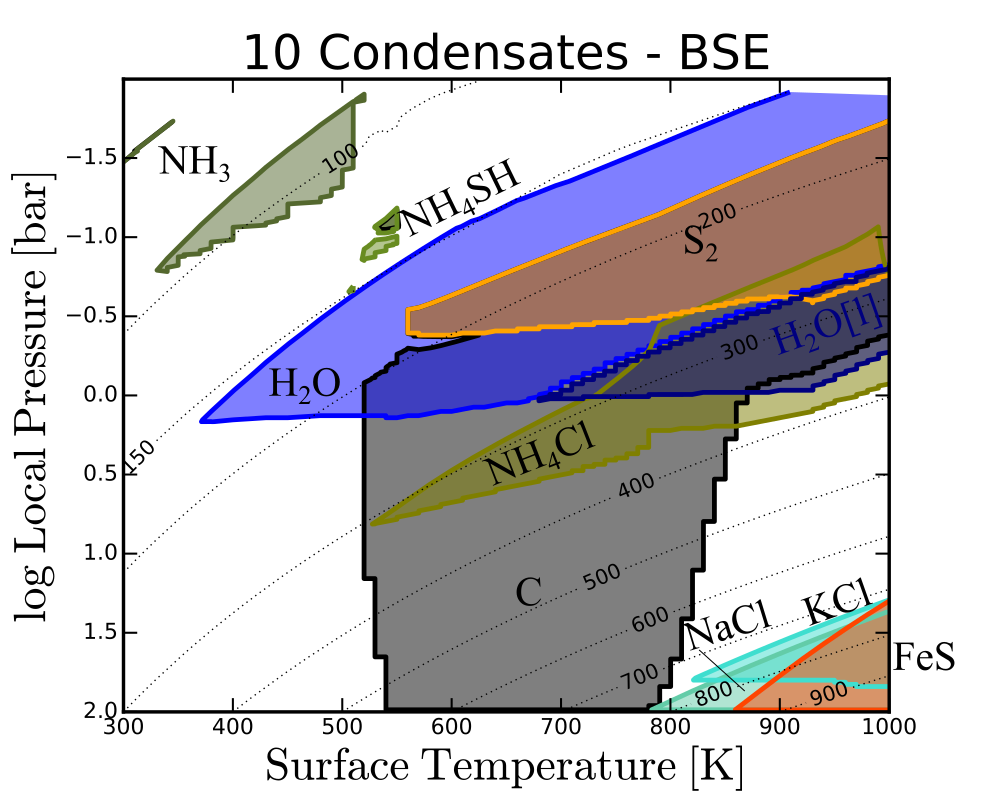}
\includegraphics[width = .32\linewidth, page=1]{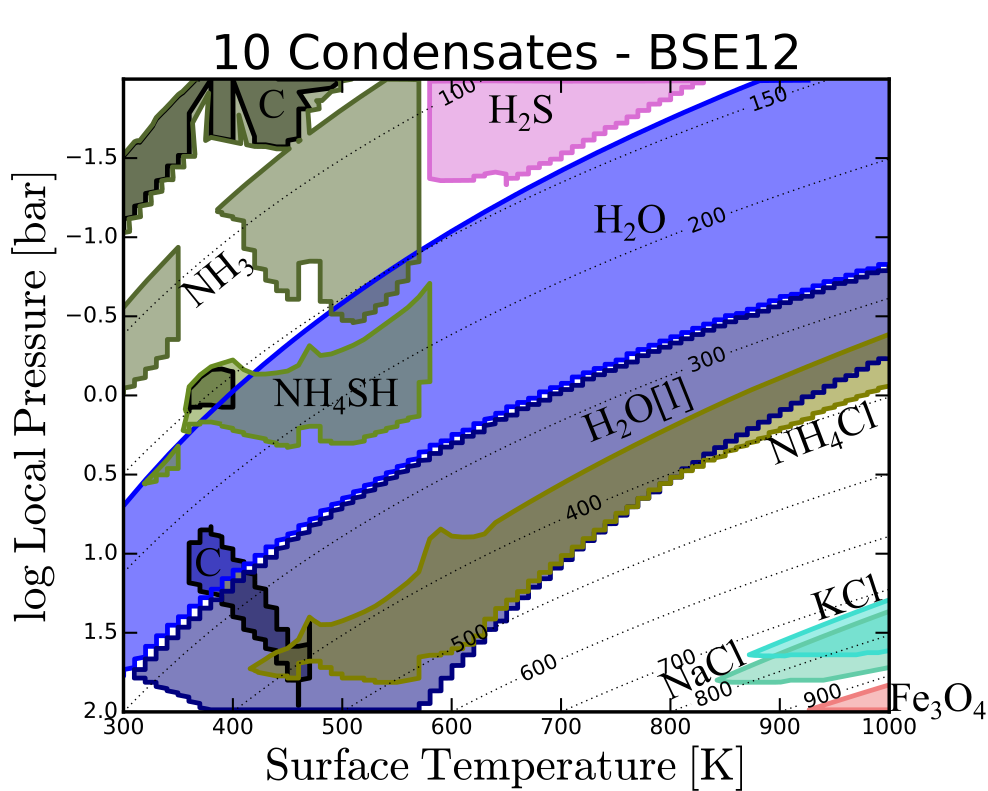}
\includegraphics[width = .32\linewidth, page=1]{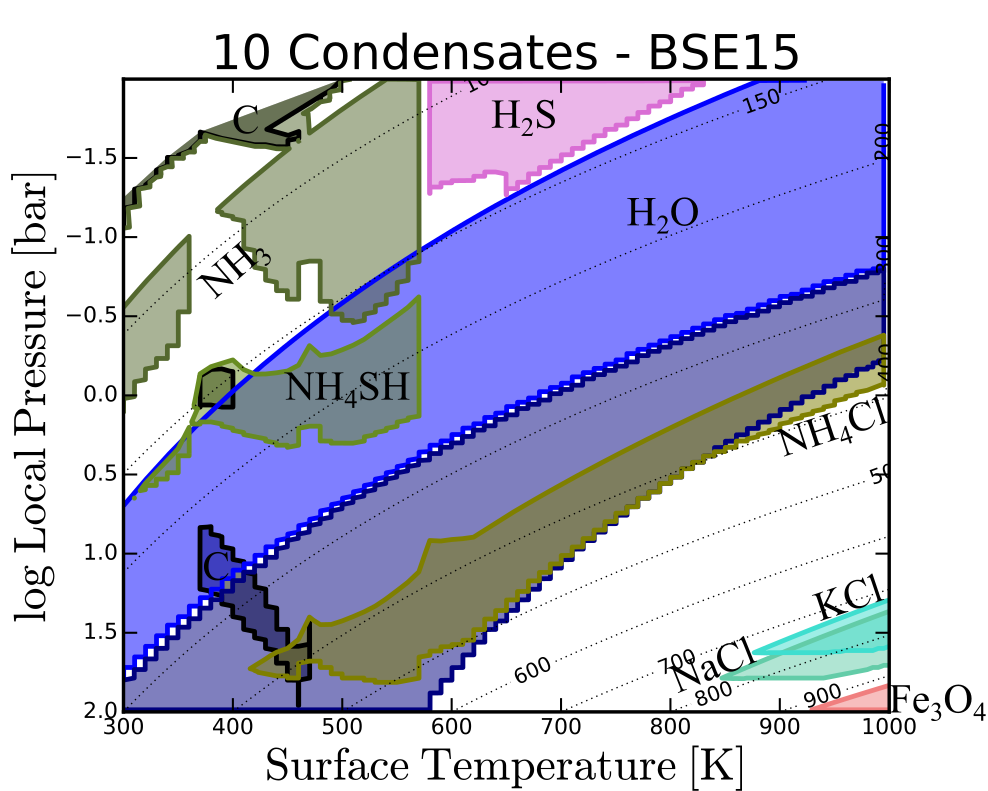}\\
\includegraphics[width = .32\linewidth, page=1]{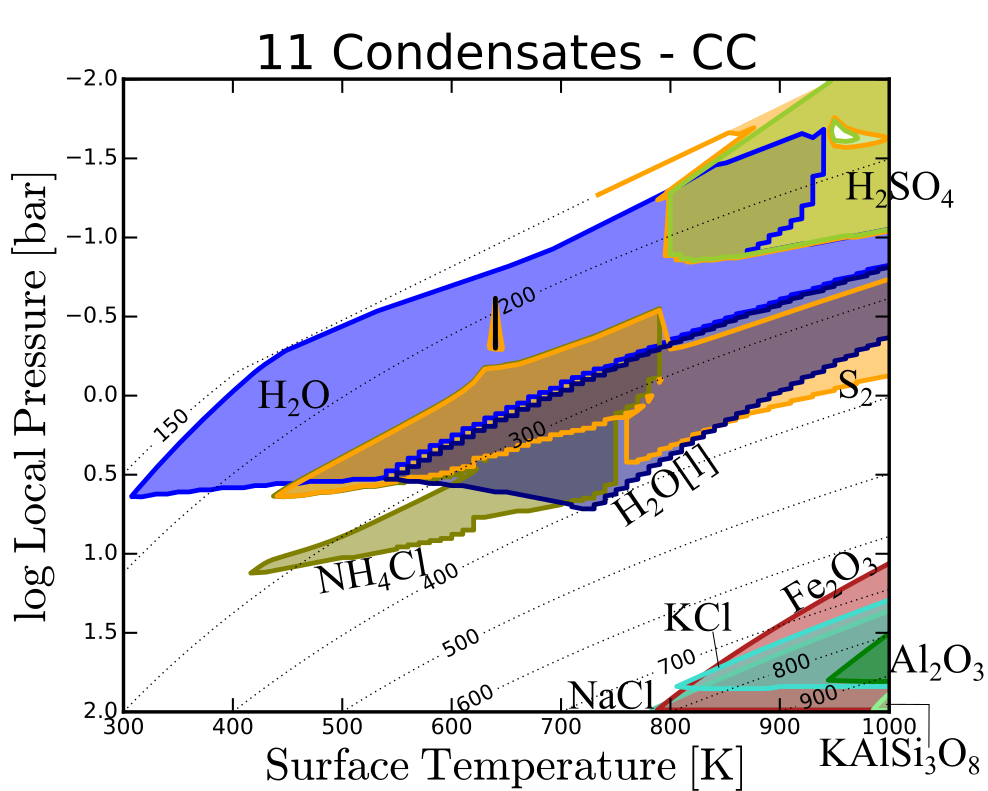}
\includegraphics[width = .32\linewidth, page=1]{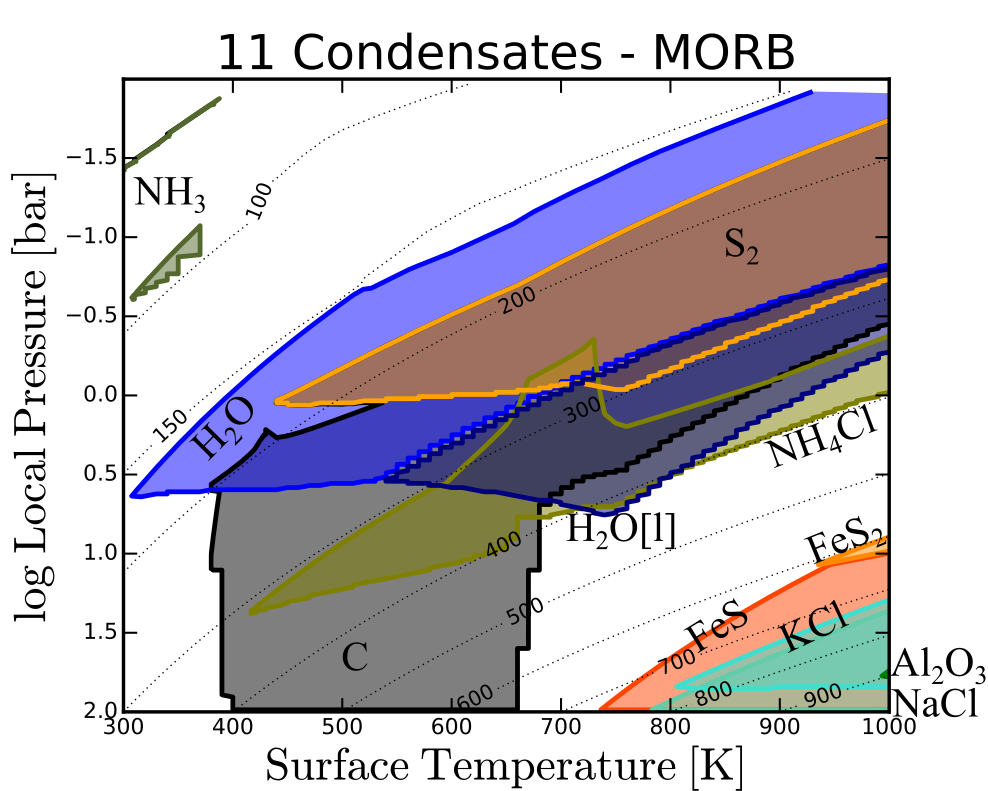}
\includegraphics[width = .32\linewidth, page=1]{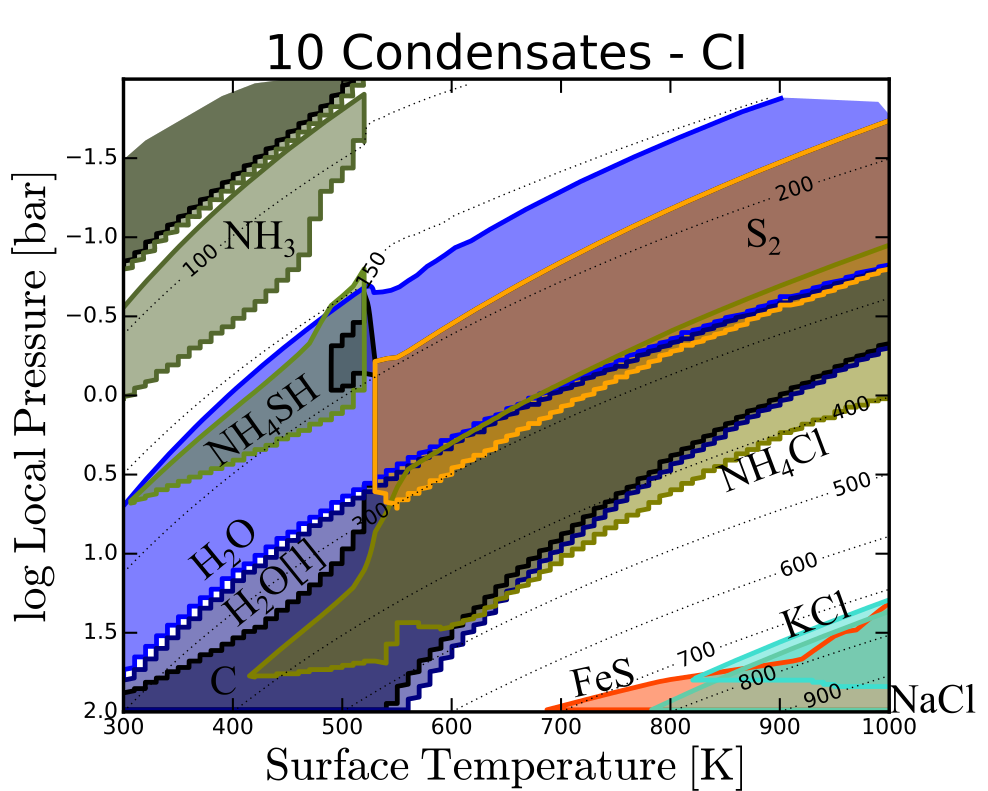}\\
\includegraphics[width = .32\linewidth, page=1]{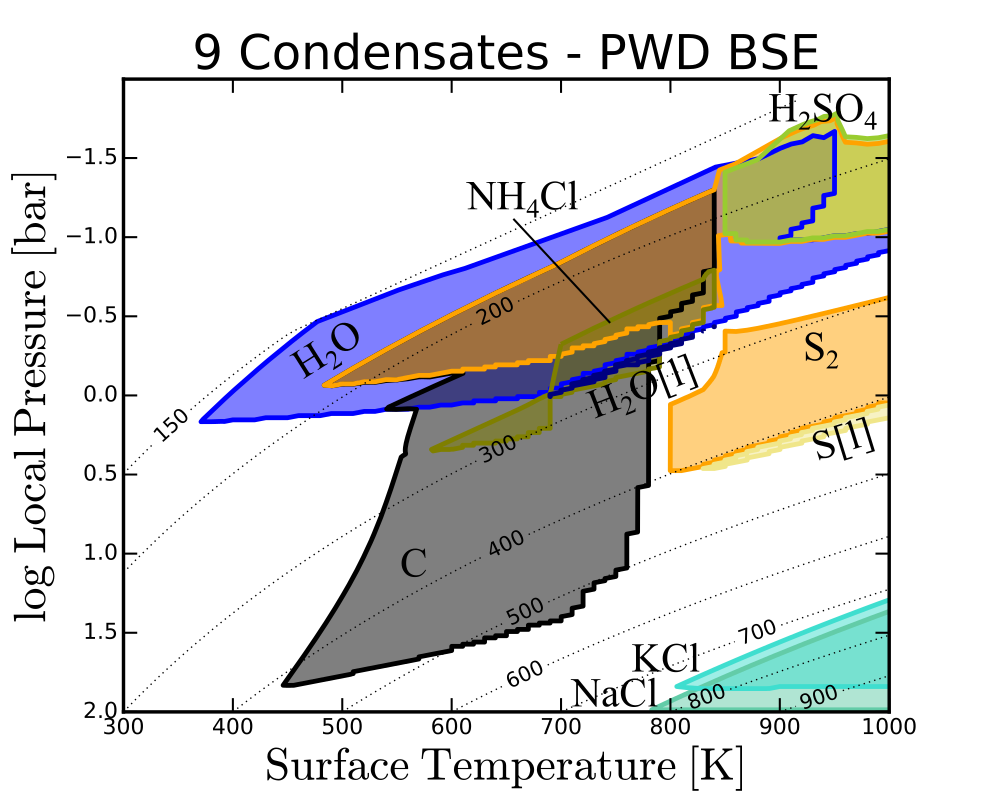}
\includegraphics[width = .32\linewidth, page=1]{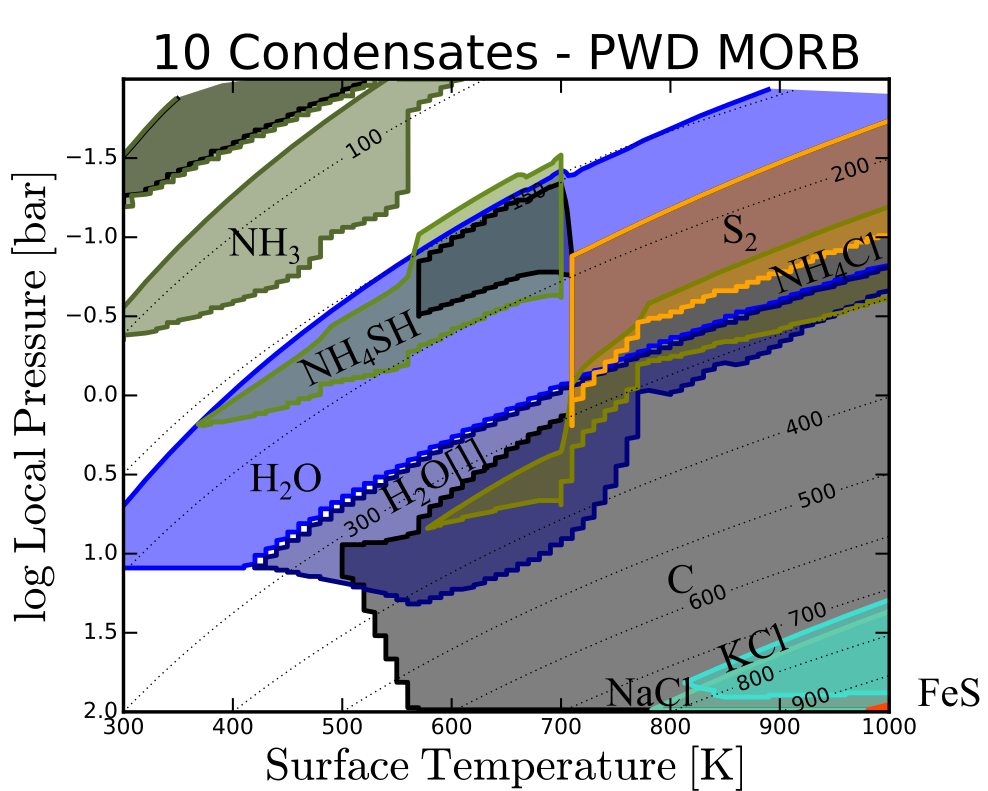}
\includegraphics[width = .32\linewidth, page=1]{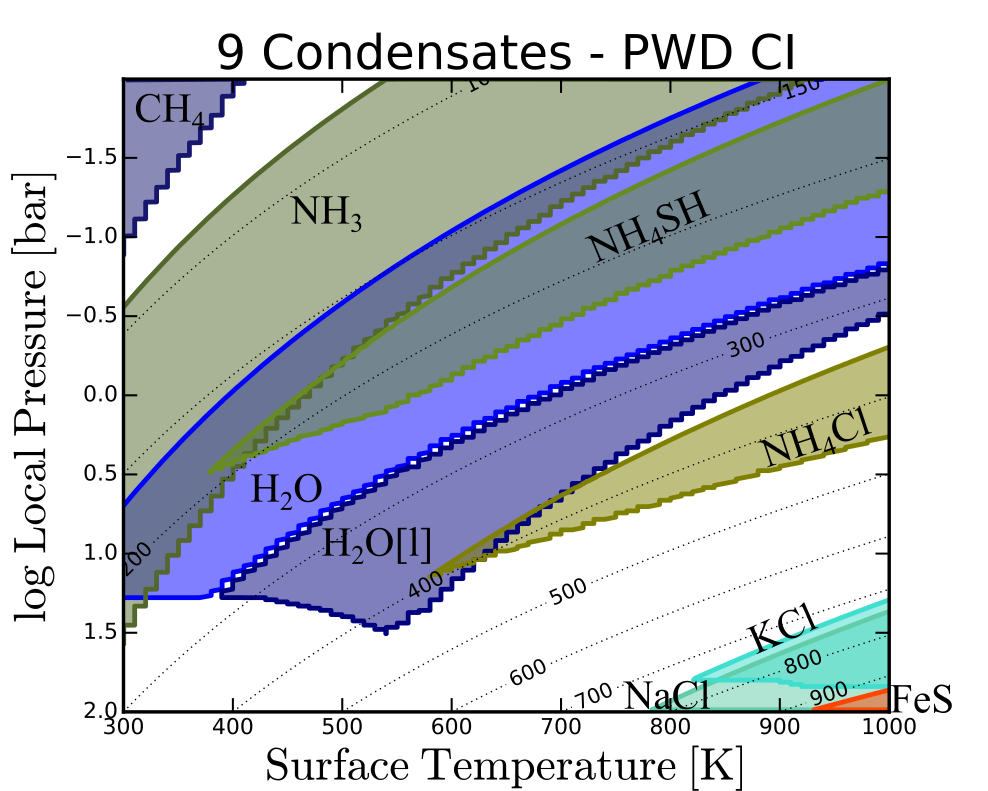}\\
\includegraphics[width = .32\linewidth, page=1]{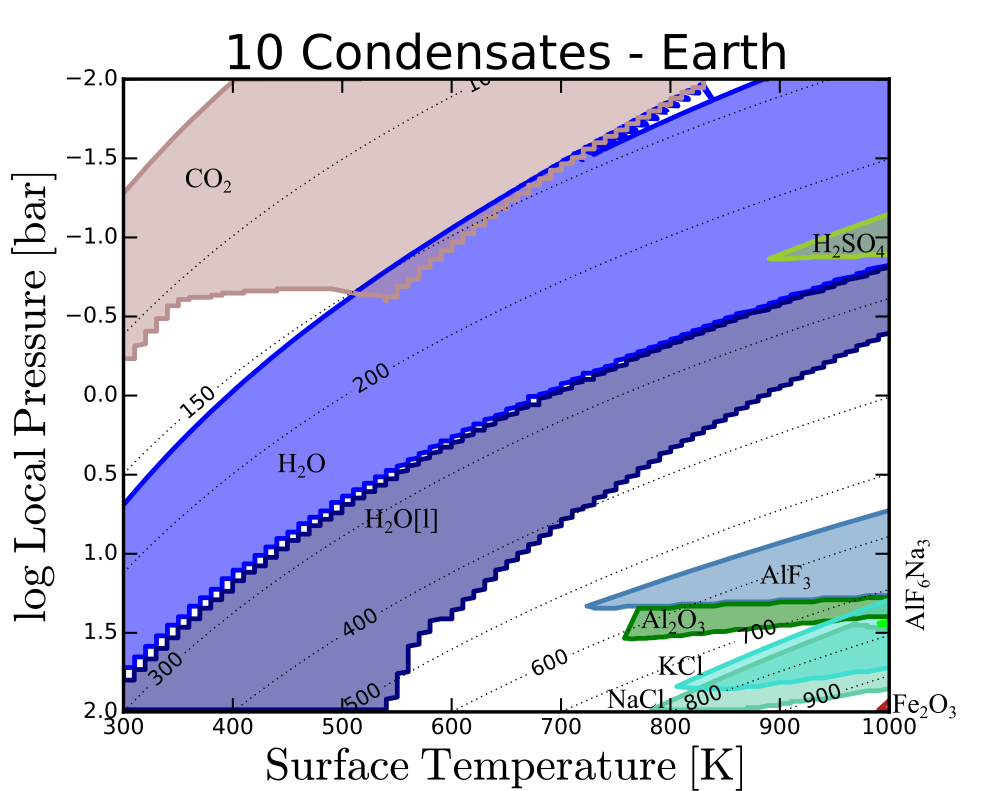}
\includegraphics[width = .32\linewidth, page=1]{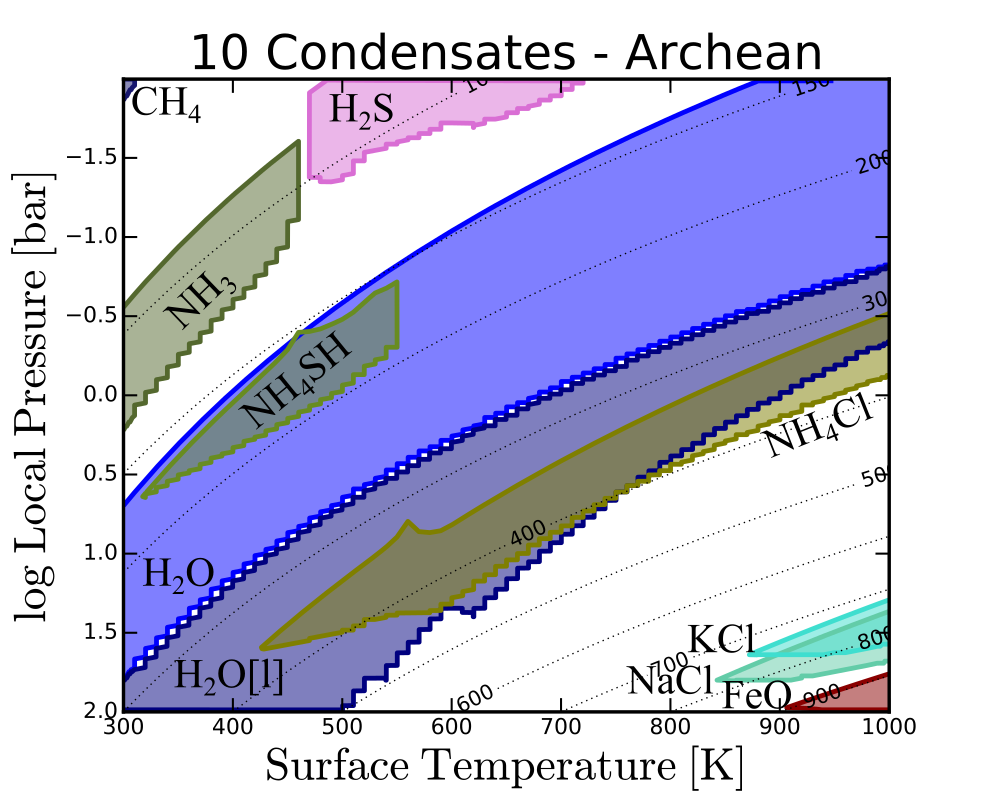}
\includegraphics[width = .32\linewidth, page=1]{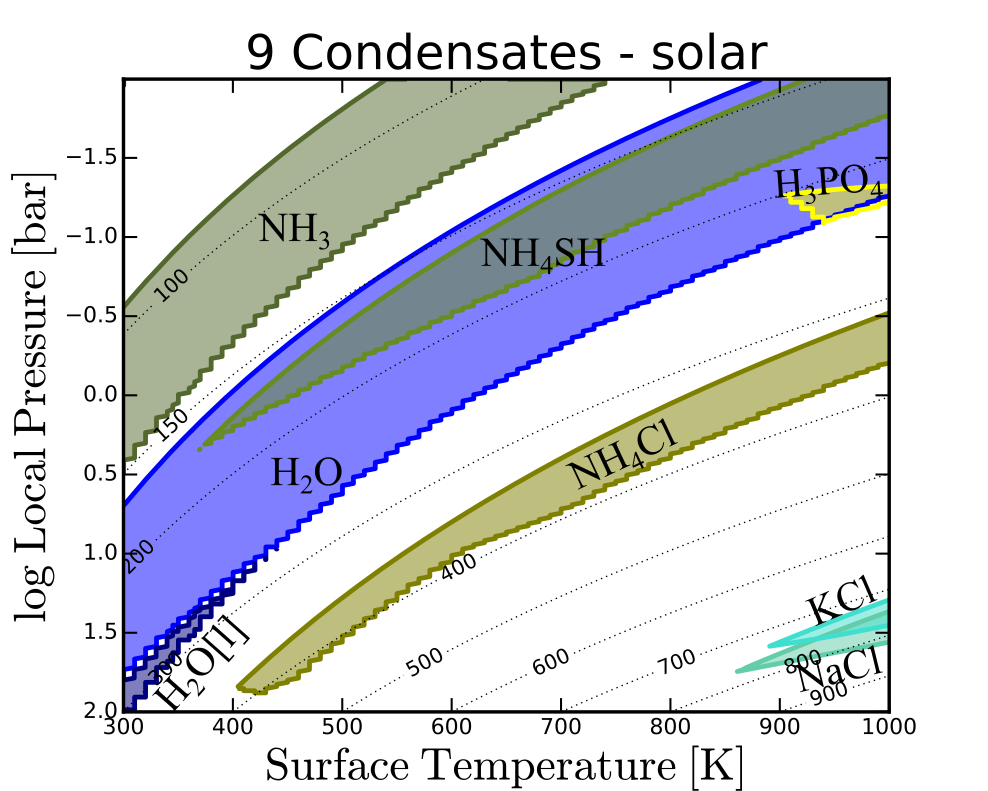}
\caption{As Fig.~\ref{fig:CloudsAll-10}, but for $p_\mathrm{surf}=100\,\mathrm{bar}$.
All models are calculated for $\gamma = 1.25$.
All thermally stable condensates with normalised number densities of $n_\text{cond}/n_\text{tot}>10^{-10}$ locally are shown.}
\label{fig:Cloudspress}
\end{figure*}

\subsection{The effect of surface pressure on cloud composition} \label{ssec:press}
The planetary  surface pressure can currently not be measured directly and therefore leaves a degeneracy for the surface conditions if only the conditions of the upper atmosphere of a planet are known. 
For atmospheres with a polytropic index of $\gamma = 1.25$, the $(p_\mathrm{gas},T_\mathrm{gas})$ profile of an atmosphere with $T_\mathrm{surf}= 300\,\mathrm{K}$ and  $p_\mathrm{surf}=1\,\mathrm{bar}$ and an atmosphere with $T_\mathrm{surf}= 750\,\mathrm{K}$ and  $p_\mathrm{surf}=100\,\mathrm{bar}$ are indistinguishable for $P_\mathrm{gas}<1\,\mathrm{bar}$.
In our solar system this can be seen in the comparison of Earth and Venus.
Both show clouds at comparable $(p_\mathrm{gas},T_\mathrm{gas})$, but Venus has a high pressure atmosphere.
Only observations at wavelength, where the atmosphere is optically thin, the atmospheric height, and therefore the surface pressure, can be determined.
This means that optical wavelength are limiting us to the upper atmosphere of Venus while radio wavelength enable us to investigate the surface of Venus.
In this section we investigate in how far an increased surface pressures of $p_\mathrm{surf}\!=\!100\,\mathrm{bar}$ affects the thermal stability of cloud condensates.

In Fig.~\ref{fig:Cloudspress} all thermally stable cloud condensates with $n_\text{cond}/n_\text{tot}>10^{-10}$ are shown for 12 different element abundances for a rocky planetary surface.
Overall, Fig.~\ref{fig:Cloudspress} demonstrates that, similarly to the $p_\mathrm{surf}=1\,$bar models (Sect.~\ref{ssec:elementabundances}), the thermally stable cloud materials can be separated into high and low local temperature regimes.
Whereas the temperature for the lowest cloud base of the low temperature clouds is located at  $T_\mathrm{gas}\approx400\,$K for all models with $p_\mathrm{surf}=1\,$bar, the corresponding temperature for the $p_\mathrm{surf}=100\,$bar models changes with different $T_\mathrm{surf}$ from $T_\mathrm{gas}\approx550\,$K at $p_\mathrm{gas}=100\,$bar for atmospheric models with $T_\mathrm{surf} = 550\,$K to $T_\mathrm{gas}\approx450\,$K at $p_\mathrm{gas}\approx2\,$bar for atmospheric models with$T_\mathrm{surf} = 1000\,$K

In comparison to the ${p_\mathrm{surf}=1\,\mathrm{bar}}$ models, a total of 17 further condensates are thermally stable if $n_\text{cond}/n_\text{tot}>10^{-20}$.
All of the condensates that reach normalised number densities of $n_\text{cond}/n_\text{tot}>10^{-10}$ (\ce{H2S}[s], \ce{H2SO4}[s], \ce{S}[l], \ce{CO2}[s], \ce{CH4}[s], and \ce{H3PO4}[s]) are part of the low temperature cloud condensates. 
\ce{H2S}[s], \ce{CO2}[s], and \ce{CH4}[s] form form only for very low temperatures of $T_\mathrm{gas}\lesssim150\,$K, which is not in the parameter space for the  $p_\mathrm{surf}=1\,$bar atmospheres.
Of these low temperature cloud condensates, only \ce{S}[l] has a type 2 transition to anther condensate (\ce{S}[l] to \ce{S2}[s]) and therefore forms a condensate chain.

Our models show that some cloud condensates only form under specific atmospheric compositions and therefore these condensates can be used as indicators for these atmospheric types.
For the low temperature cloud condensates we find sets of mutually exclusive cloud condensates that do not coexist with each other.
One mutually exclusive sets is \ce{NH3}[s], \ce{H2S}[s], and \ce{H2SO4}[s].
\ce{NH3}[s] only forms in N-rich type A atmospheres, while \ce{H2SO4}[s] only forms in type B atmospheres.
\ce{H2S}[s] can form in type A and type B atmospheres.
Another pair of mutually exclusive condensates is \ce{S2}[s] and \ce{NH4SH}[s], with the latter only forming in H-rich environments.

The investigation of the cloud condensates with $n_\text{cond}/n_\text{tot}>10^{-20}$ shows that the gap between the high and low temperature cloud layers  is filled with thermally stable condensates.
Further 2 condensates (\ce{FeCl3}[s], \ce{TiF4}[s]) reach  $n_\text{cond}/n_\text{tot}>10^{-15}$, while 9 condensates reach  $n_\text{cond}/n_\text{tot}>10^{-20}$.
These can be divided into different croups of pyhllosilicates (\ce{KAl3Si3O12H2}[s], \ce{FeAl2SiO9H2}[s], \ce{Mg3Si2O0H4}[s], \ce{KMg2Al3Si2O12H2}[s]), hydroxides (\ce{MgO2H2}[s], \ce{FeO2H}[s]), carbonates (\ce{FeCO3}[s]), and metal oxides (\ce{MgTiO3}[s], \ce{Mg2TiO4}[s]).
Only \ce{FeO2H}[s], \ce{FeCl3}[s], and \ce{TiF4}[s] are stable in the low temperature cloud regime, while all other cloud condensates are thermally stable relatively close to the surface.

The lesser abundant cloud condensates enhance the condensate chains discussed in Sect.~\ref{sssec:clouddiversity}.
Every condensate that is only found in the high $p_\mathrm{surf}=100\,$bar models is shaded in grey in Fig.~\ref{fig:CloudTransitions}.

With the increased surface pressure, the boiling point of water also increases, allowing liquid water as a stable condensate for higher temperatures as well.
This becomes especially noticeable for the model with a BSE12 surface composition, where there is no water stable at the surface with $p_\mathrm{surf}=1\,$bar, but for $p_\mathrm{surf}=100\,$bar \ce{H2O}[l] is stable for ${T\mathrm{surf}\geq370\,\mathrm{K}}$ and ${T\mathrm{surf}\leq570\,\mathrm{K}}$.
This shows that liquid water could be found on planets with high pressure although their element abundances do not allow water condensates at Earth like $(p_\mathrm{surf},T_\mathrm{surf})$ conditions.

\begin{figure}
\centering
\includegraphics[width = .95\linewidth]{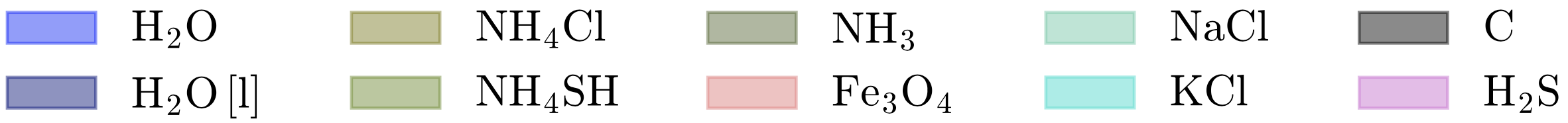}\\
\includegraphics[width = .99\linewidth, page=1]{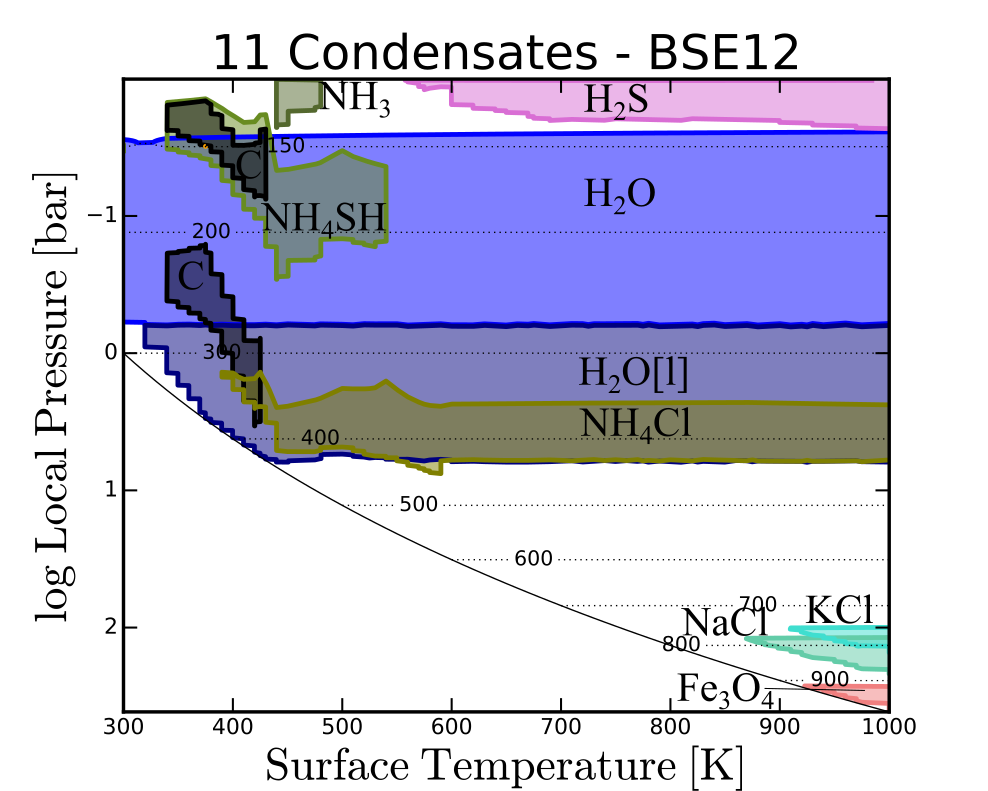}\\
\includegraphics[width = .99\linewidth]{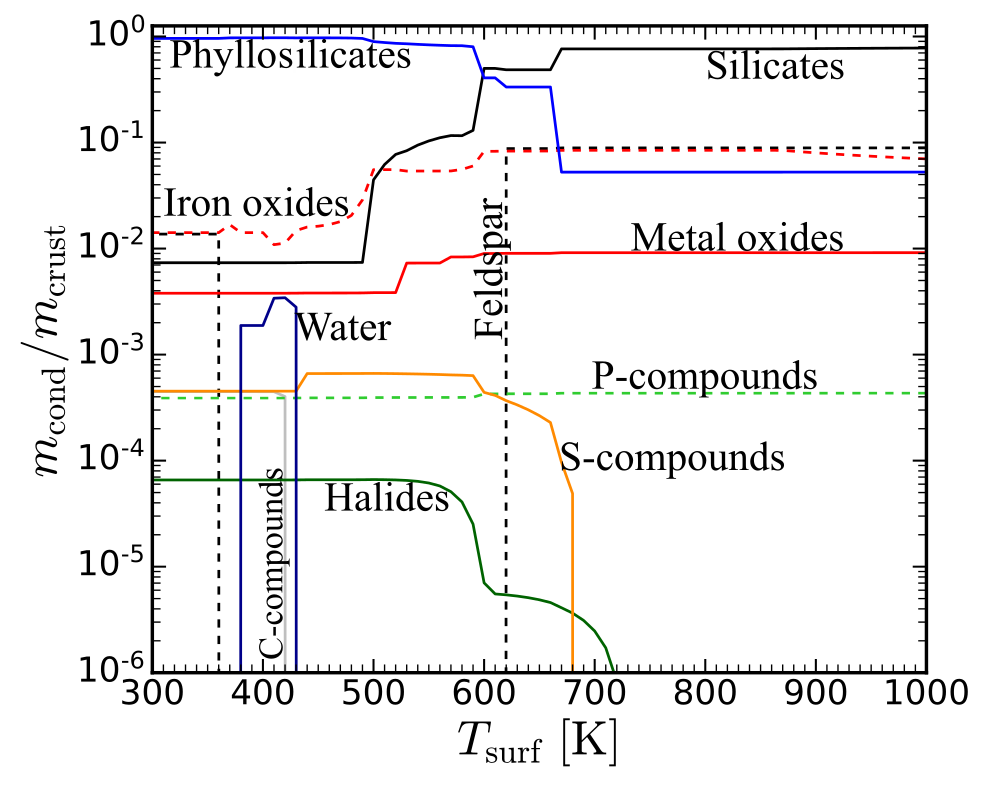}\\
\caption{\textbf{Upper panel:} Thermally stable cloud condensates with $n_\text{cond}/n_\text{tot}>10^{-10}$ for overlapping atmospheric $(p_\mathrm{gas},T_\mathrm{gas})$ profiles with varying surface conditions $(p_\mathrm{surf},T_\mathrm{surf})$. All atmospheric profiles go through $(p_\mathrm{gas}=1\,\mathrm{bar},T_\mathrm{gas}=300\,\mathrm{K})$ and have a BSE12 like crust composition.
$(p_\mathrm{surf}$ and $T_\mathrm{surf})$ for each model are indicated by the solid black line. The polytropic index is kept at $\gamma = 1.25$ for all models.
\textbf{Lower panel:} Crust composition for the atmospheric models in the upper panel. The condensates belonging to each group are listed in Table~\ref{tab:CrustComp}.}
\label{fig:clouds_pTchange}
\end{figure}

\begin{table}[!t]
\caption{Explanation of the crust condensate grouping from Fig. \ref{fig:clouds_pTchange} for the BSE12 like crust composition.}
\label{tab:CrustComp}
\centering
\vspace*{-2mm}
 \begin{tabular}{lll}\hline
Group       & Included condensates & Condensate name\\ \hline

Silicates   & \makecell[tl]{\ce{Mg2SiO4}[s]\\ \ce{Fe2SiO4}[s]\\ \ce{MgSiO3}[s]\\ \ce{CaTiSiO5}[s]\\ \ce{CaMgSi2O6}[s]\\ \ce{Mn3Al2Si3O12}[s]} & \makecell[tl]{Fosterite\\Fayalite\\Enstatite\\Sphene\\Diopside\\Spessartine}\\ \hline

Feldspar    & \makecell[tl]{\ce{NaAlSi3O8}[s]\\ \ce{CaAl2Si2O8}[s]} & \makecell[tl]{Albite\\Anorthide}\\ \hline

Metal oxides& \makecell[tl]{\ce{MgCr2O4}[s]\\ \ce{FeTiO3}[s]\\ \ce{Cr2O3}[s]} & \makecell[tl]{Picrochromite\\Ilmenite\\Eskolaite}\\ \hline

Iron Oxides & \makecell[l]{\ce{Fe3O4}[s]} & Magnetide\\ \hline

Phyllosilicates& \makecell[tl]{\ce{CaAl2Si2O10H4}[s]\\ \ce{Fe3Si2O9H4}[s]\\ \ce{CaAl4Si2O12H2}[s]\\ \ce{Ca2FeAlSi3O12H2}[s]\\ \ce{Mg3Si2O9H4}[s]\\ \ce{Ca2Al3Si3O13H}[s]\\ \ce{FeAl2SiO7H2}[s]\\ \ce{Ca2FeAl2Si3O13H}[s]\\ \ce{Mg3Si4O12H2}[s]\\ \ce{KMg3AlSi3O12H2}[s]\\ \ce{NaMg3AlSi3O12H2}[s]} & \makecell[tl]{Lawsonite\\Greenalite\\Margarite\\Ferri-Prehnite\\Lizardite\\Clinozoisite\\Fe-Chloritoid\\Epidote\\Talc\\Phlogopite\\Sodaphlogopite}\\ \hline

Water       & \makecell[tl]{\ce{H2O}[l]} & Liquid water\\ \hline

Halides     & \makecell[tl]{\ce{CaCl2}[s]\\ \ce{MgF2}[s]\\ \ce{CaF2}[s]\\ \ce{NaCl}[s]} & \makecell[tl]{Ca-Dichloride\\Mg-Flouride\\Flourite\\Halite}\\ \hline

P-compounds & \ce{Ca5P3O12F}[s] & Flourapatite\\ \hline
S-compounds & \makecell[tl]{\ce{FeS2}[s]\\ \ce{FeS}[s]} & \makecell[tl]{Pyrite\\Troilite}\\ \hline
C-compounds & \makecell[tl]{\ce{CaCO3}[s]} & Calcite\\ \hline
\end{tabular} 
\end{table}

Only reflected light observations from rocky exoplanets can provide direct information on surface properties \citep[e.g.][]{Madden2018, AsensioRamos2021}.
Transmission spectroscopy on the other hand is limited to the higher atmosphere and potentially a higher cloud deck \citep{Komacek2020}, but can provide estimates about the $(p_\mathrm{gas},T_\mathrm{gas})$ conditions.
However, these high atmosphere conditions are ambiguous with regard to the surface pressure and therefore, assuming a  surface temperature.
One $(p_\mathrm{gas},T_\mathrm{gas})$ point can be used with Eq.~\ref{eq:TpropKappa} to find   $(p_\mathrm{surf},T_\mathrm{surf})$.
The resulting polytropic atmosphere shows the exact same atmospheric structure in the higher atmosphere, but can be calculated to different surface conditions.

All models in Fig.~\ref{fig:clouds_pTchange} follow atmospheric profiles of ${T=300\,\mathrm{Kbar^{-0.2}}\cdot p^{0.2}}$ with different $(p_\mathrm{surf},T_\mathrm{surf})$.
In the upper panel of Fig.~\ref{fig:clouds_pTchange}, all thermally stable cloud condensates that reach normalised number densities of $n_\text{cond}/n_\text{tot}>10^{-10}$ for these atmospheric profiles are shown.
The crust composition is BSE12 like.
Although the $(p_\mathrm{gas},T_\mathrm{gas})$ in the higher atmosphere are by construction the exact same for the different models, the changes in the surface pressure level result in different cloud condensates becoming thermally stable.
\ce{NH4Cl}[s] and \ce{H2O}[l,s] are thermally stable at similar temperatures, independent of the surface conditions..
However, for models with higher $(p_\mathrm{surf},T_\mathrm{surf})$ \ce{H2S}[s] becomes thermally stable, while for the lower $(p_\mathrm{surf},T_\mathrm{surf})$ \ce{C}[s], \ce{NH4SH}[s], and \ce{NH3}[s] are thermally stable condensates in the higher atmosphere.
This is a result of different condensate makeup of the crust (lower panel of Fig.~\ref{fig:clouds_pTchange}).
The different crust forming condensates are grouped according to their nature (see Table~\ref{tab:CrustComp}).
\ce{H2S}[s] is only a cloud condensate, if enough sulphur and hydrogen is available in the near crust atmosphere.
This can also be seen by sulphur compounds only becoming stable in the crust for $T_\mathrm{surf}\leq650\,\mathrm{K}$.
For lower surface pressures, the sulphur remaining in the atmosphere is condensing and becomes a stable condensate in form of \ce{NH4SH}[s].
Furthermore, the phyllosilicates become more important as well. 
A similar effect can be seen for the condensation of \ce{C}[s], which - in the case of BSE12 crust composition - only occurs once carbon is a stable crust condensate ($T_\mathrm{surf}\leq400\,\mathrm{K}$).

The results from Fig.~\ref{fig:clouds_pTchange} underline that assuming the right surface conditions is crucial for understanding the atmosphere. 
Although elements that undergo high temperature condensation are depleted during the atmospheric buildup, the resulting upper atmosphere is affected by the removal of different condensates in the atmosphere with respect to the formation in the crust for a lower surface pressure. 
This is a result of the full availability of elements at the crust, and not the depleted element abundances in the gas phase.
Especially Si is forming crust condensates and is thus very depleted in the near crust atmosphere.
This in turn results in Silicate clouds only being stable close to the crust, if at all.

\section{Implications for observable planets}\label{sec:planets}
In this section we use the material presented in Sect.~\ref{sec:results} to discuss the possible implications of observations by future space missions and ground based instruments.
The observations of a specific exoplanet can in principle provide three different types of information.
First, the molecular composition of the gas above a cloud layer, which can be observed via absorption lines, probing the optically thin part of the atmosphere \citep[e.g.][]{Helling2019a, Samra2020, Young2020}. 
Second, information about the chemical composition of the cloud layer can be obtained by absorption bands of solid materials \citep[e.g.]{Wakeford2015, Pinhas2017, Barstow2020, Gao2020}, and third, potentially, the temperature and pressure at the cloud deck \citep[e.g.][]{Line2016, Mai2019, Fossati2020}.

The detection of certain atmospheric gases (\ce{H2O}, \ce{CO2}, \ce{CH4}, \ce{H2}, and \ce{O2}) can be used to identify the atmospheric type of atmospheres \citep[see][]{Woitke2020a}.
This atmospheric type of the optically thin atmosphere is the same as the atmospheric type at the near-crust atmosphere, as the atmospheric type does not change with height, despite the condensation of clouds. 
Additionally, the detection of certain cloud condensates can constrain the element abundance ratio between O and H.
For example, \ce{S2}[s] can only be thermally stable, when O>H.
Whereas \ce{NH4SH}[s] is only thermally stable for H>O.

Furthermore, the chemical composition of the clouds at a given $(p_\mathrm{gas}, T_\mathrm{gas})$ is related to the crust composition, although conclusions can be ambiguous (see Tables \ref{tab:Planets} and \ref{tab:Planets-hot}).
For example, the detection of \ce{S2}[s] at ($p_\mathrm{gas}\approx\,30$\,mbar, $T_\mathrm{gas}\approx\,200\,$K) constrains the surface conditions to $T_\mathrm{surf}\approx\,400$\,K and $T_\mathrm{surf}\approx\,1000$\,K, under the assumption of a polytropic atmosphere with surface pressures of $T_\mathrm{gas}\approx\,200\,$K and $p_\mathrm{gas}\approx\,30$\,mbar, respectively.
Our study shows, that under these conditions, the cloud condensate \ce{S2}[s] constrains the crust composition to be CC or CI like for the 1\,bar atmosphere. 
For $p_\mathrm{surf}=100\,$bar, the crust compositions is more ambiguous, and could be similar to BSE, CC, MORB, CI, PWD BSE, or PWD MORB.
However, additionally ruling out the presence of water clouds can further constrain the crust composition to only CC or PWD BSE.

In order to use cloud condensates to constrain crust compositions of already known planets, we created a list of known transiting planets with planetary masses $M_{\rm{P}}<8\,\rm{M}_{\oplus}$, planetary radius $R_{\rm{P}}<2\,\rm{R}_{\oplus}$ and equilibrium temperature $250\,\mathrm{K}<T_{\rm{eq}}<1050\,$K.
In Tables \ref{tab:Planets} and \ref{tab:Planets-hot} we provide an overview of which cloud condensates suggest a certain crust composition.
For this classification, we assume a polytropic atmosphere with $p_\mathrm{surf}=1\,$bar and $T_\mathrm{surf}=T_\mathrm{eq}$. 
Where $T_\mathrm{eq}$ is the reported equilibrium temperature of the respective planets.

One of the fundamental assumptions for the determination of the crust composition is the surface pressure, which cannot directly be determined, leaving a degeneracy if only the upper atmosphere is constrained \citep[see also][]{Keles2018, Yu2021}.
Our study showed that the detection of \ce{H2S}[s] or \ce{H2SO4}[s] condensates would suggest a high surface pressure.

Although our study does show the thermal stability of \ce{H2O}[l,s] clouds high in the atmosphere for all crust compositions, water as a stable condensate on the surface of the respective rocky exoplanets is rare (see Sect.~\ref{ssec:surfacewater}). 
Only if the water cloud base is close to the surface, water can be stable at the planetary surface.
Therefore, a detection of water clouds alone, without further constrains on the height and extent of those clouds, does not necessarily imply that liquid water is also present on the surface.
This is underlined by the solar abundance model, which shows that even for planets without a surface in the sense of rocky exoplanet, water clouds can be thermally stable in the planets atmosphere.

Our work shows that multiple cloud condensates can be thermally stable in one atmosphere.
Some of which are connected via condensate chains (Fig.~\ref{fig:CloudTransitions}).
Although not always all condensates of a condensate chain appear as thermally stable condensates in a given atmosphere, the condensate chains itself are robust against changes in crust composition as well as $(p_\mathrm{surf}, T_\mathrm{surf})$.
Based on this, the detection of a condensate in a planetary atmosphere can suggest further condensates in lower atmospheric layers. 
An example would be the detection of \ce{FeS2}[s], which suggests the presence of  \ce{FeS}[s] in lower atmospheric layers.
As these clouds would most likely be situated in the optically thick part of the atmosphere, their detection will be a great challenge.

\section{Summary}\label{sec:discussion}
This paper investigated the stability of a large collection of cloud species in the tropospheres of rocky planets made of various elemental compositions.
We considered a solid planetary surface in chemical and phase equilibrium with the near-crust atmosphere, where the deposition and the outgassing rates balance each other for every condensed species.
Other geological (e.g. tectonics and volcanism) or biological fluxes are disregarded.  
Our cloud condensate models are then based on a hydrostatic, polytropic $(p_\mathrm{gas}, T_\mathrm{gas})$ profile in the lower (tropospheric) part of the atmosphere, where we assumed the polytropic index to be $\gamma = 1.25$ in all models.
The model uses a bottom-up algorithm where we determine, in each new layer, which condensates become thermally stable and successively remove the elements contained in these condensates, until chemical and phase equilibrium is again established.

This fast and simple model provides first insights into the sequence of cloud layers to be expected in the atmospheres of rocky exoplanets, depending on surface temperature, surface pressure, and crust composition.
These equilibrium cloud condensation sequences should be regarded as a starting point for further investigations in kinetic cloud formation models.
However, models that include boundary fluxes, both at the bottom and the top of the atmosphere, for example geological and biological fluxes from below, or atmospheric escape rates at the top, as well as non-equilibrium processes like photochemistry are beyond the scope of this paper.

The thermally stable cloud condensates in our model can be divided into low and high temperature cloud condensates, with a dividing temperature of about $T\!\approx\!400\,$K.
Below that dividing temperature, the overall most abundant condensates are \ce{H2O}[l,s], \ce{C}[s], and \ce{NH3}[s].
Further condensates at low temperatures are \ce{S2}[s] or contain ammonia, in particular \ce{NH4Cl}[s] and \ce{NH4SH}[s].
Only in models with high surface pressures, we find \ce{H2SO4}[s] to become another stable condensate in high atmospheric layers.
For very low temperatures ($T\!\lesssim\!150\,$K) \ce{CO2}[s], \ce{CH4}[s], \ce{NH3}[s], and \ce{H2S}[s] clouds can be present.
\ce{H2S}[s] is also only thermally stable for higher surface pressures.

In our solar system some of these cloud condensate are present in various planets.
\ce{H2O}[s,l] on Earth. On Mars clouds of \ce{CO2}[s] \citep{2007JGRE..11211S90M} and \ce{H2O}[s] ice \citep{Curran1973} are present.
For the clouds on Venus, \ce{H2SO4}[s] is discussed as the main cloud constituent \citep[e.g.][]{Rimmer2021}.
The four giant planets in our solar system are thought to have multiple cloud layers consisting of \ce{H2O}[s], \ce{NH4SH}[s] and \ce{NH3}[s] \citep[e.g.][]{Atreya1999, Wong2015}.
Furthermore, \ce{CH4}[s] clouds are found in the atmospheres of Uranus, Neptune and Saturn's moon Titan \citep[e.g.][]{Brown2002, Sromovsky2011, Lellouch2014}.
\ce{N2}[s] has been been suggested for clouds on Neptune's moon Triton \citep{Tokano2017}.

At higher temperatures $T\!\gtrsim\!500\,$K, the most relevant cloud species are found to be halides (\ce{KCl}[s], \ce{NaCl}[s]), iron sulphides (\ce{FeS}[s], \ce{FeS2}[s]) and iron oxides (\ce{FeO}[s], \ce{Fe2O3}[s] \ce{Fe3O4}[s]).
Contrasting to previous research focusing on giant planets, which are hydrogen rich, we do not find any \ce{Na2S}[s] in our models \citep{Visscher2006,Parmentier2016,Gao2020} but rather \ce{NaCl}[s] instead. 
For temperatures higher than investigated in this paper ($T_\mathrm{surf}\!\gtrsim\!1250\,$K), when S is more abundant than Cl, we find \ce{Na2S}[s] as a stable cloud condensate \citep[see also][]{Woitke2020}.


The only condensate that is thermally stable in the temperature range $400\,\mathrm{K}\lesssim T_\mathrm{gas}\!\lesssim\!500\,$K and also high in abundance is \ce{C}[s] (i.e. graphite, soot clouds).
It is the only condensate which can be stable throughout the atmospheric parameter range investigated in this work.
In agreement to this, previous studies have shown that carbon is thermally stable in the form of graphite  for carbon rich stars, but also giant planets with C/O$\geq 0.85$ \citep[e.g.][]{Sharp1988, Lodders1997, Lodders2002, Moses2013, Mbarek2016}.

Looking at all condensed species, including those with trace abundances, we show that for the various different element abundances and surface temperatures from 300\,K to 1000\,K considered in this paper, a total of 55 condensates are found to be thermally stable in models with 1\,bar surface pressure, plus additional 17 condensates for the 100\,bar models.
The gap between the low and high temperature cloud condensates is filled by the high temperature condensates at lower abundances.

The different cloud condensates are not independent of each other but follow a sequence of mostly type~2 phase transitions as explained in \citet{2018A&A...614A...1W}, charaterised by sudden changes of the stable condensates. Since we are only considering models with $T\!<\!1000\,K$ in this paper, type~1 phase transitions, where condensates are slowly building up from the gas phase, are relatively rare. 
It is remarkable that these condensation sequences, or condensation chains as shown in Fig.~\ref{fig:CloudTransitions}, are relatively robust against changes of the surface element abundances,
even for the hydrogen-rich models considered in this paper (Archean and solar element abundances).
These models can be seen as a chemical composition of atmospheres either of very young or of very massive planets, which are able to keep the light elements in the atmosphere over longer times.

All atmosphere models in this paper can be classified by being either a member of type~A (H-rich), type~B (O-rich), or type~C (coexistence of \ce{CO2} and \ce{CH4}) according to  \citet{Woitke2020a}.
An important conclusion from this paper is that the atmospheric type does not change as function of height, despite the changing element abundances due to condensation, whereas other characteristics like the carbon to oxygen ratio (C/O) or the metallicity (O/H) can change significantly. 
This finding emphasises the robustness of the classification scheme presented in \citet{Woitke2020a}.

The results of this bottom to top atmospheric model are a step towards solving the inverse problem of inferring surface conditions based on cloud condensates and the higher atmospheric composition.
In general, the inferred crust composition based on cloud condensates is degenerate (see Tables~\ref{tab:Planets} and \ref{tab:Planets-hot}).
Showing again how challenging and ambiguous the task of inferring surface conditions from the high atmosphere conditions can be.
However, certain condensates like the sulphur containing clouds at the low temperature regime do constrain the surface conditions. For a further discussion on the degeneracies see Appendix~\ref{sec:Degen}.
For example, \ce{NH4SH}[s] and \ce{S2}[s] are only thermally stable in reducing and oxidising environments, respectively.
Whereas \ce{H2SO4}[s] and \ce{H2S}[s] have only been found for our polytropic atmospheres with higher surface pressure.
This increased $p_\mathrm{surf}$ results in an increased $T_\mathrm{surf}$, which is incompatable with liquid water at the planetary surface. 
\citet{Loftus2019} found that thick \ce{H2SO4}[s] clouds and hazes are incompatible with a substantial water ocean, as the sulphur compound will dissolve in the ocean.
It has been suggested that under oxygen rich conditions in sulphur will condense in the form of \ce{S8}[s] \citep[e.g.][]{Hu2013}.
We on the other hand find the allotrope \ce{S2}[s] to form instead.

\ce{H2O}[l,s] is a thermally stable cloud condensate for every crust composition investigated in this paper.
However, the pressure level of the \ce{H2O}[l,s] cloud base is varying substantially among the models and is a consequence of the hydration level of the surface rocks.
Only for a fully hydrated crust where all hydrated minerals have formed, water clouds extend down to the atmosphere-crust interaction layer, which is equivalent to \ce{H2O}[l,s] as a stable part of the planetary crust.
This underlines, that the detection of water vapour in a planetary atmosphere or even water condensates do not necessarily imply the presence of surface water.

Although the crust compositions without \ce{H2O}[l] are not habitable for water based life, we show that \ce{H2O}[l] can be stable in parts of these planetary atmospheres.
These conditions can be used for thoughts on aerial biospheres \citep{Seager2021}.
With our model, we can show, that not only liquid water is stable in some atmospheres, but other condensates, which can provide some nutrients for biological processes (e.g. \ce{NH4SH}[s] or \ce{Mg3P2O8}[s]).
Additionally some gas phase molecules like \ce{CH4} or \ce{NH3} could provide energy sources for life to thrive.

\begin{acknowledgements}
O.H.\ acknowledges the PhD stipend form the University of St Andrews' Centre for Exoplanet Science. P.W.\ and Ch.H.\ acknowledge funding from the European Union H2020-MSCA-ITN-2019 under Grant Agreement no. 860470 (CHAMELEON).
\end{acknowledgements}

\begin{landscape}
\begin{table}[!ht]
\caption{Confirmed transiting planets with planetary mass $M_{\rm{P}}<8\,\rm{M}_{\oplus}$, planetary radius $R_{\rm{P}}<2\,\rm{R}_{\oplus}$ and equilibrium temperature $250\,\mathrm{K}<T_{\rm{eq}}<700\,$K. Sorted by equilibrium temperature. 
For different abundant cloud condensates in the high atmospheres, as predicted by out atmospheric model, we list the corresponding crust element abundances.
We assume $T_{\rm{eq}}=T_{\rm{surf}}$ and an atmospheric pressure of $p_{\rm{surf}}=1\,$bar.
The pressure level at which the clouds are present vary with condensate (see Fig.~\ref{fig:CloudsAll-10}).}
\label{tab:Planets}
{
\vspace*{-2mm}
\begin{tabular}{c|ccccc|cccccc}
\hline
$T_\text{surf}$ [K] &Planet &$	M_{\rm{P}} [\rm{M}_\oplus] $&$	R_{\rm{P}} [\rm{R}_\oplus]$&$	T_{\rm{eq}} [\rm{K}]$&Ref.& \ce{H2O}[l,s]&\ce{C}[s]&\ce{NH4Cl}[s]&\ce{NH4SH}[s]&\ce{NH3}[s]&\ce{S2}[s]\\
\hline <350 
&TRAPPIST-1 e	&$	0.6	\pm	0.6 $&$	0.92	\pm	0.04	$&$	251	\pm	5	$&	(1)&all&&&BSE12&PWD CI&\\
&K2-3 d      	&$	2.8	    	$&$	1.6	    \pm	0.3	    $&$	282.0      	$&	(2)&&&&BSE15&&\\
&TRAPPIST-1 d	&$	0.4	\pm	0.3	$&$	0.77	\pm	0.04	$&$	288	\pm	6	$&	(1)&&&&CI&&\\
&TOI-700 d	    &$	2.3	\pm	0.7	$&$	1.19	\pm	0.11	$&$	300	\pm 60	$&	(3)&&&&Archean&&\\
&TRAPPIST-1 c	&$	1.4	\pm	0.7	$&$	1.056	\pm	0.004	$&$	341	\pm	7   $&	(1)&&&&&&\\
&K2-3 c	        &$	2.1	\pm	1.1 $&$	1.9	    \pm	0.3	    $&$	344.0    	$&	(2)&&&&&&\\ 
\hline 350-400
&TOI-237 b	    &$	2.9	\pm	1.5 $&$	1.44	\pm	0.12	$&$	355.0    	$&	(4)&all&BSE12&		&BSE12	&		&	\\
&K2-155 d    	&$	4.9	\pm	1.2 $&$	1.9 	\pm	0.5	    $&$	381	\pm	40	$&	(5)&&BSE15&&BSE15&&\\
&K2-240 c	    &$	4.7	\pm	0.5	$&$	1.8 	\pm	0.3	    $&$	389	\pm	18	$&	(6)&&CI&&CI&&\\
&K2-239 d    	&$	1.4	\pm	0.5	$&$	1.10	\pm	0.11	$&$	399	\pm	1	$&	(6)&&&&PWD MORB&&\\
&TRAPPIST-1 b	&$	0.9	\pm	0.8	$&$	1.086	\pm	0.004	$&$	400	\pm	8   $&	(1)&&&&PWD CI&&\\
\hline 400-450
&K2-239 c	    &$	0.9	\pm	0.3 $&$	1.00	\pm	0.11	$&$	427	\pm	22	$&	(6)&all	&CI	&&\makecell[t]{BSE12\\ BSE15\\ PWD CI}	&&	CC, CI\\
\hline 450-500
&K2 42 d        &$ <0.9         $&$ 0.57    \pm 0.18    $&$ 450 \pm 45  $&  (7)&all &\makecell[t]{MORB\\CI}&&\makecell[t]{BSE12\\BSE15}&&\makecell[t]{CC, CI\\MORB}\\ 
\hline 500-550
&K2-239 b	    &$	1.4	\pm	0.5	$&$	1.10	\pm	0.11	$&$	502.0       $&	(6)&all&BSE&	solar	&		&		&CC, CI\\
&TOI-776 b	    &$	4.7	\pm	1.0 $&$	1.83	\pm	0.11	$&$	513	\pm	12	$&	(8)&& MORB&&&&MORB\\
&GJ 357 b	    &$	2.1	\pm	0.4	$&$	1.17	\pm	0.04	$&$	525	\pm	10	$&	(9)&&CI&&&&\\
&TOI-270 b   	&$	1.9	\pm	1.1	$&$	1.25	\pm	0.09	$&$	528	\pm	50	$&	(10)&&PWD BSE&&&&\\
&Kepler-167 d	&$	1.9	\pm	1.0	$&$	1.20	\pm	0.05	$&$	536.0    	$&	(11)&&&&&&\\
\hline 550-600
&Kepler-62 c	&$	4.131	    $&$	0.54	\pm	0.04	$&$	578	\pm	40	$&	(12)&all&BSE&PWD MORB&&&CC, CI\\
&LHS 1478 b	    &$	2.3	\pm	0.20$&$	1.24	\pm	0.06	$&$	595	\pm	10	$&	(13)&but solar&\makecell[t]{MORB, CI\\PWD BSE\\PWD MORB}&solar&&&PWD MORB\\
\hline 600-650
&GJ 9827 d   	&$	3.4	\pm	0.7	$&$	1.96	\pm	0.08	$&$	605	\pm	20	$&	(14)&all, but&BSE&BSE&&&CC\\
&LHS 1815 b	    &$	4.2	\pm	1.5	$&$	1.09	\pm	0.07	$&$	617	\pm	90	$&	(15)&solar&MORB, CI&solar&&&\\
&Kepler-114 c	&$	2.9	\pm	0.7	$&$	1.60	\pm	0.19	$&$	623.0       $&	(16)&PWD BSE&PWD BSE&&&&\\
&K2 244 b       &$	5.7 \pm	2.4 $&$	1.75	\pm	0.13    $&$	638	\pm	7	$&	(17)&&PWD MORB&&&&\\
&K2 252 b       &$	5.6	\pm	2.4	$&$	1.74	\pm	0.17    $&$	639	\pm	7	$&	(17)&&&&&&\\
\hline 650-700
&GJ 486 b	    &$	2.8	\pm	0.13$&$	1.31	\pm	0.07	$&$	700	\pm	13	$&	(18)&	\makecell[t]{BSE, BSE12\\ BSE15, CC\\ MORB, CI\\ Earth, Archean} 	&	\makecell[t]{BSE\\ MORB, CI\\ PWD BSE\\ PWD MORB}	&	\makecell[t]{BSE, BSE12\\BSE15, MORB\\ PWD CI\\ Archean\\ solar}	&		&		&	CC\\ \hline
\end{tabular}}
\\{\footnotesize \tablebib{ (1) \citet{Gillon2017}; (2) \citet{Sinukoff2016}; (3) \citet{Suissa2020}; (4)\citet{Waalkes2020}; (5) \citet{Hirano2018}; (6) \citet{Alonso2018}; (7) \cite{Muirhead2012}; (8) \citet{Luque2021}; (9) \citet{Luque2019}; (10) \citet{Guenther2019}; (11) \citet{Kipping2016}; (12) \citet{Borucki2013}; (13) \citet{Soto2021} ; (14)\citet{Niraula2017}; (15) \citet{Gan2020}; (16) \citet{Batalha2013}; (17) \citet{Livingston2018}; (18) \citet{Trifonov2021} }}
\end{table}
\end{landscape}
\begin{landscape}
\begin{table}[!ht]
\caption{Confirmed transiting planets with planetary mass $M_{\rm{P}}<8\,\rm{M}_{\oplus}$, planetary radius $R_{\rm{P}}<2\,\rm{R}_{\oplus}$ and equilibrium temperature $700<T_{\rm{eq}}<1050\,$K. Sorted by equilibrium temperature.
For different abundant cloud condensates in the high atmospheres, as predicted by out atmospheric model, we list the corresponding crust element abundances.
We assume $T_{\rm{eq}}=T_{\rm{surf}}$ and an atmospheric pressure of $p_{\rm{surf}}=1\,$bar.
The pressure level at which the clouds are present vary with condensate (see Fig.~\ref{fig:CloudsAll-10}).}
\label{tab:Planets-hot}
{
\vspace*{-2mm}
\begin{tabular}{c|ccccc|cccccccc}
\hline
$T_\text{surf}$ [K] &Planet	&$	M_{\rm{P}} [\rm{M}_\oplus] $&$	R_{\rm{P}} [\rm{R}_\oplus]$&$	T_{\rm{eq}} [\rm{K}]$&Ref.& \ce{C}[s]&\ce{NH4Cl}[s]&\ce{S2}[s]&\ce{FeS}[s]&\ce{FeS2}[s]&\makecell[t]{\ce{KCl}[s]\\ \ce{NaCl}[s]}&\ce{AlF3}[s]&\ce{Fe2O3}[s]\\
\hline 700-750
&GJ 143 c	    &$	3.69919	    $&$	0.89	\pm	0.07	$&$	703	\pm	24	$&	(19)&CI&BSE&CC	&	&	&	&Earth	&	\\
&K2 155 b    	&$	4.7	\pm	0.4	$&$	1.81	\pm	0.16	$&$	708	\pm	40	$&	(5) &PWD MORB& BSE12&PWD BSE&&&&&\\
&K2	154 b       &$	6.8	\pm	2.5	$&$	1.99	\pm	0.21	$&$	715	\pm	9	$&	(17)&&BSE15&&&&&&\\
&GJ 9827 c	    &$	1.9	\pm	0.5	$&$	1.20	\pm	0.05	$&$	722	\pm	25	$&	(20)&&MORB&&&&&&\\
&Kepler-114 b	&$	7	\pm	4	$&$	1.26	\pm	0.14	$&$	722.0   	$&	(15)&&\makecell[t]{PWD CI\\ Archean\\ solar}&&&&&&\\
\hline 750-800
&KOI-1599.01	&$	4.6	\pm	0.3	$&$	1.9 	\pm	0.4 	$&$	755.0   	$&	(21)&	PWD MORB	&	BSE&	CC	&		&&&	Earth	&	\\
&GJ 3473 b   	&$	1.9	\pm	0.3	$&$	1.264	\pm	0.006	$&$	773 \pm	16	$&	(22)&&BSE12&PWD BSE&&&&&\\
&TOI-1235 b	    &$	5.9	\pm	0.7	$&$	1.69	\pm	0.08	$&$	775	\pm	13	$&	(23)&&BSE15&&&&&&\\
&K2	257 b       &$	0.7	\pm	0.5	$&$	0.83	\pm	0.06	$&$	789	\pm	14	$&	(17)&&MORB&&&&&&\\
&K2	254 b       &$	5.2	\pm	2.6	$&$	1.63	\pm	0.28	$&$	791	\pm	26	$&	(16)&&\makecell[t]{PWD CI\\Archean\\solar}&&&&&&\\
\hline 800-900
&L 168-9 b   	&$	4.6	\pm	0.6	$&$	1.390	\pm	0.09	$&$	820	\pm	160	$&	(24)&PWD MORB&PWD CI&CC&BSE&MORB&BSE&Earth&\\
&TOI-178 c	    &$	4.8	\pm	0.7	$&$	1.67	\pm	0.11	$&$	873	\pm	18	$&	(25)&&&PWD BSE	&&&	BSE12	&&\\
&LTT 3780 b	    &$	3.1	\pm	0.6	$&$	1.33	\pm	0.08	$&$	892.0   	$&	(26)&&&&&& \makecell[t]{BSE15\\Archean\\ solar}&&\\
\hline 900-1000
&HD 136352 b	&$	4.6	\pm	0.5	$&$	1.48	\pm	0.06	$&$	911	\pm	18	$&	(27)&PWD MORB&&PWD BSE&BSE&MORB&BSE&Earth&CC\\
&K2	228 b       &$	3.2	\pm	1.7	$&$	1.21	\pm	0.10	$&$	914	\pm	14	$&	(17)&&&&&&BSE12&&\\
&TOI-1634 b	    &$	4.9	\pm	0.7	$&$	1.79	\pm	0.09	$&$	924	\pm	22	$&	(28)&&&&&&BSE15&&\\
&K2 158 c	    &$	3.6	\pm	2.0	$&$	1.28	\pm	0.13	$&$	948	\pm	13	$&	(17)&&&&&&PWD BSE&&\\
&K2	224 b       &$	4.9	\pm	2.4	$&$	1.56	\pm	0.16	$&$	1002\pm	11	$&	(17)&&&&&&PWD CI&&\\
&Kepler-93 b	&$	4.5	\pm	0.9	$&$	1.57	\pm	0.11	$&$	1037.0    	$&	(29)&&&&&&Archean&&\\
&K2-233 b	    &$	3	\pm	3	$&$	1.40	\pm	0.07	$&$	1040\pm	27	$&	(30)&&&&&&solar&&\\
&TOI-178 b	    &$	1.5	\pm	0.5	$&$	1.15    \pm	0.08	$&$	1040\pm	22	$&	(25)&&&&&&&&\\
&GJ 9827 b   	&$	4.9	\pm	0.4	$&$	1.53    \pm	0.06	$&$	1040\pm	40	$&	(20)&&&&&&&&\\
 \hline
\end{tabular}}
{\footnotesize \tablebib{(5) \citet{Hirano2018}; (16) \citet{Batalha2013}; (17) \citet{Livingston2018}; (19) \citet{Dragomir2019}; (20) \citet{Niraula2017}; (21)\citet{Panichi2019}; (22) \citet{Kemmer2020}; (23) \citet{Bluhm2020}; (24) \citet{Astudillo_Defru2020}; (25) \citet{Leleu2021}; (26) \citet{Cloutier2020}; (27) \citet{Kane2020}; (28) \citet{Cloutier2021}; (29) \citet{Ballard2014}; (30) \citet{LilloBox2020}}}
\end{table}
\end{landscape}

\bibliography{library} 

\begin{appendix}
\section{The Earth model}\label{Sect:EarthModel}
The composition of the surface of the Earth and its atmosphere are strongly affected by geological processes like plate-tectonics and volcanism, as well as biological activity \citep[see e.g.][]{Claire2006, Mills2014, 2018A&ARv..26....2L}.
It is nevertheless an interesting question whether or not a simple model can be constructed, in which crust and atmosphere are in chemical and phase equilibrium with each other, to better discuss and quantify the influence of the non-equilibrium processes.   
In this appendix, we demonstrate that it is indeed possible to find a set of total (condensed $+$ gas) element abundances which simultaneously fit the measured molecular abundances in the Earth atmosphere and the condensed mass fractions in the Earth's crust when assuming chemical and phase equilibrium.

Our starting point of this model are the mass fractions $f_k$ of the 18 elements H, C, N, O, F, Na, Mg, Al, Si, P, S, Cl, K, Ca, Ti, Cr, Mn, and Fe as measured in the Earth's continental crust \citep{Schaefer2012}, see also column "CC Schaefer" in table A.2 in \citet{Herbort2020}.  These are the abundances which represent the condensed matter in the model.

Next, we (i) convert these mass fractions $f_k$ to nuclei particle ratios, (ii) multiply with a large, arbitrary factor $\Delta=1000$ to account for the dominance of condensed over gaseous matter, and (iii) add the observed gas abundances in the Earth atmosphere.
\begin{equation}
  \epsilon_k^{\rm tot} = \Delta\,\frac{f_k}{m_k} + \sum_i s_{i,k}\, (c_i^{\rm gas} + \delta_i)
\end{equation}
where $m_k$ is the mass of element $k$, $s_{i,k}$ is the stoichiometric factor of element $k$ in molecule $i$, $c_i^{\rm gas}$ is the particle concentration of molecule $i$ in the Earth atmosphere, and $\delta_i$ are corrections as discussed below.

Based on these total element abundances $\epsilon_k^{\rm tot}$ we ran a {\sc GGchem} model for the mean conditions at the surface of Earth $p\!=\!1.013$\,bar and $T\!=\!288.15$\,K. The results include an identification of the stable condensates, the condensed mass fractions (which are very close to $f_k$ because of the large factor $\Delta$), and the molecular concentrations in the atmosphere.

Without corrections $\delta_i$, this experiment results in an almost pure \ce{N2} atmosphere ($\sim\!99$\%) with traces of noble gases, but all other molecules are incorporated in the condensed species. For example, the gaseous \ce{H2O} we put disappears in phyllosilicates and the \ce{CO2} is incorporated in form of carbonates.  In order to arrive at a better fit, we add
\begin{itemize}
    \item more \ce{O2} to make the crust extremely oxidised, until a small leftover of \ce{O2} is contained in the atmosphere,
    \item more \ce{H2O} to completely saturate the phyllosilicates, and then even more to achieve an atmosphere that is saturated with liquid water,
    \item more \ce{CO2} to completely saturate the carbonates in the crust, until we have some leftover \ce{CO2} in the atmosphere. 
\end{itemize}
After some trial and error, leading to $\delta_{\rm O2}\!\approx\!4.92$, $\delta_{\rm H2O}\!=\!160$, and $\delta_{\rm CO2}\!\approx\!162.8$, we arrived at a model that roughly fits the composition of the Earth atmosphere (Table~\ref{tab:Earth1}) and makes interesting predictions about the Earth crust composition in phase equilibrium (Table~\ref{tab:Earth2}).

\begin{table*}
\caption{Comparison of Earth atmospheric data \citep{Muralikrishna2017} with a {\sc GGchem} phase-equilibrium model at $p\!=\!1.013$\,bar and $T\!=\!288.15$\,K based on total element abundances adjusted to fit the data, see text.}
\label{tab:Earth1}
\vspace*{-2mm}
\resizebox{\textwidth}{!}{\begin{tabular}{c|ccccccccccc}
\hline
&&&&&&&\\[-2ex]
& \ce{N2} & \ce{O2} & \ce{H2O} 
     & Ar & \ce{CO2} & Ne 
     & He & \ce{CH4} & \ce{H2} 
     & \ce{SO2} & \ce{HNO3} \\
\hline
&&&&&&&\\[-2ex]
observed ($c_i$) 
& 77\% & 20.5\% & 10\,ppm\,$-$\,5\% 
& 0.93\% & 400\,ppm & 18\,ppm
& 5.2\,ppm & 1.8\,ppm & 550\,ppb
& 75\,ppb & ? \\
&&&&&&&\\[-2ex]
GGchem model     
& 77\% & 20.5\% & 1.7\%
& 0.93\% & 400\,ppm & 18\,ppm
& 5.2\,ppm & $\ll$ & $\ll$
& $\ll$ & 2\,ppb \\
\hline 
\end{tabular}}\\[1mm]
\footnotesize The measured concentrations of \ce{H2O} depend on latitude, height, and weather conditions. The value of 1.7\% corresponds to saturated liquid water (100\% humidity) at 288.15\,K in the model. Molecules marked with "$\ll$" and all other molecules not listed have concentrations $<10^{-10}$ in the model.
\end{table*}

\begin{table}
\caption{Composition of the Earth crust as predicted from the {\sc GGchem} phase equilibrium model described in Table\ref{tab:Earth1}. }
\label{tab:Earth2}
\vspace*{-2mm}
\begin{tabular}{rll}
\hline
formula & name & mass fraction\\
\hline
&&\\[-2ex]
\ce{NaAlSi3O8[s]} & {\sl albite}       & 24.2\% \\
\ce{SiO2[s]}      & {\sl quartz}       & 22.8\% \\
\ce{Al2Si2O9H4[s]}& {\sl kaolinite}    & 16.3\% \\
\ce{CaMgC2O6[s]}  & {\sl dolomite}     & 15.0\% \\
\ce{KAlSi3O8[s]}  & {\sl microcline}   & 13.8\% \\
\ce{FeO2H[s]}     & {\sl goethite}     & 6.25\% \\
\ce{TiO2[s]}      & {\sl rutile}       & 0.61\% \\
\ce{Ca5P3O12F[s]} & {\sl fluorapatite} & 0.38\% \\
\ce{CaSO4[s]}     & {\sl anhydrite}    & 0.27\% \\
\ce{Mn2O3[s]}     & {\sl bixbyite}     & 0.094\% \\
\ce{H2O[l]}       & {\sl liquid water} & 0.077\% \\
\ce{NaCl[s]}      & {\sl halite}       & 0.070\% \\
\ce{MgF2[s]}      & {\sl Mg-fluoride}  & 0.048\% \\
\ce{Cr2O3[s]}     & {\sl eskolaite}    & 0.017\% \\
\ce{CaF2[s]}      & {\sl fluorite}     & 0.009\% \\
\hline
\end{tabular}
\end{table}

The resulting mass fractions of the elements included in this paper are given in Table~\ref{tab:abundances}.
The noble gases show no condensate phase in the parameter space investigated and are therefore not included in this work. 
The resulting mass fractions necessary to match the Earth atmosphere are mfrac(He)=2.462E-10, mfrac(Ne)=4.306E-09, and mfrac(Ar)=4.379E-06.

It is noteworthy that, when we choose a ten times larger value for $\Delta$, i.e.\ 10000 instead of 1000, we arrive at results that are very close to those listed in Tables \ref{tab:Earth1} and \ref{tab:Earth2}, when the corrections $\delta_i$ are also increased by a factor of 10. Hence, given an initial set of condensed element fractions (here $f_k$) and the necessity to fit the abundances of the main molecules in the atmosphere (here $c_i$) leads to unique conclusions.

The results in Table~\ref{tab:Earth1} show that it is indeed possible to fit the concentrations of the main molecules \ce{N2}, \ce{O2}, \ce{H2O} and \ce{CO2} (type B atmosphere), and the noble gases, but it is entirely impossible to fit other molecules with ppb-concentrations, in particular \ce{CH4}, which is a well-known non-equilibrium indicator for biological activity \citep[see e.g.][]{Cicerone1988, Sterzik2012, GuzmanMarmolejo2013}.  
Sulphur is found to be practically absent from the gas phase, the most relevant sulphur molecule is \ce{H2SO4} with a concentration of $\sim\!10^{-15}$ in this model.

Table~\ref{tab:Earth2} shows a number of well-known minerals that are indeed found in large quantities on the Earth surface.  Remarkably, we find {\sl goethite} which is a weathering product of iron that is abundantly found in soil and other low-temperature environments such as sediment. 
{\sl Kaolinite} is a common clay mineral with chemical composition \ce{Al2Si2O5(OH)4}. It is a soft, earthy mineral produced by the chemical weathering of aluminium silicate minerals like feldspar.  
Sulphur is found to be completely locked up in {\sl anhydrite} in the model. However, more abundantly found on Earth is {\sl gypsum} with chemical formula \ce{CaSO4}$\cdot\,$2\,\ce{H2O}, the weathering product of {\sl anhydrite}, but unfortunately, {\sl gypsum} is not included in {\sc GGchem}'s data collection.
We therefore consider the results of our simple {\sc GGchem} model as continental crust after complete weathering, i.e. after bringing the material in intensive contact with atmosphere and water, in the limiting case $t\!\to\!\infty$.  
All other condensates not listed in Table~\ref{tab:Earth2} have exactly zero mass fractions in the model, because they are transformed into thermodynamically even more stable condensates in equilibrium.

\section{Degeneracies in obtaining the surface composition}\label{sec:Degen}
The model presented in this paper works towards understanding the link between the surface and the atmospheric composition as well as potential cloud condensates in these atmospheres.
One application for this model is the reverse problem of inferring the surface composition and conditions from measurements of high atmospheric composition and cloud condensates. 
This inverse problem is becoming highly important for future missions like JWST, but especially ARIEL, LUVOIR, and HabEx, which are especially designed to detect and characterize atmospheres of rocky exoplanets.
Our model helps towards identifying and understanding some of the degeneracies implied by the concept.

The approach of an atmosphere and crust in chemical and phase equilibrium yields a unique composition of the gas and condensate phase under the assumption of a known total element abundance of the entire system. 
However, even the best future measurement will not provide the entire mass element abundance, but rather the gas composition, which is only a result of the constrains of the supersaturation ratio $S_j\leq1$ for all condensates $j$.
From the gas phase abundances, a list of thermally stable condensates can be inferred.

Any given set of element abundances ($\epsilon_\text{tot}$) results in solutions for the gas composition ($\epsilon_\text{gas}$) and crust condensates ($\epsilon_\text{crust}$) for all elements $k$.
That is
\begin{align}
\epsilon_\text{tot}^k = \epsilon_\text{gas}^k + \epsilon_\text{crust}^k.
\end{align}
However, for one given st of gas phase abundances $\epsilon_\text{gas}$, the total element abundances are not unambiguously defined.
Only the the list $\{c_j | j=1,...,N\}$ of $N$ stable condensates $c_j$ with supersaturation ratios $S_j=1$ is well defined.
If we add an amount of elements to $\epsilon_\text{tot}^k$ that corresponds to the stoichiometry ($s_{j,k}$) of a chosen stable condensate, that condensate will simply fall out again.
This does not change the supersaturation ratio and therefore has no influence on the resulting gas abundances.  
The modified total element abundances $\tilde\epsilon_\text{tot}^k$ can be obtained by
\begin{align}
\tilde\epsilon_\text{tot}^k = \epsilon_\text{tot}^k + X_j s_{j,k}.
\end{align}
Where $X_j$ can be any number for a stable condensate, as long as the abundance of the corresponding condensate remains positive.
The gas phase results are the same for both $\epsilon_\text{tot}^k$ and $\tilde\epsilon_\text{tot}^k$, resulting in n $N$-dimensional manifold for the crust composition, if the gas phase is known.
Thus, we can predict the mixture of stable condensates, and determine for example whether or not water is stable on the surface of an exoplanet.
However, we cannot predict how much of these stable condensates are present, when only the gas properties are known.

\onecolumn{
\section{Gas phase composition plots}\label{App:Figs}
Here we show the gas phase composition plots for the lower temperature models (Fig.~\ref{fig:AtmoOH} and \ref{fig:AtmoOC}).
\begin{figure*}[!h]
\centering
\includegraphics[width = .32\linewidth, page=1]{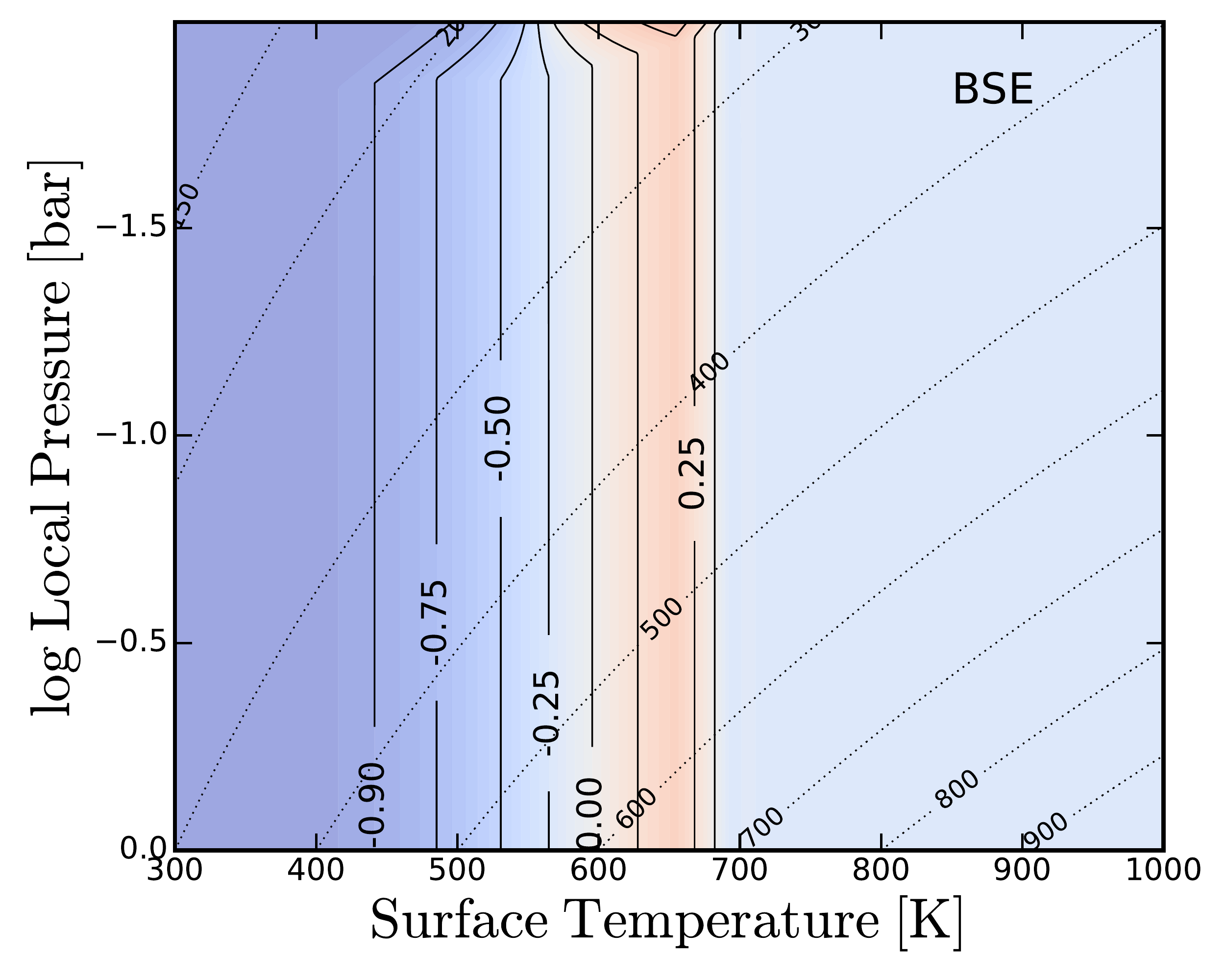}
\includegraphics[width = .32\linewidth, page=1]{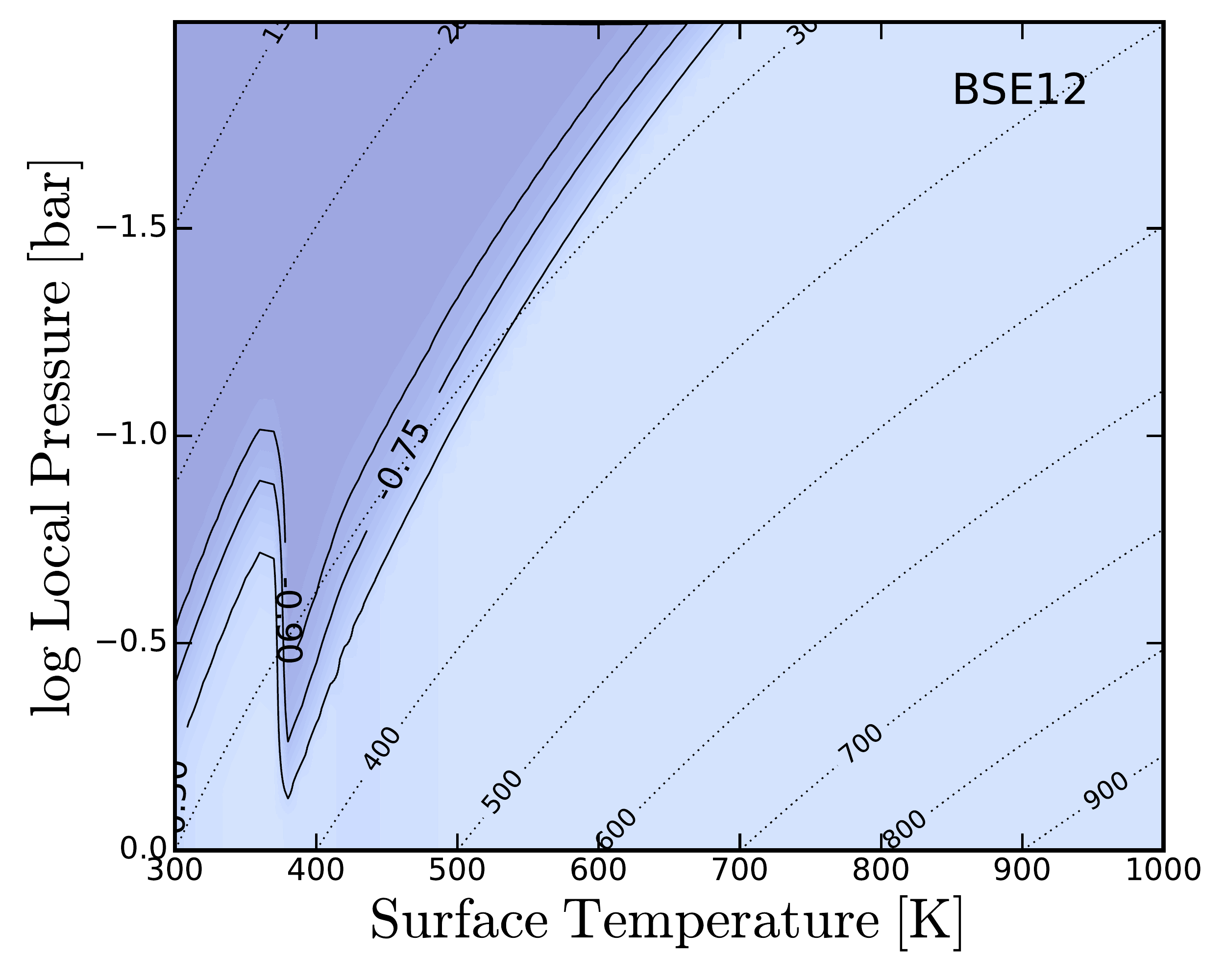}
\includegraphics[width = .32\linewidth, page=1]{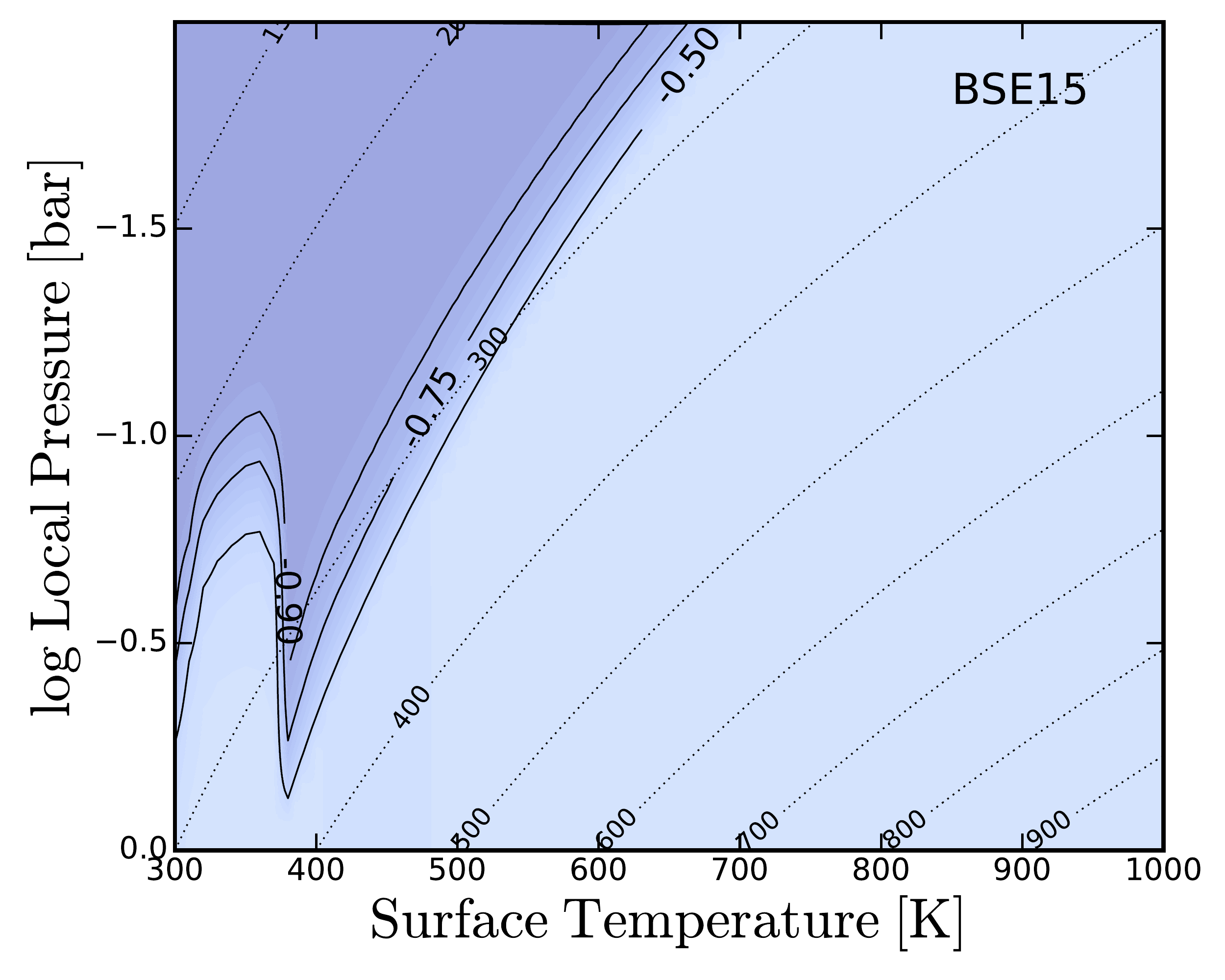}\\
\includegraphics[width = .32\linewidth, page=1]{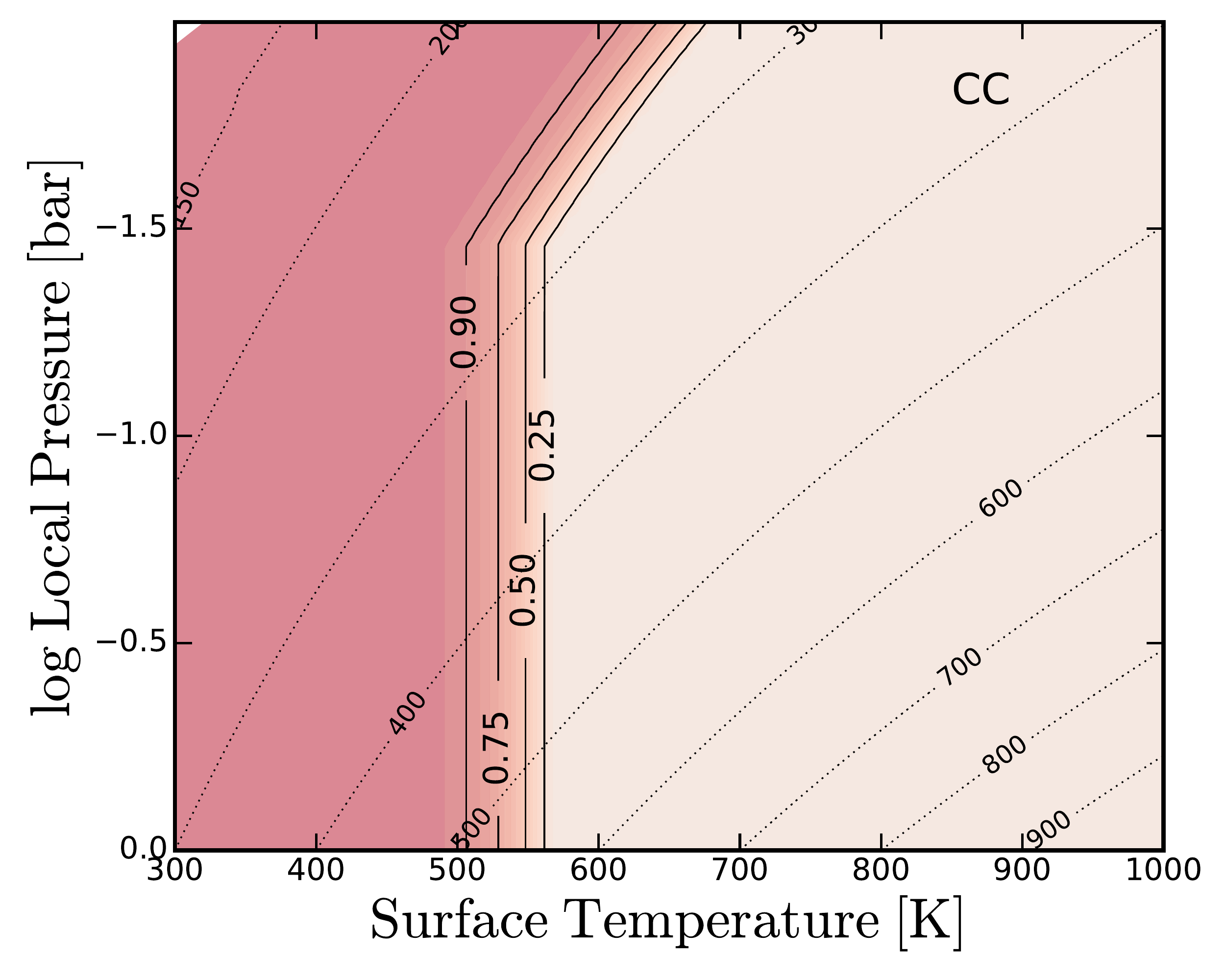}
\includegraphics[width = .32\linewidth, page=1]{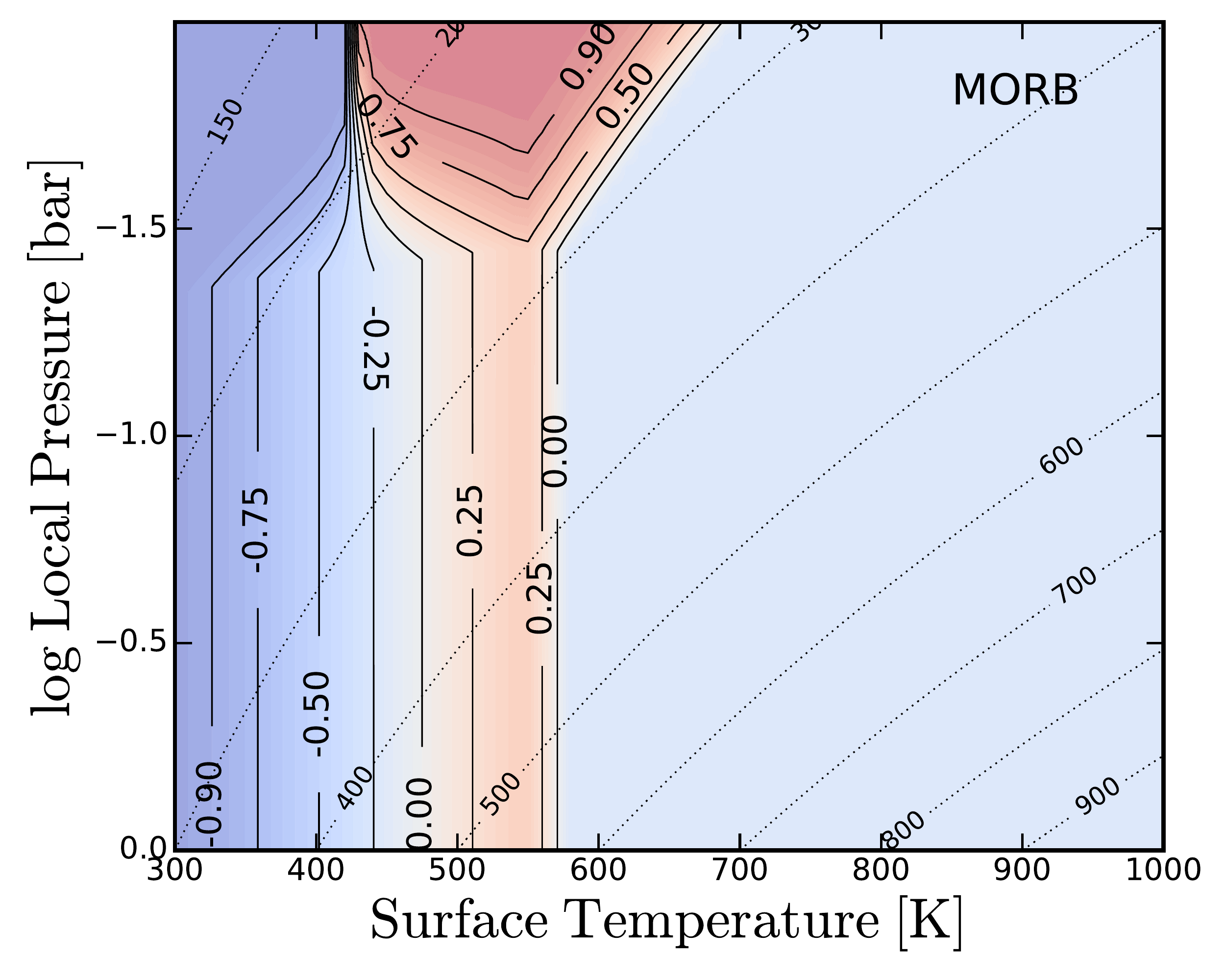}
\includegraphics[width = .32\linewidth, page=1]{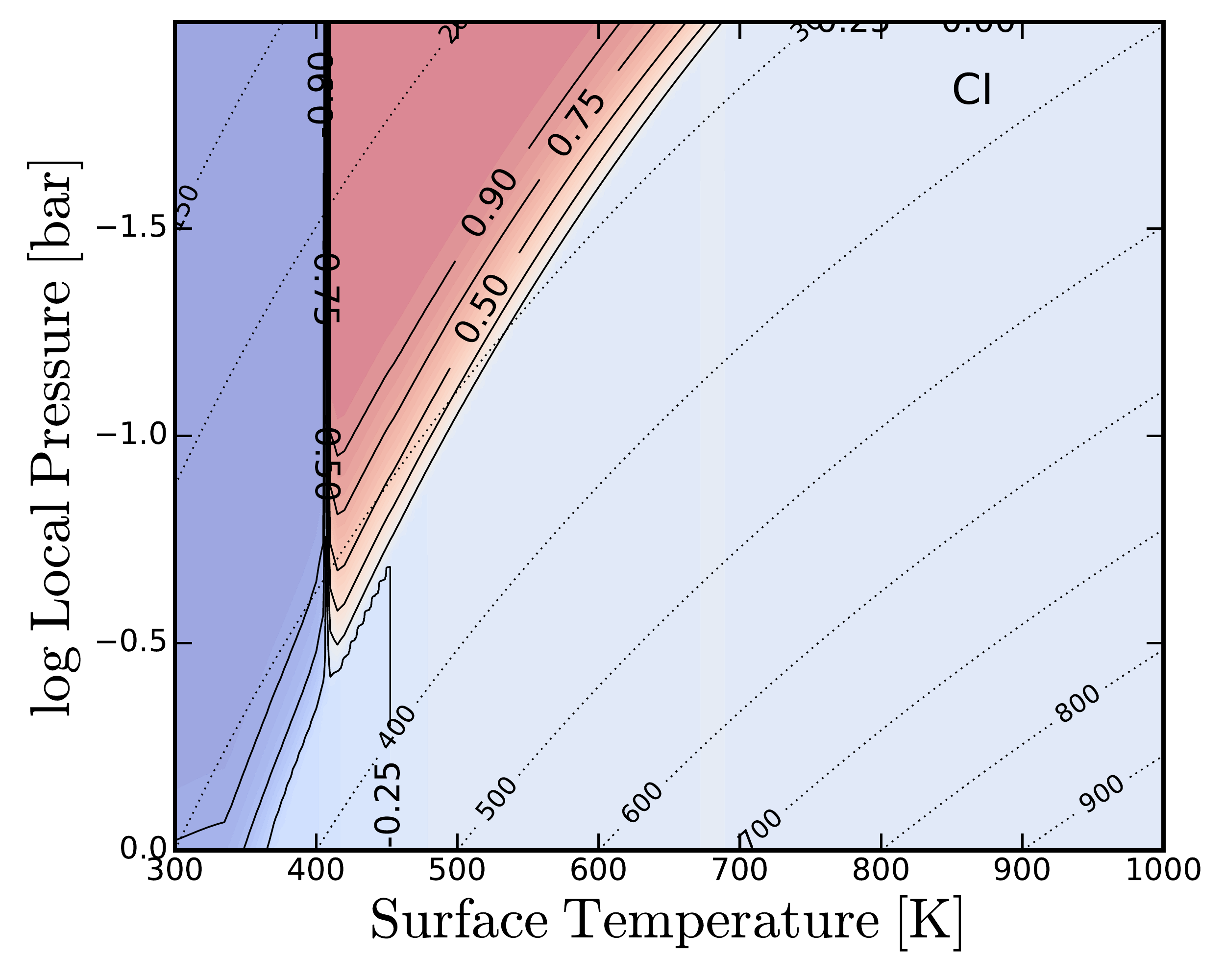}\\
\includegraphics[width = .32\linewidth, page=1]{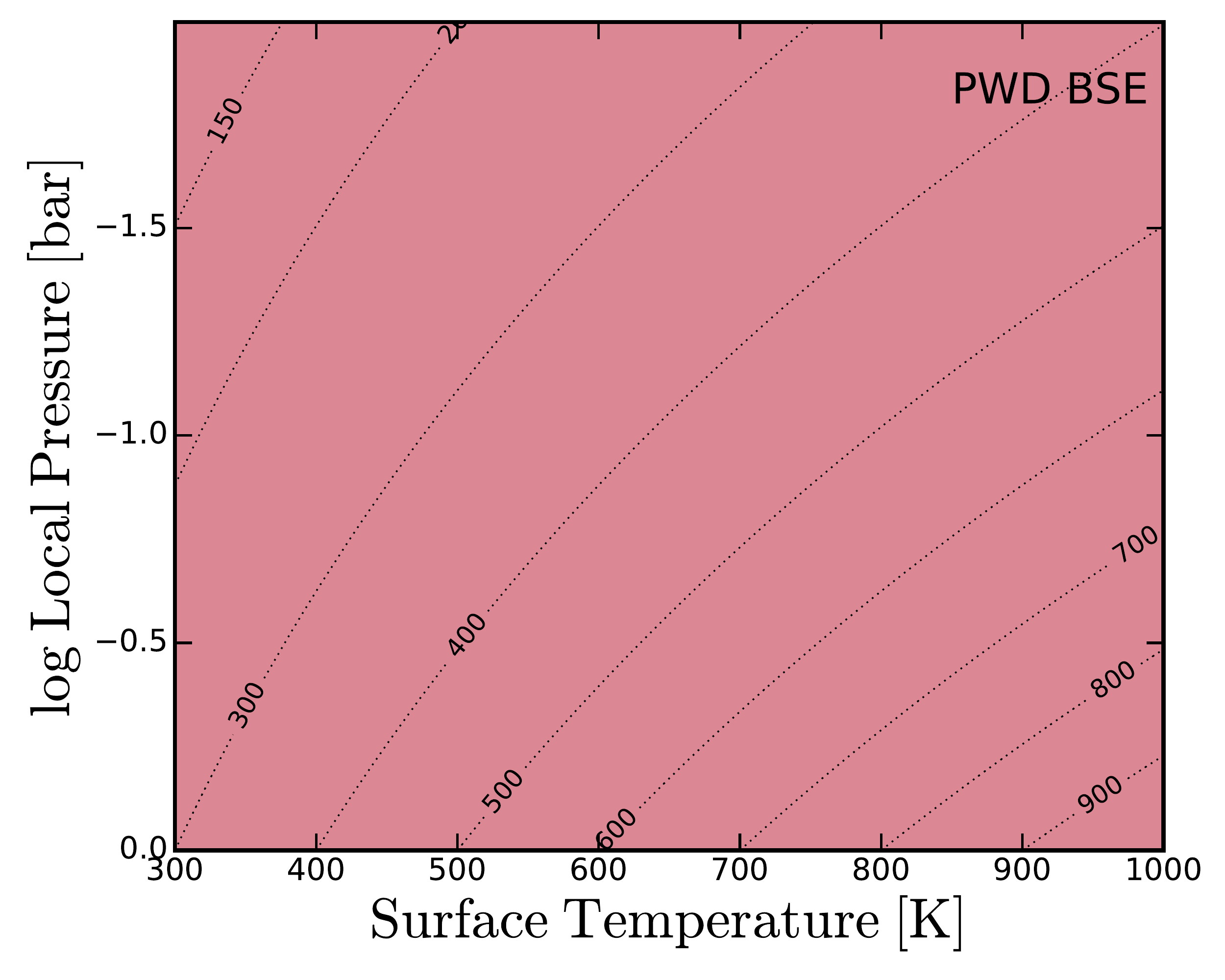}
\includegraphics[width = .32\linewidth, page=1]{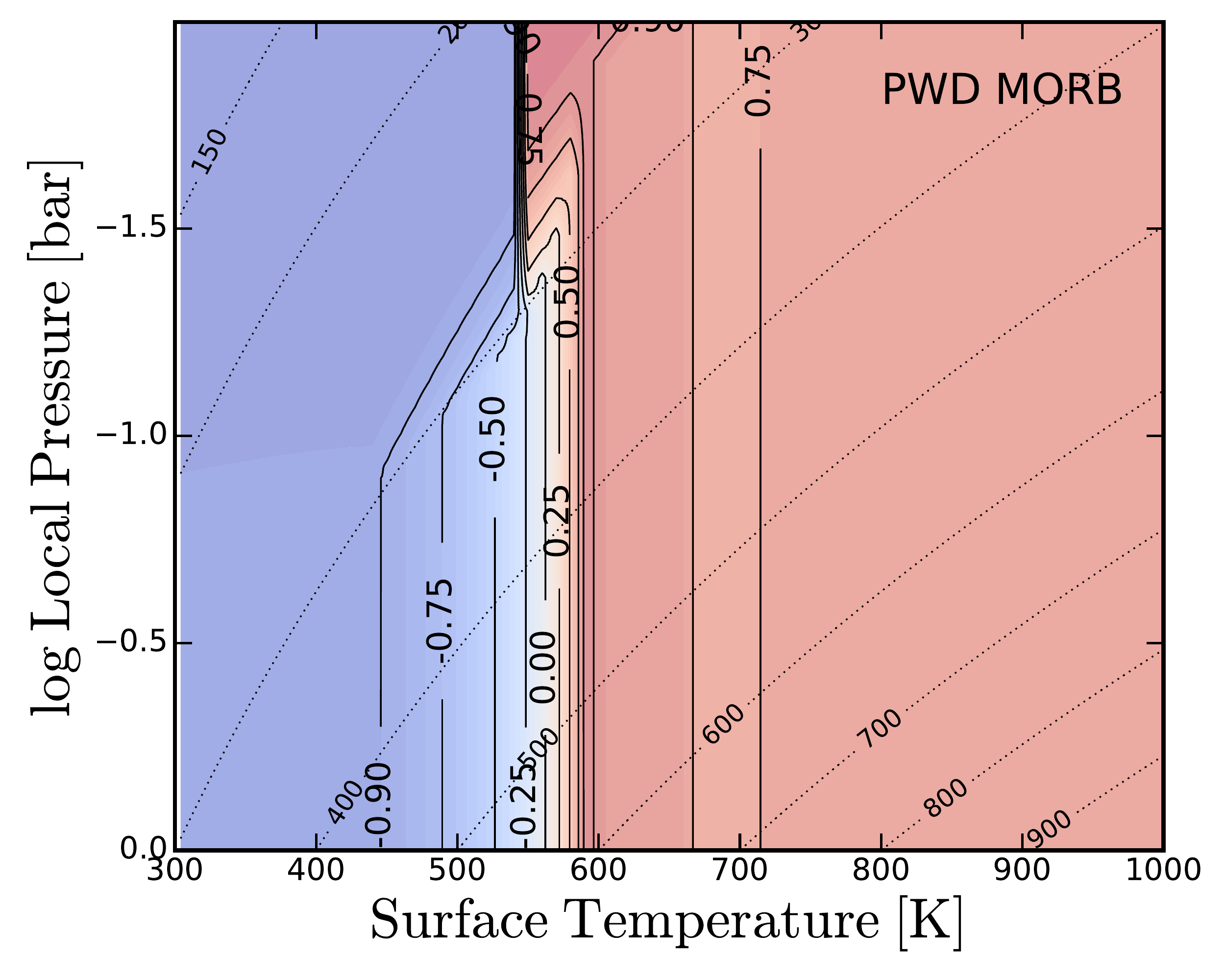}
\includegraphics[width = .32\linewidth, page=1]{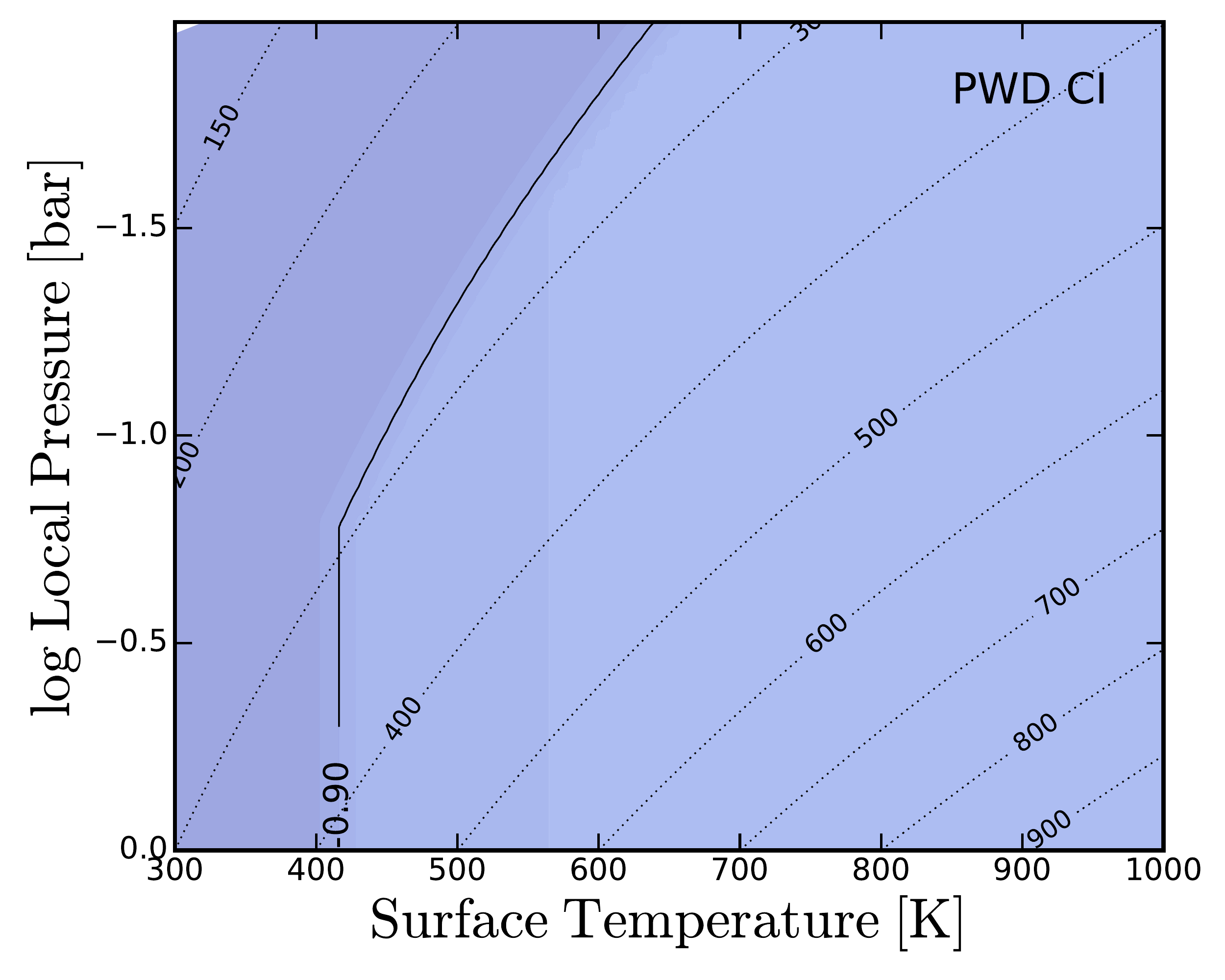}\\
\includegraphics[width = .32\linewidth, page=1]{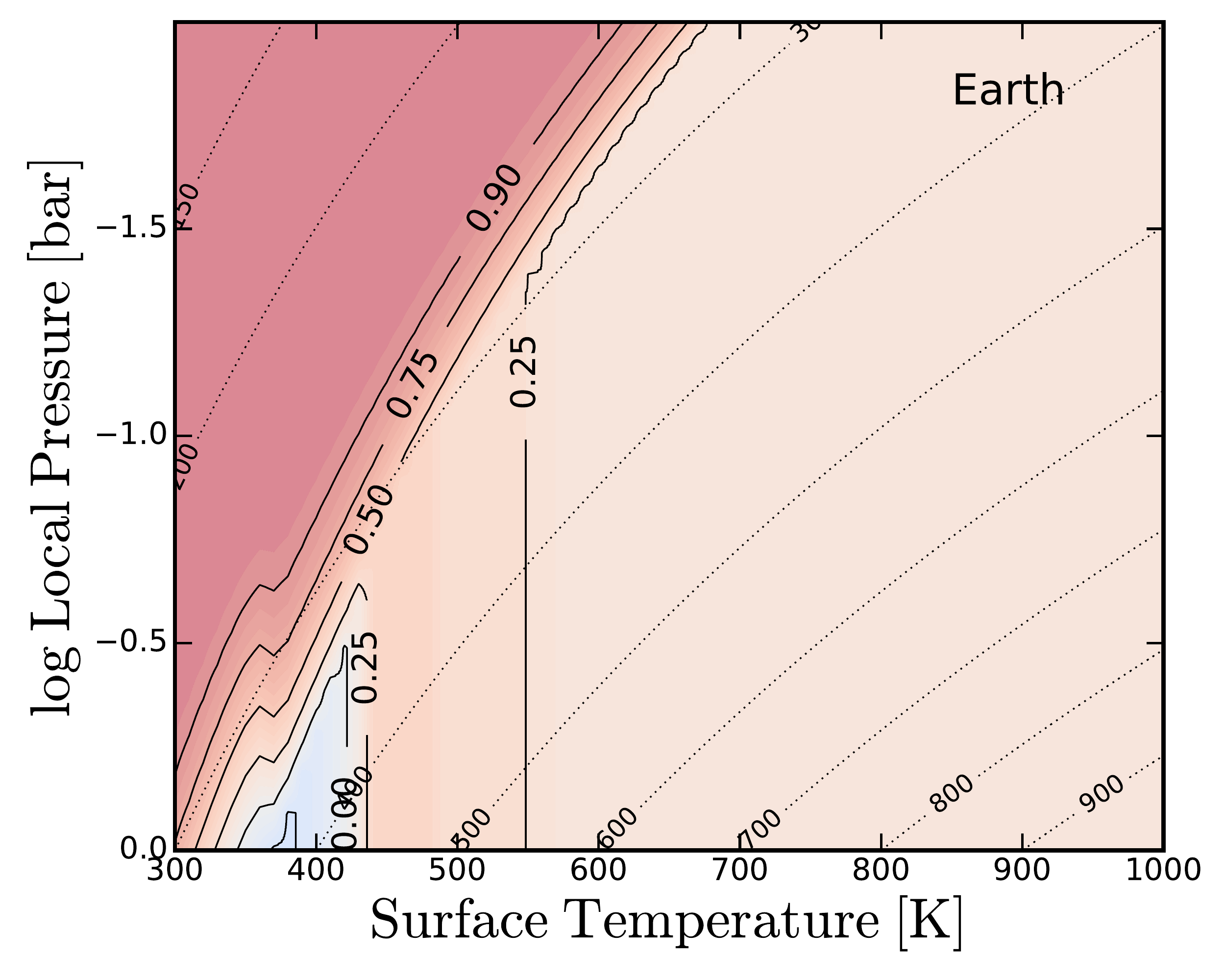}
\includegraphics[width = .32\linewidth, page=1]{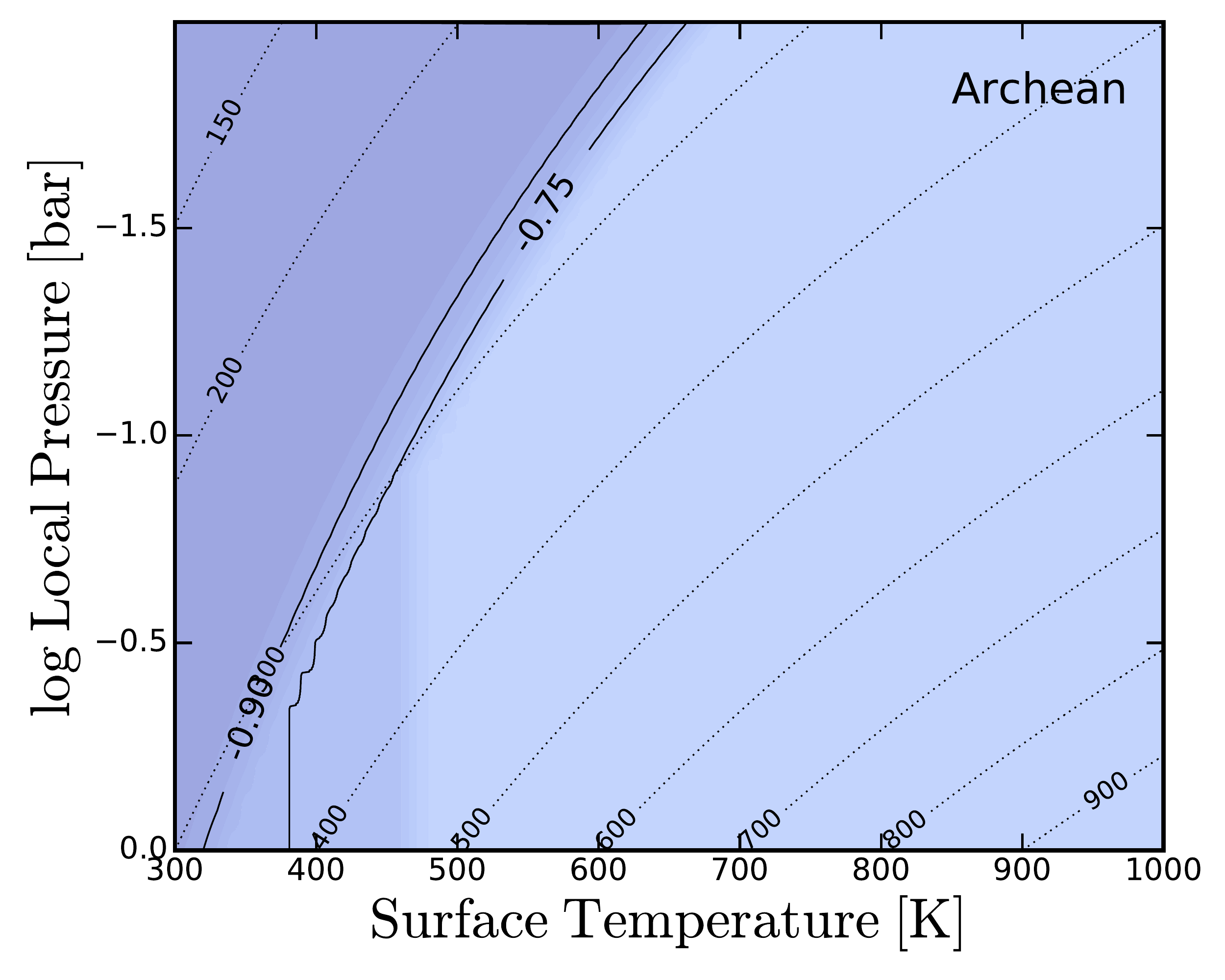}
\includegraphics[width = .32\linewidth, page=1]{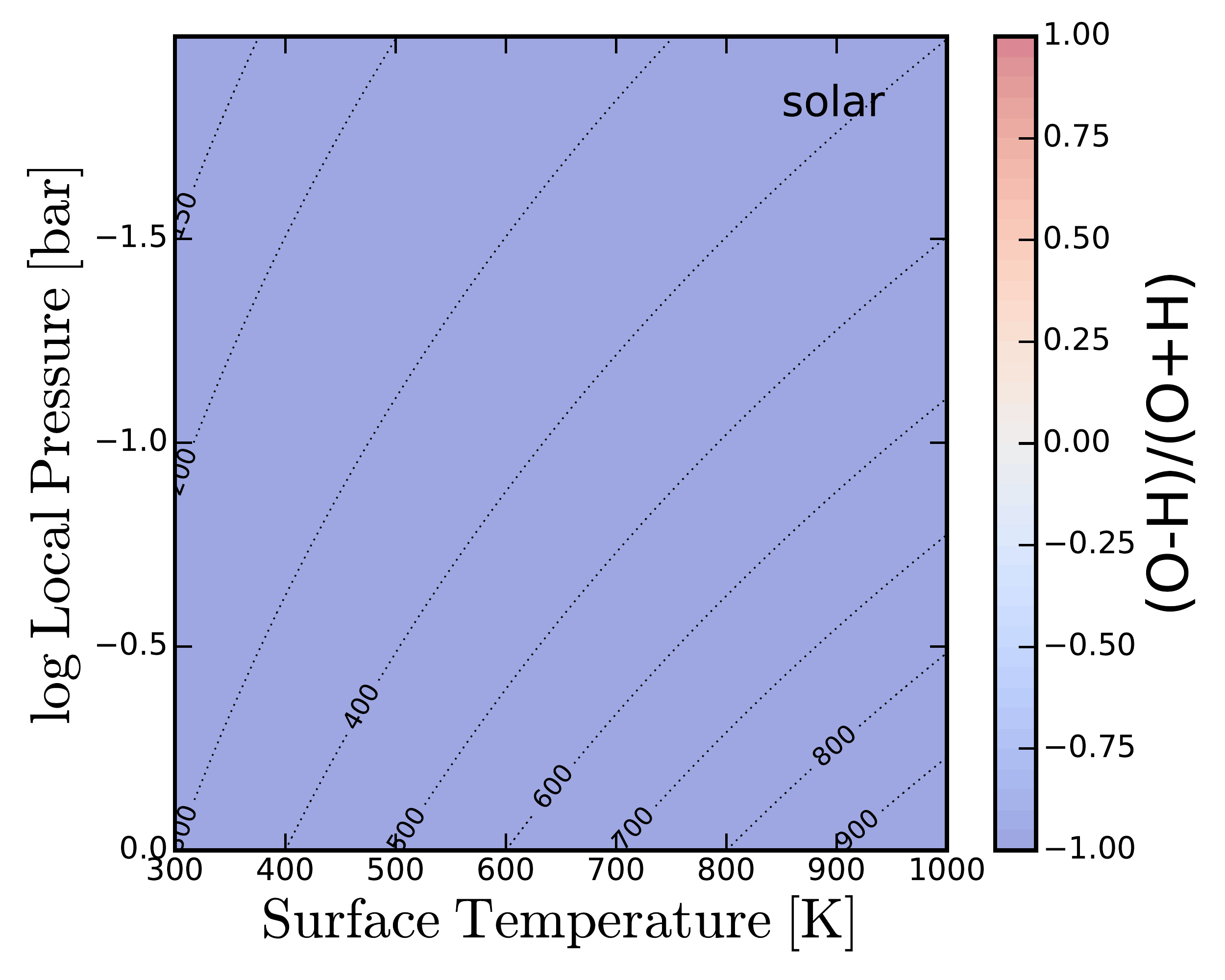}
\caption{\textbf{Comparison of the gas phase O and H abundances, expressed as the (O-H)/(O+H) ratio. Oxygen is dominant in the red regions, whereas carbon dominance is indicated in blue regions. As in Fig.~\ref{fig:CloudsAll-10} the dotted lines indicate the local gas temperatures.}}
\label{fig:AtmoOH}
\end{figure*}
\newpage

\begin{figure*}
\centering
\includegraphics[width = .32\linewidth, page=2]{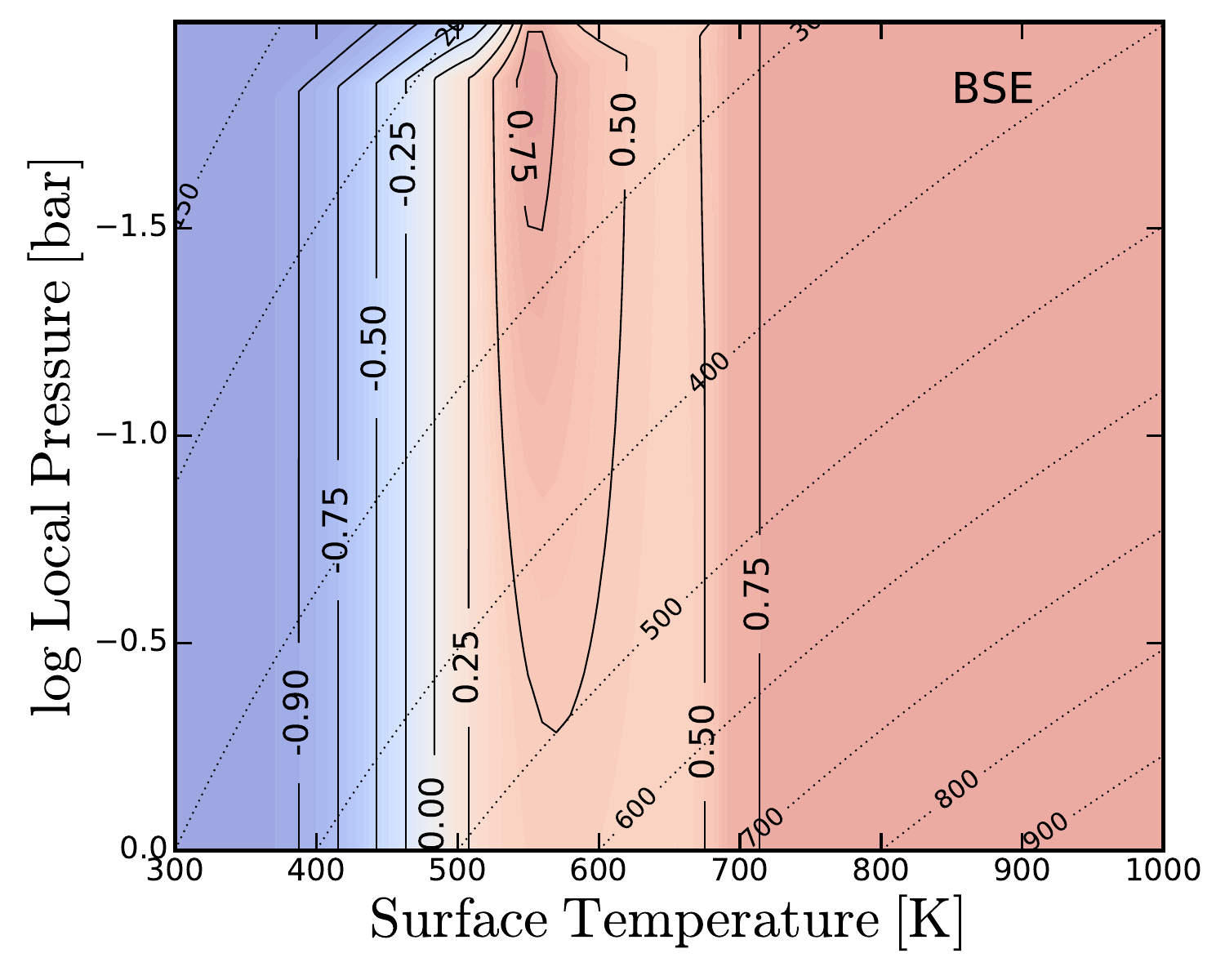}
\includegraphics[width = .32\linewidth, page=2]{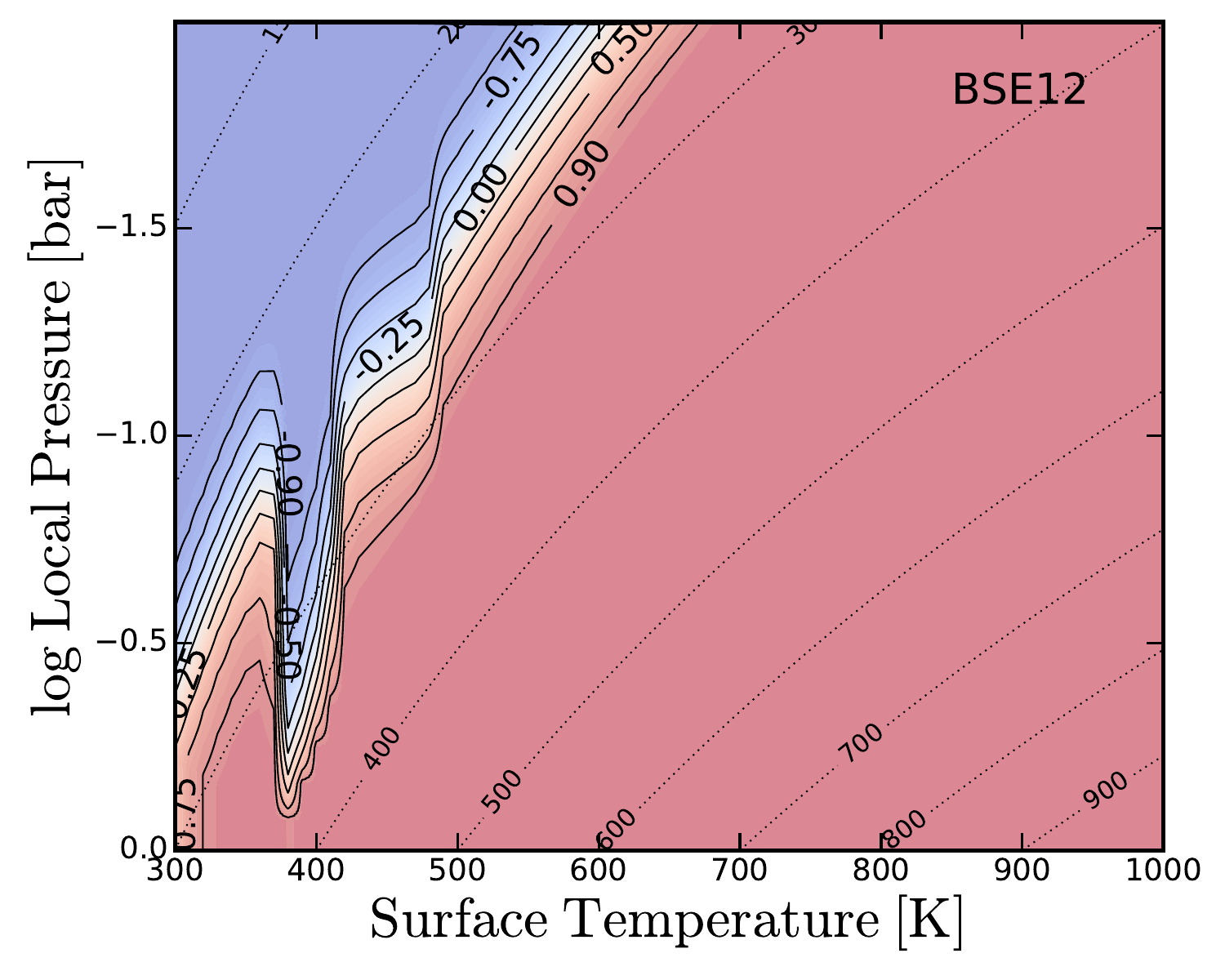}
\includegraphics[width = .32\linewidth, page=2]{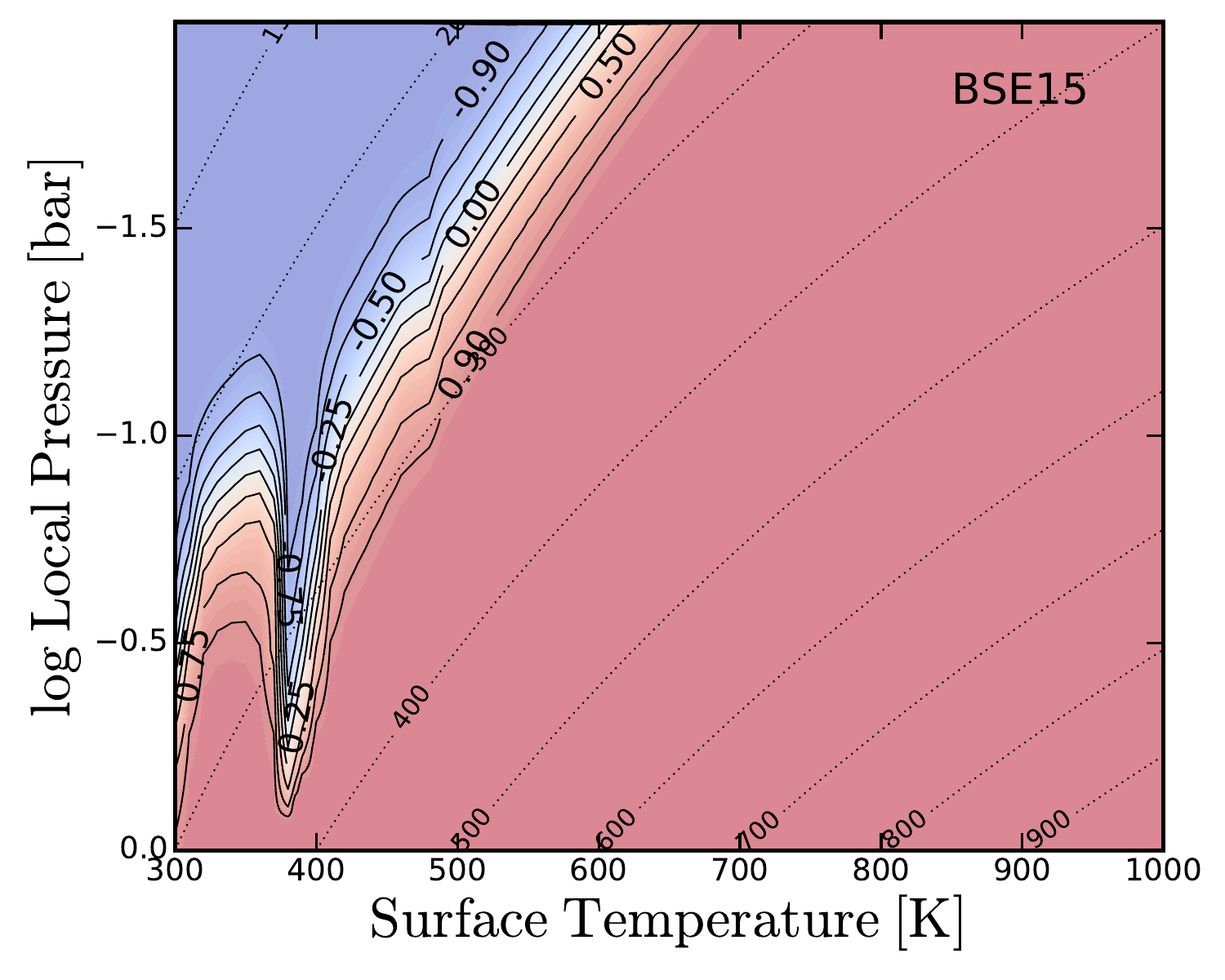}\\
\includegraphics[width = .32\linewidth, page=2]{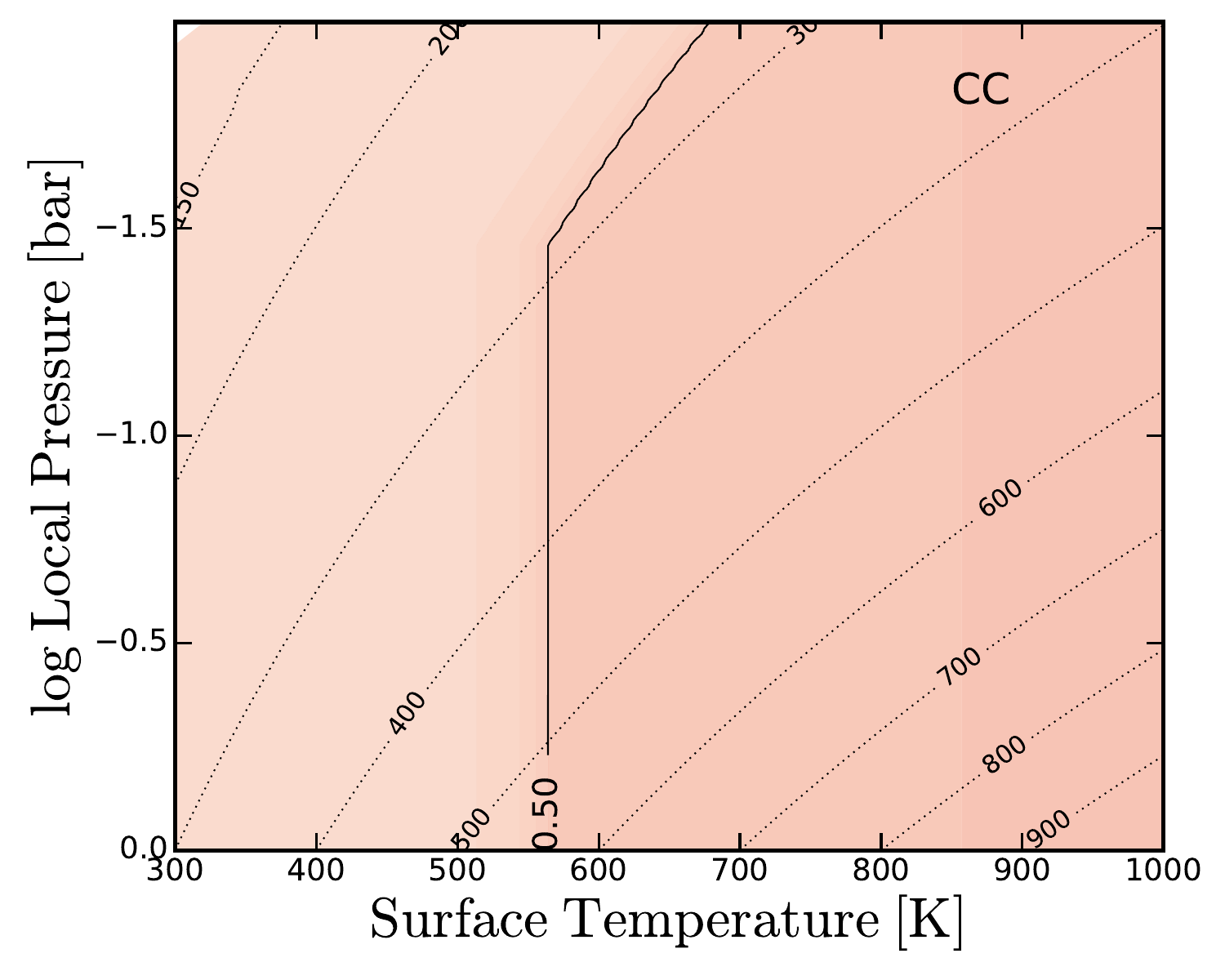}
\includegraphics[width = .32\linewidth, page=2]{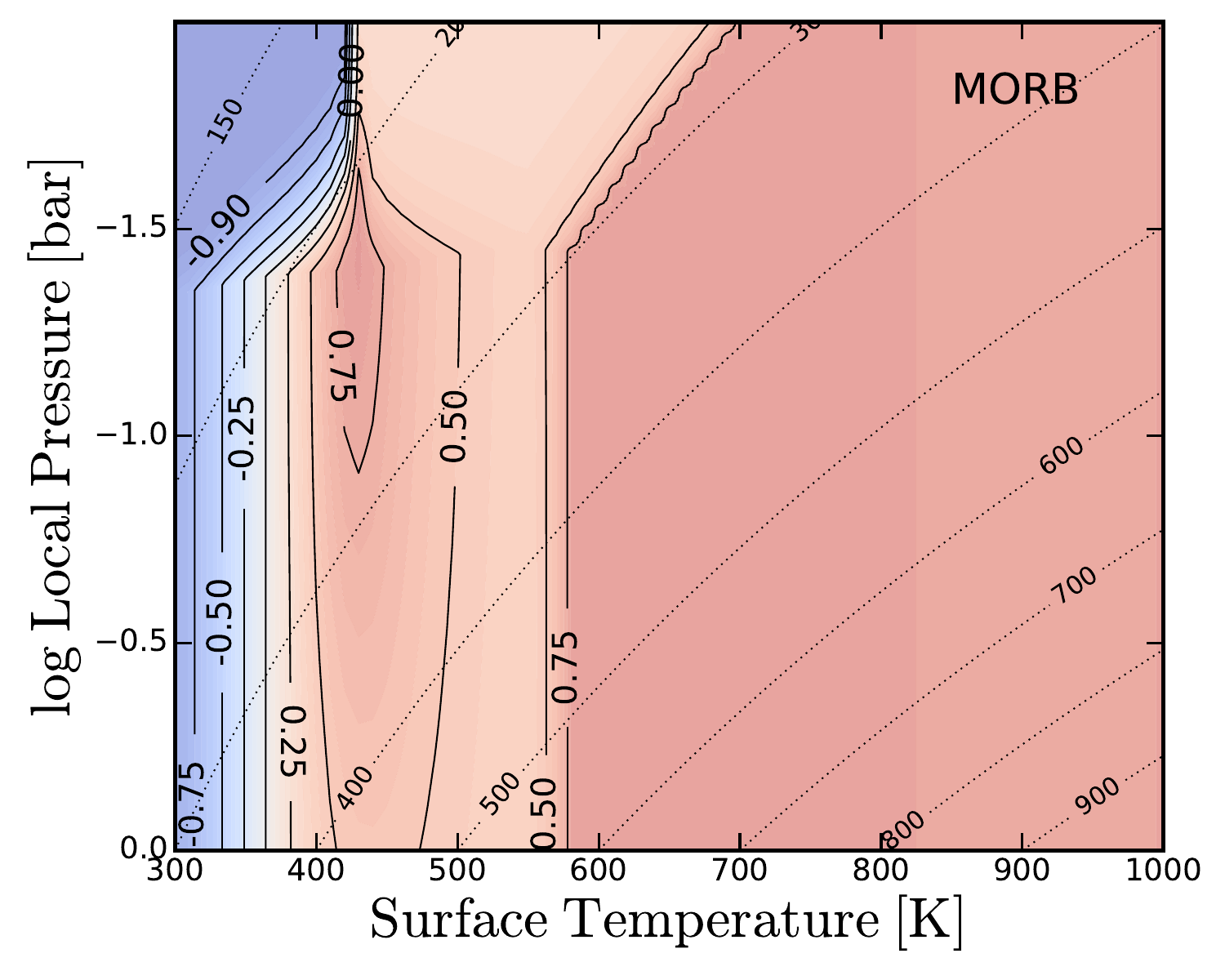}
\includegraphics[width = .32\linewidth, page=2]{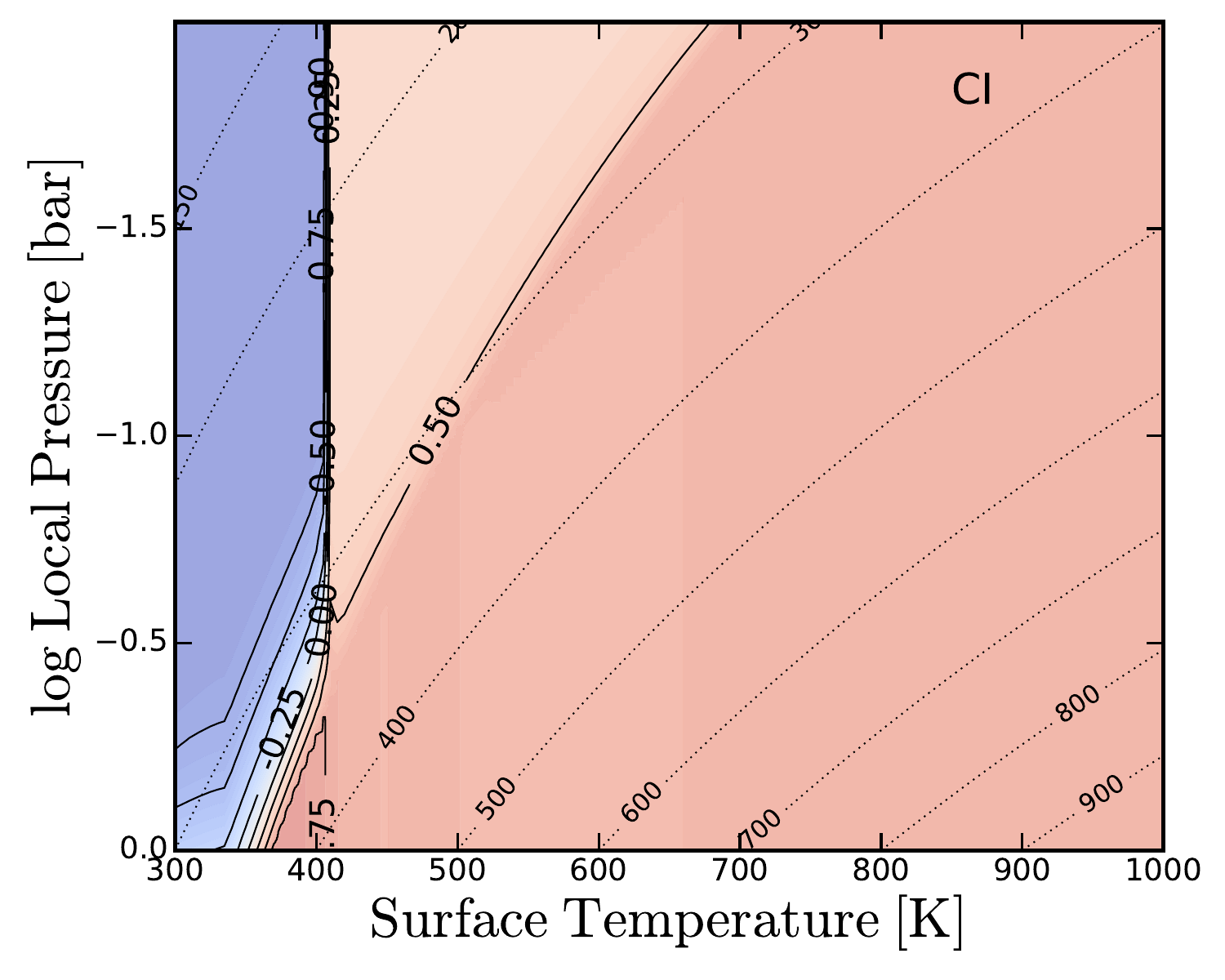}\\
\includegraphics[width = .32\linewidth, page=2]{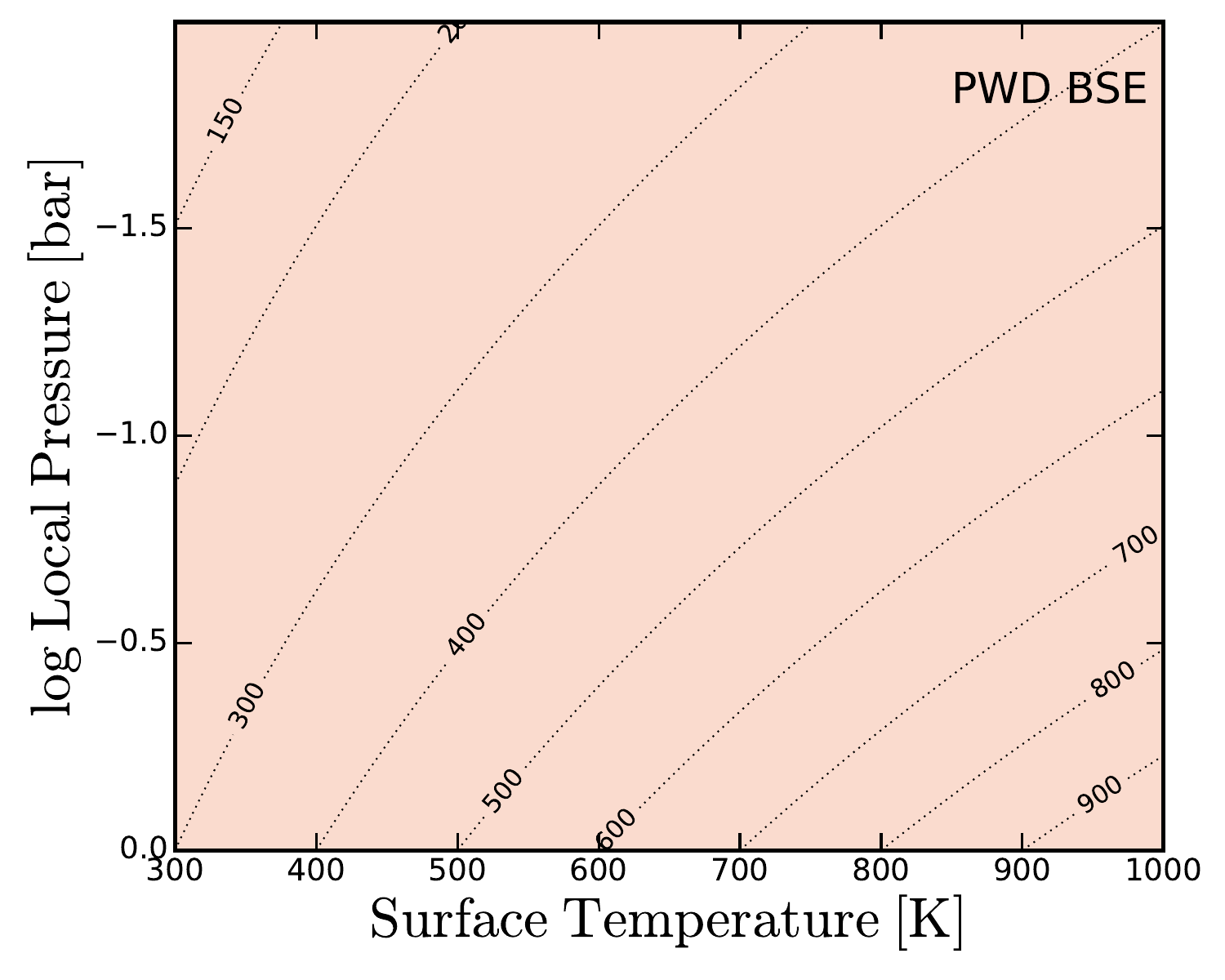}
\includegraphics[width = .32\linewidth, page=2]{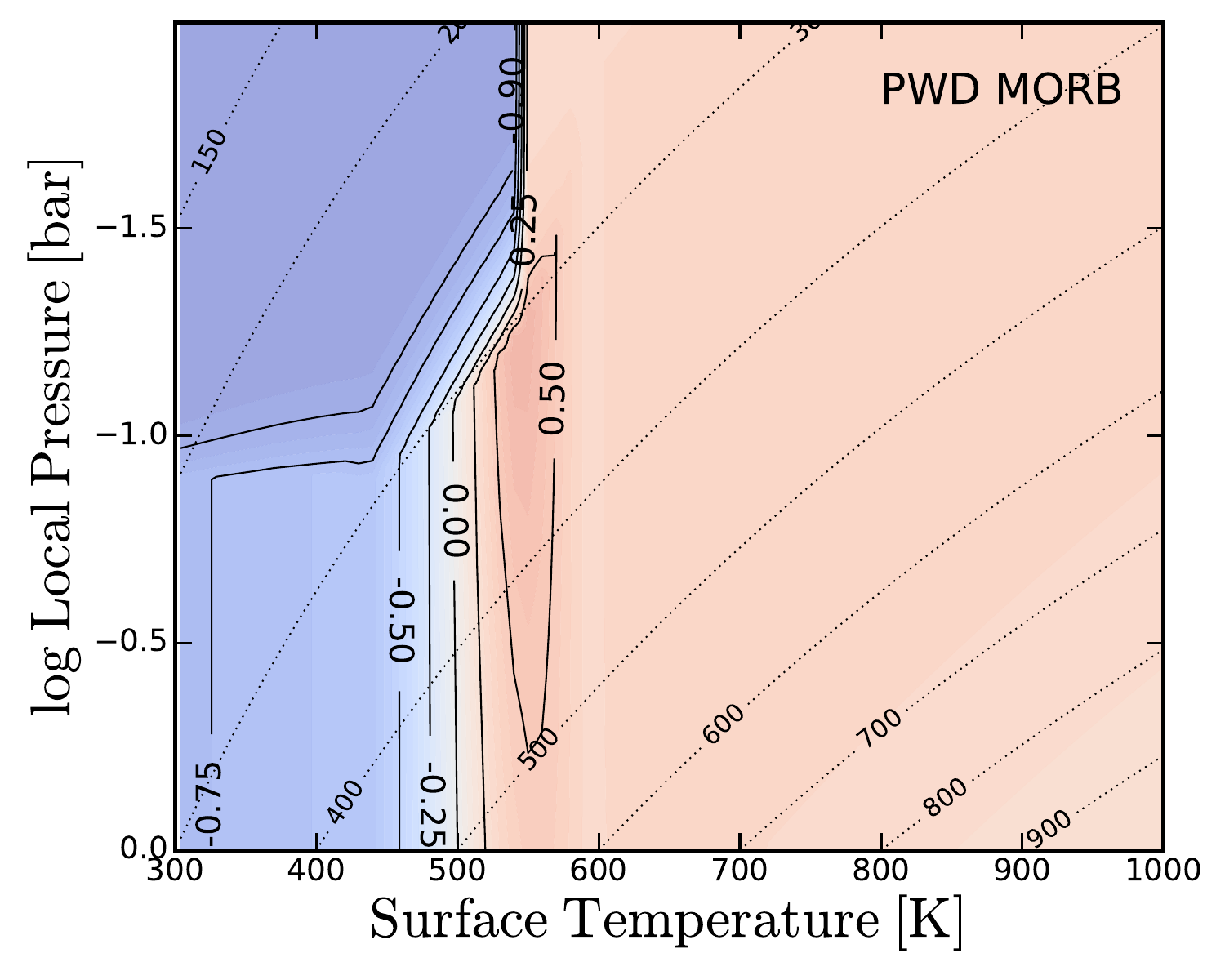}
\includegraphics[width = .32\linewidth, page=2]{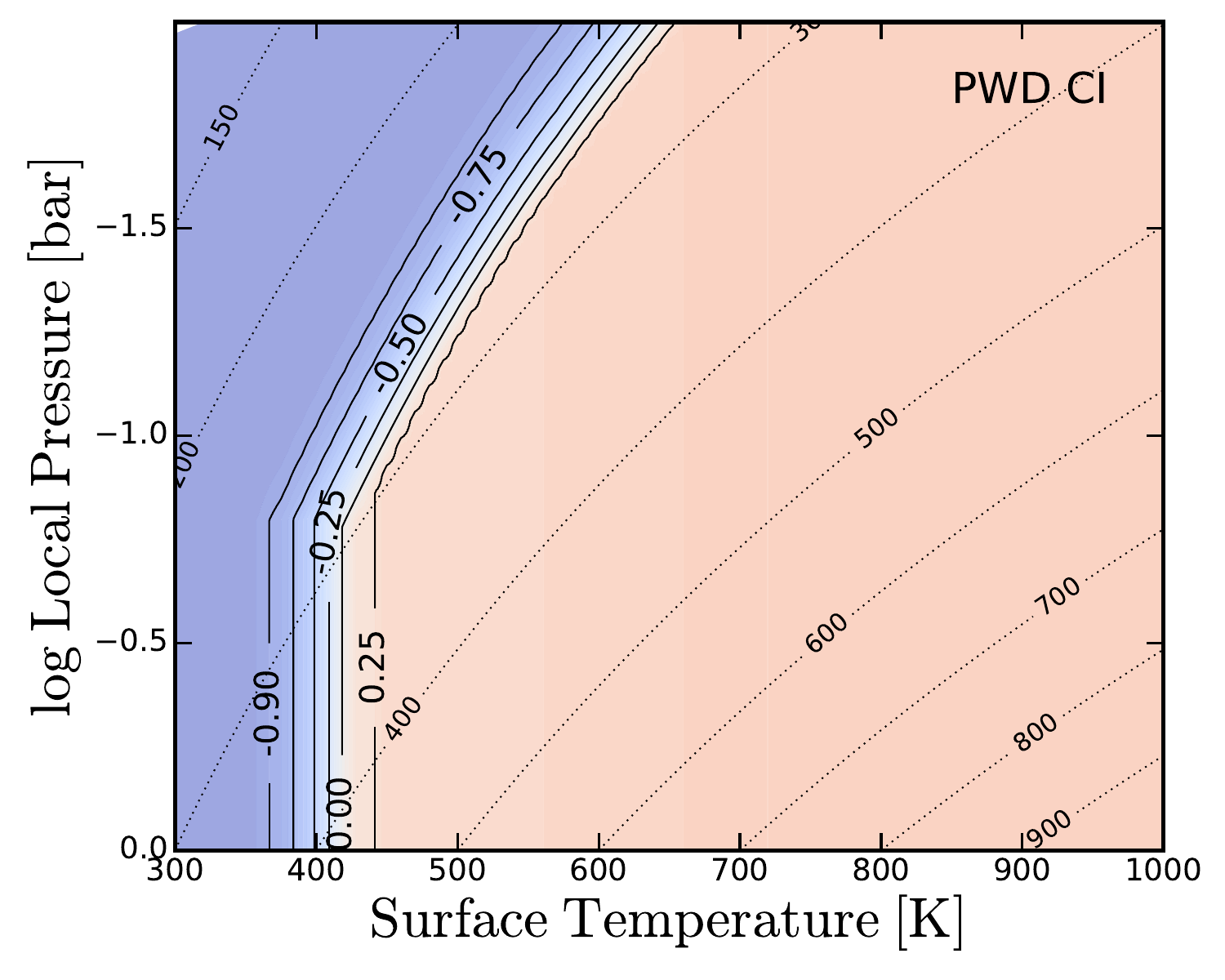}\\
\includegraphics[width = .32\linewidth, page=2]{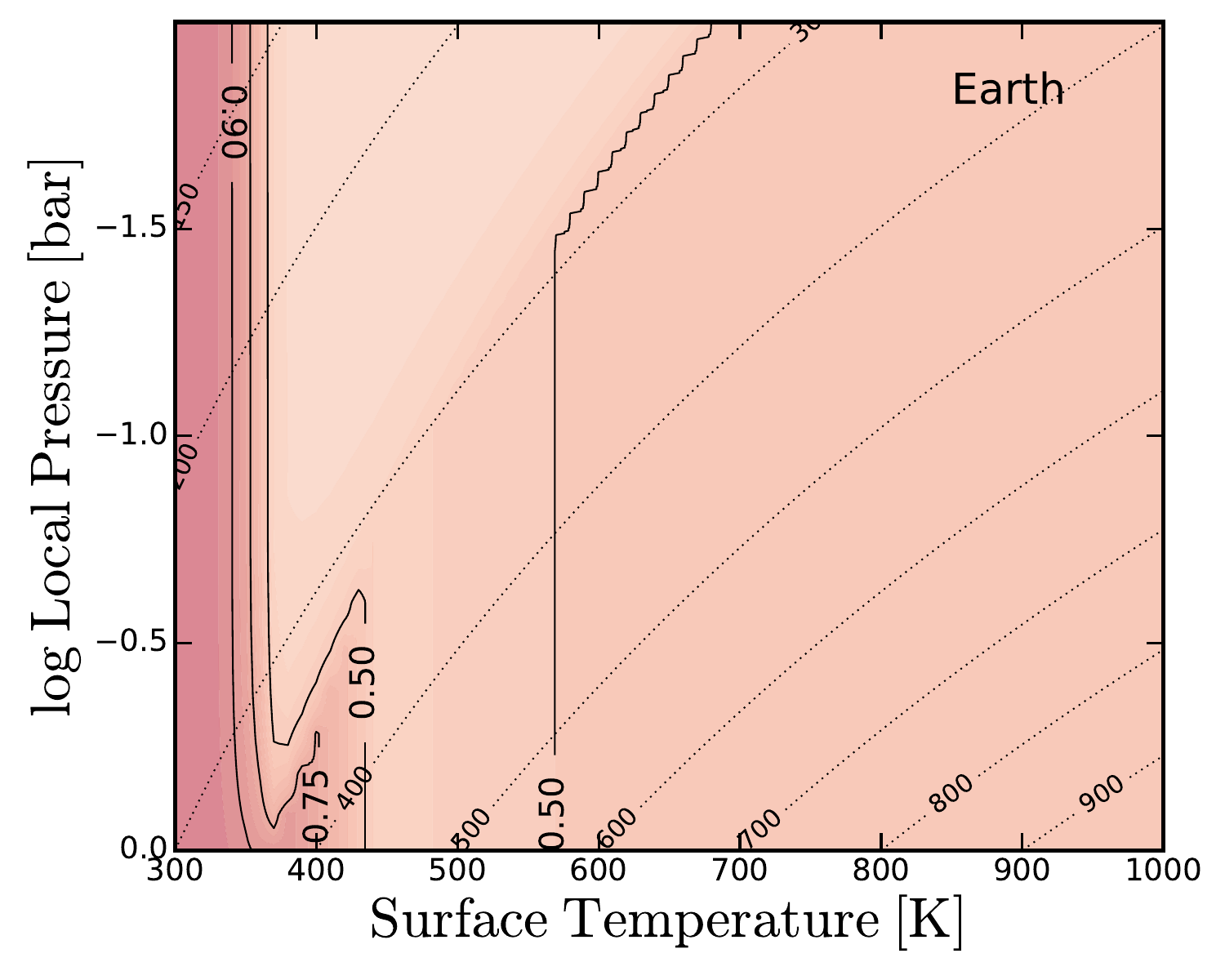}
\includegraphics[width = .32\linewidth, page=2]{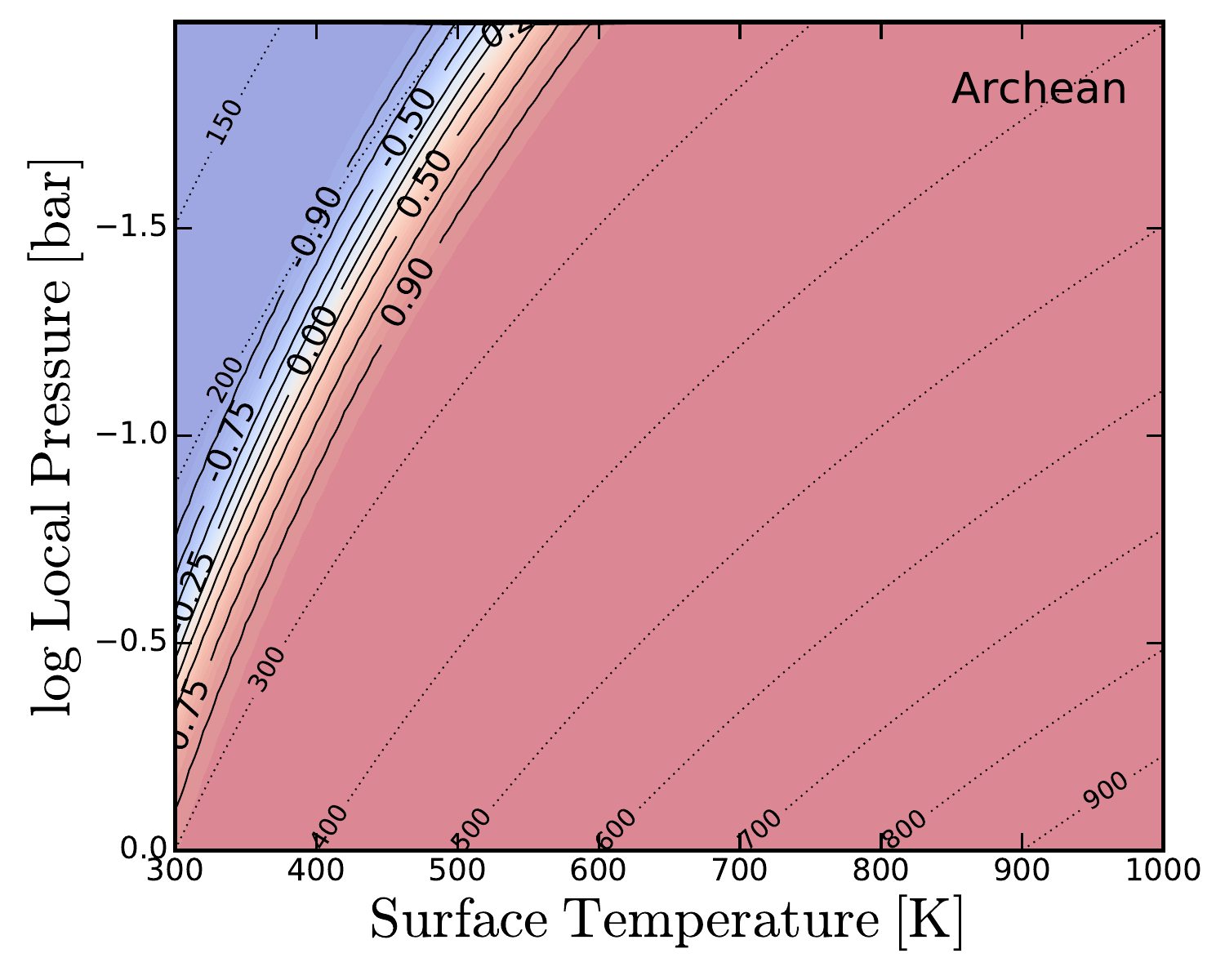}
\includegraphics[width = .32\linewidth, page=2]{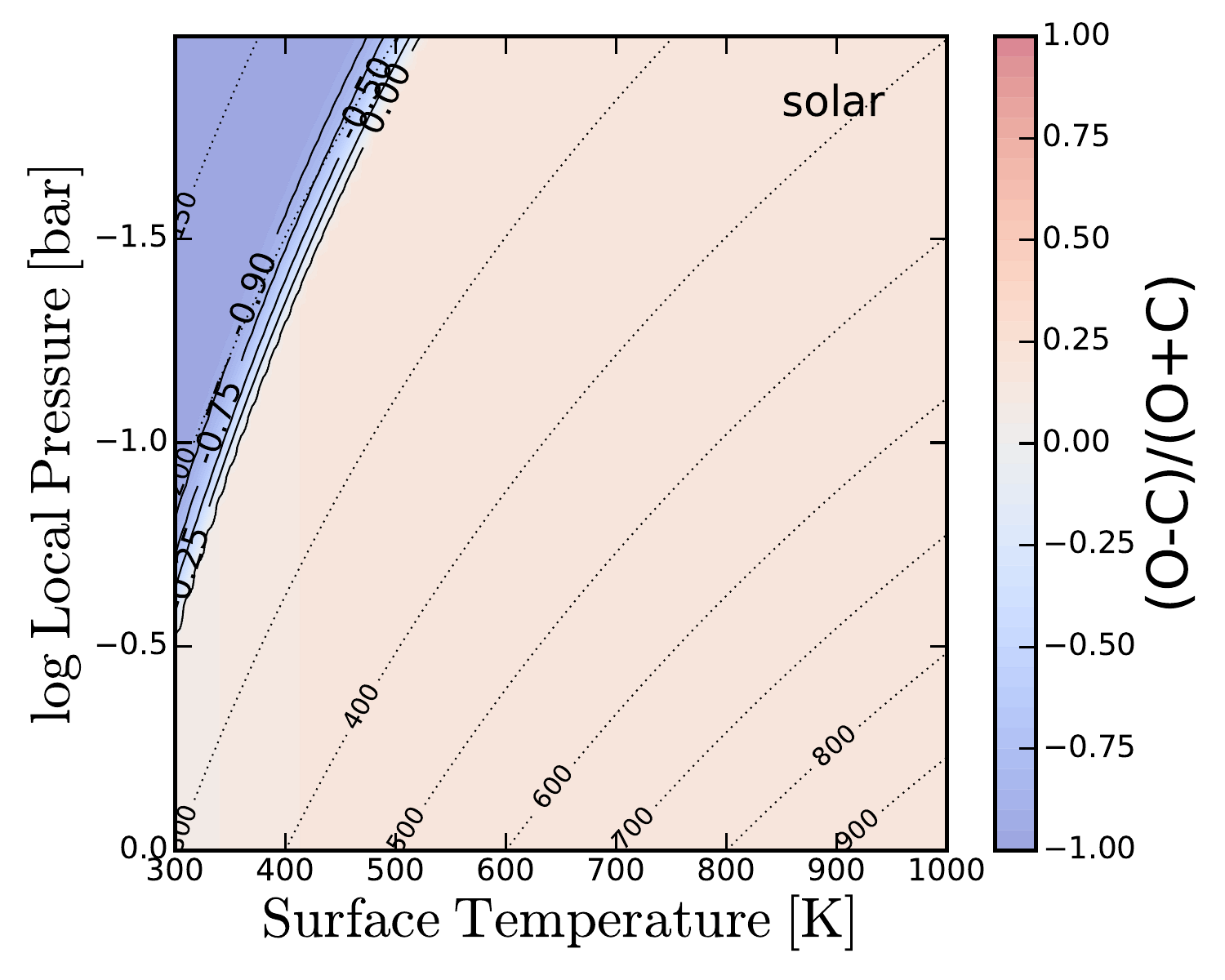}
\caption{\textbf{Comparison of the gas phase O and C abundances, expressed as the (O-C)/(O+C) ratio. Oxygen is dominant in the red regions, whereas carbon dominance is indicated in blue regions. The colour scheme is consistent for all panels. As in Fig.~\ref{fig:CloudsAll-10} the dotted lines indicate the local gas temperatures.}}
\label{fig:AtmoOC}
\end{figure*}

}
\end{appendix}
\end{document}